\documentclass[11pt,a4paper]{article}
\pdfoutput=1
\usepackage{jheppub}
\usepackage[utf8]{inputenc}

\usepackage{color}
\usepackage[dvipsnames]{xcolor}
\usepackage{soul}
\usepackage{amsmath}
\usepackage{amssymb}
\usepackage{mathtools}
\usepackage{esint}
\usepackage{pifont}
\usepackage{bbold}
\usepackage{graphbox}

\usepackage{feynmf}

\usepackage{bbm}
\usepackage{verbatim}   
\usepackage{subfigure}
\usepackage{caption}
\usepackage{acronym}
\usepackage[export]{adjustbox}

\usepackage{amsfonts}
\usepackage{mathrsfs}
\usepackage{graphicx}
\usepackage{multirow}
\usepackage{slashed}

\usepackage{caption}

\newcommand{\newc}{\newcommand}

\DeclareMathOperator\arctanh{arctanh}

\definecolor{mypurple}{RGB}{164,64,214}

\newc{\pd}[1]{{\bf\color{blue}~#1}}
\newc{\bl}[1]{{\it\color{purple}#1}}

\newc{\fpi}{f_{\pi}}
\newc{\etap}{\eta^{\prime}}
\newc{\llll}{\langle\lambda\lambda\rangle}
\newc{\FFd}{F^a\tilde F^a}
\newc{\qbar}{{\overline q}}
\newc{\TR}{{\rm Tr}}
\newc{\Kahler}{K\"ahler }
\newc{\Zbb}{{\mathbb Z}}
\newc{\Rt}{{\mathbb R}^3}
\newc{\Rf}{{\mathbb R}^4}
\newc{\So}{{\mathbb S}^1}
\newc{\zt}{{\mathbb Z}_2}
\newc{\RtSo}{{\mathbb R}^3\times{\mathbb S}^1}
\newc{\scriminus}{{\cal I}^-}
\newc{\scriplus}{{\cal I}^+}
\newc{\mpl}{M_p}
\newc{\Ricci}{\mathcal{R}}
\newc{\bv}{\phi}
\newc{\calU}{{\cal U}}
\newc{\calK}{K}
\newc{\calUi}{{\cal U}^{-1}}
\newc{\calG}{{\cal G}}
\newc{\calM}{{\cal M}}
\newc{\calL}{{\cal L}}
\newc{\calO}{{\cal O}}
\newc{\calR}{{\cal R}}
\newc{\calQ}{{\cal Q}}
\newc{\calI}{{\cal I}}
\newc{\calOb}{{\cal O}^\dagger}
\newc{\hphi}{{\hat\phi}}

\newcommand{\abs}[1]{\ensuremath{\left|#1\right|}}

\renewcommand{\eqref}[1]{Eq.~(\ref{#1})}
\DeclareMathOperator\arcsinh{arcsinh}

\title{de Sitter Decays to Infinity}

\author[a]{Patrick Draper,}
\emailAdd{pdraper@illinois.edu}
\author[b]{Isabel Garcia Garcia,}
\emailAdd{isabel@kitp.ucsb.edu}
\author[a]{and Benjamin Lillard}
\emailAdd{blillard@illinois.edu}

\affiliation[a]{Department of Physics, University of Illinois, Urbana, IL 61801}
\affiliation[b]{Kavli Institute for Theoretical Physics, University of California, Santa Barbara, CA 93106, USA}

\abstract{
Bubbles of nothing are a class of vacuum decay processes present in some theories with compactified extra dimensions. We investigate the existence and properties of bubbles of nothing in models where the scalar pseudomoduli controlling the size of the extra dimensions are stabilized at  positive vacuum energy, which is a necessary feature of any realistic model. We map the construction of bubbles of nothing to a four-dimensional Coleman-De Luccia problem and establish necessary conditions on the asymptotic behavior of the scalar potential for the existence of suitable solutions. We perform detailed analyses in the context of five-dimensional theories with metastable $\text{dS}_4 \times S^1$ vacua, using analytic approximations and numerical methods to calculate the decay rate. We find that bubbles of nothing sometimes exist in potentials with no ordinary Coleman-De Luccia decay process, and that in the examples we study, when both processes exist, the bubble of nothing decay rate is typically faster. Our methods can be generalized to other stabilizing potentials and internal manifolds.
}

\begin{document}

\maketitle

\section{Introduction}
\label{sec:intro}

Our Universe is entering a phase of dark energy domination. To date, cosmological observations are consistent with an epoch of late-time acceleration driven by a small and positive cosmological constant \cite{Perlmutter:1998np,Percival:2009xn,Aghanim:2018eyx}. 

Although an epoch of accelerated expansion can be easily accommodated within general relativity, building full-fledged de Sitter vacua in the context of string theory --- while seemingly not impossible \cite{Kachru:2003aw,Balasubramanian:2005zx} --- has proven surprisingly hard \cite{Maldacena:2000mw,Townsend:2003qv,Hertzberg:2007wc,Wrase:2010ew,Shiu:2011zt,Bena:2014jaa,Kutasov:2015eba,Andriot:2016xvq,Moritz:2017xto,Sethi:2017phn,Danielsson:2018ztv,Hamada:2018qef,Hamada:2019ack,Carta:2019rhx,Gao:2020xqh,Basile:2020mpt}. This has led to conjecture and debate about whether metastable de Sitter vacua can exist in consistent theories of quantum gravity \cite{Obied:2018sgi,Garg:2018reu,Ooguri:2018wrx}, and if so, whether there is an upper bound on their lifetimes \cite{Bedroya:2019snp}. For example, it is well-known that in theories with compactified extra dimensions, if the effective potential for the geometric moduli admits minima of positive energy density, these minima are at best local. In turn, a four-dimensional de Sitter vacuum may `spontaneously decompactify,' decaying into a Universe with zero vacuum energy and spatial extra dimensions of infinite size \cite{Dine:1985he,Giddings:2003zw,Giddings:2004vr}. Depending on the details of the higher-dimensional construction, additional decay channels can also be present and are generally described by either the Coleman-De Luccia (CDL) \cite{Coleman:1980aw} or Hawking-Moss (HM) \cite{Hawking:1981fz} instantons.

\medskip

In this paper, we focus on an alternative instability: the generalization of Witten's bubble of nothing (BON)~\cite{Witten:1981gj} to spacetimes with four-dimensional de Sitter factors and compact internal manifolds with stabilized moduli.
Witten's bubble describes an instability of the Kaluza-Klein (KK) vacuum $\mathbb{M}^4 \times S^1$. The gravitational instanton is the five-dimensional Euclidean Schwarzschild solution, and the endpoint of the decay is the destruction of the spacetime. Via dimensional reduction, Witten's bubble can be recast as a solution to a CDL problem of a real scalar field minimally coupled to gravity in four dimensions. The scalar field plays  the role of the radial modulus setting the size of the $S^1$, and Witten's solution is recovered in the case of a vanishing scalar potential. This four-dimensional description, however, makes it natural to consider generalizations of the BON in the presence of a potential for the moduli, as first suggested in \cite{Dine:2004uw}. In this way, non-perturbative instabilities associated with moduli stabilization, including generalizations of the bubble of nothing, can be studied from the bottom-up.

A number of studies have explored relatives of Witten's bubble from a top-down perspective, including in string compactifications~\cite{Fabinger:2000jd,Brill:1991qe,DeAlwis:2002kp,Acharya:2019mcu,GarciaEtxebarria:2020xsr}, in higher-dimensional models~\cite{Dibitetto:2020csn}, and in flux compactifications with stabilized moduli~\cite{Horowitz:2007pr,BlancoPillado:2010df,Brown:2010mf,BlancoPillado:2010et,Blanco-Pillado:2016xvf,Ooguri:2017njy}. In particular, Refs.~\cite{BlancoPillado:2010et,Brown:2010mf} consider BON instabilities in 6D models with explicit mechanisms for moduli stabilization at a positive value of the vacuum energy.

Our approach is complementary to previous work, attempting to factorize as much as possible the specific details of moduli stabilization from the existence and construction of  bubbles of nothing. Our goals are to describe, from the four-dimensional perspective, existence criteria for BON solutions in theories with stabilized moduli; to develop analytic approximations for the BON action in the presence of a scalar potential;  to compare the corresponding decay rate to those of the ordinary CDL and HM channels; and to study numerically those examples that are analytically intractable. This paper is a longer, more detailed companion to~\cite{Draper:2021gmq}, where some of the main ideas and techniques were presented and illustrated with an eye toward minimal, streamlined presentation. Here we review and extend the results of~\cite{Draper:2021gmq}, deriving some of the assertions, expanding the  range of models studied, elaborating on the broader context of the work, and undertaking a more extensive numerical survey.

 We perform our most detailed analysis in the context of a five-dimensional  model with a range of moduli potentials. Depending on the  features of the  potential, the BON may be the only accessible instability, or it may exist alongside more conventional CDL or HM bounces. In the latter case, the BON decay rate is always the fastest in the examples we study.

Unlike the familiar CDL and HM solutions, the CDL solution corresponding to Witten's bubble exhibits singular boundary conditions at the center of the bounce, with divergent values for the scalar and its derivatives. In the ordinary CDL problem, the scalar field satisfies equations of motion analogous to those of a particle moving on an inverted potential. The solution can then be found via an overshoot-undershoot method, with the `particle' starting at rest at the center of the bounce, and asymptotically rolling over to the (inverted) false vacuum. This picture can be extended to bounces that satisfy BON boundary conditions. In this case, the particle does not begin at rest, but rather it starts moving  `from infinite distance with infinite velocity,' corresponding to the singular boundary conditions. Implementing an overshoot-undershoot algorithm appropriate to accommodate these boundary conditions allows us to find BON instabilities of four-dimensional de Sitter vacua numerically, verifying and going beyond our analytic results.

\medskip

This paper is organized as follows. In Section~\ref{sec:review}, we review the CDL formalism, the properties of Witten's bubble most relevant to our  analysis, and some of the familiar sources of moduli potentials that can  be present in higher-dimensional models. In Section~\ref{sec:bounceofnothing} we study BON decays of four-dimensional de Sitter vacua. We find necessary conditions on the asymptotic form of the scalar potential for the CDL equations to admit a BON bounce solution, and, under some assumptions, we obtain an approximate analytic solution for the bounce. We evaluate the action of these instantons, showing that it remains finite in the limit of vanishing false vacuum energy density, and that for a large class of models it is mostly independent of the behavior of the scalar potential in the compactification regime. We  discuss a qualitatively different type of instanton that also satisfy BON boundary conditions in Section~\ref{sec:exotic}, and argue that these may be thought of as superpositions of a BON together with an ordinary CDL or HM bounce. In Section~\ref{sec:numerics} we present a detailed numerical study of BON bounce solutions and their Euclidean actions for a wide range of parameter space. We conclude in Section~\ref{sec:conclusions} and outline directions for future work. A series of Appendices collect various additional results of a more  detailed or tangential nature and are referred to throughout the text.

\section{Vacuum Decay in the Presence of Gravity}
\label{sec:review}

In Section~\ref{sec:CDL}, we summarize the CDL formalism describing vacuum decay in the presence of gravity. We review the properties of Witten's BON instability that will be most relevant to our work in Section~\ref{sec:BONasCDL}, with special focus on how it can be recast as the solution to a four-dimensional CDL problem, as first done in \cite{Dine:2004uw}. In Section~\ref{sec:potentials} we discuss several sources of moduli potentials that are naturally present in theories with extra dimensions. We conclude  with some qualitative observations regarding the interpretation of moduli potentials in Section~\ref{sec:potentials_comments}.

\begin{figure}
\centering
\includegraphics[width=0.8\textwidth]{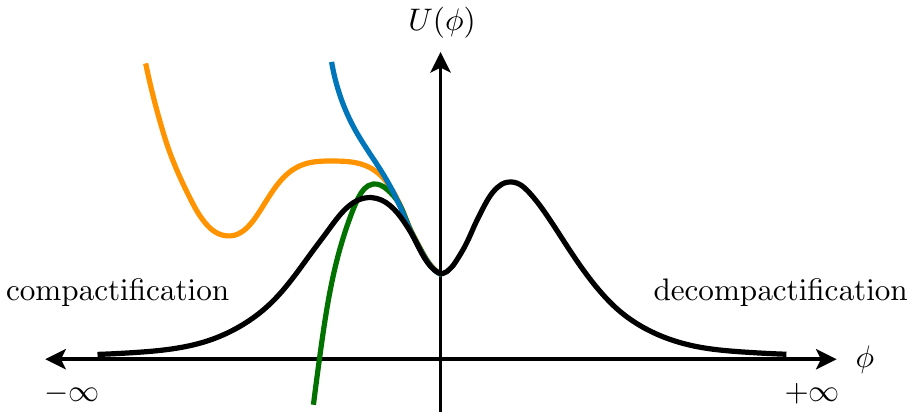}
\caption{
Theories featuring compactified extra dimensions, and moduli spaces that are stabilized at a positive value of the vacuum energy, are typically susceptible to decay into a higher-dimensional, decompactified theory, \cite{Dine:1985he,Giddings:2003zw,Giddings:2004vr}, as indicated by the $\phi \geq 0$ region in this figure. On the other hand, the behavior of the scalar potential in the compactification limit may vary widely depending on the details of the higher-dimensional construction. Our work will be concerned with instabilities of four-dimensional de Sitter vacua into the compactification regime, where $\phi \leq 0$. The location of the de Sitter false vacuum is  arbitrary, and we choose $\phi_{\rm fv} \equiv 0$ for convenience.}
\label{fig:compdecomp}
\end{figure}

As emphasized in the Introduction, our focus will be on instabilities of theories with compactified dimensions, stabilizing potentials for the geometric moduli, and local de Sitter vacua. Fig.~\ref{fig:compdecomp} shows the qualitative picture that we will have in mind throughout our work.

\subsection{Coleman-De Luccia formalism}
\label{sec:CDL}

The semiclassical treatment of vacuum decay in quantum field theory was developed in \cite{Coleman:1977py,Callan:1977pt}. To leading order, the decay rate of the false vacuum is given by the action of the $O(4)$-symmetric solution to the Euclidean field equations known as the ``bounce." In Euclidean signature the bounce interpolates between true and false vacuum, and via analytic continuation, it also provides the initial condition and real time evolution of true vacuum bubbles after nucleation. The decay rate per unit volume of the false vacuum is given by
\begin{equation}
	\Gamma \sim v^4 e^{- \Delta S} ,
\end{equation}
where $v$ is a typical energy scale, and $\Delta S$ is the action of the bounce minus the action of the Euclidean false vacuum.

Coleman and De Luccia further developed the formalism of \cite{Coleman:1977py,Callan:1977pt} to include the effects of gravitation~\cite{Coleman:1980aw}. In four dimensions, the $O(4)$ symmetry of the bounce restricts the form of the relevant metrics to those compatible with the form:
\begin{equation}
ds_4^2 = d\xi^2+\rho(\xi)^2 d\Omega_3^2 ,
\label{eq:cdlmetric}
\end{equation}
where $\xi$ is a radial coordinate and $\rho$ sets the curvature radius of the transverse 3-sphere. 
For a real scalar field minimally coupled to gravity, the corresponding field equations read
\begin{gather}
	\label{eq:EOM_phi} \phi''+\frac{3\rho'}{\rho}\phi' = \frac{\partial U (\phi)}{\partial \phi} , \\
	\label{eq:EOM_rho} \rho'^2 = 1+\frac{\rho^2}{6 M_p^2}\left(\phi'^2-2U(\phi)\right) .
\end{gather}
The relevant boundary conditions accompanying Eqs.~(\ref{eq:EOM_phi}) and (\ref{eq:EOM_rho}) depend on the topology of the solution. For a de Sitter false vacuum, and a scalar potential that is $\geq 0$, the solution has the topology of an $S^4$. In this case, $\rho$ vanishes at two values of $\xi$, corresponding to a minimum and a maximum of the radial coordinate~\cite{Guth:1982pn}. Without loss of generality, we can choose $\xi_\text{min} \equiv 0$. Then, $\xi \in [0, \xi_{\rm max}]$, and $\xi_{\rm max}$ is defined via $\rho (\xi_{\rm max}) \equiv \rho(0) = 0$. Any non-singular solution to the  equations of motion must then also obey the boundary conditions $ \phi'(0) = \phi'(\xi_{\rm max}) = 0$.

For example, the Euclidean solution describing the de Sitter false vacuum is given by
\begin{equation}
	\phi_{\rm dS}(\xi) \equiv 0 , \qquad {\rm and} \qquad
	\rho_{\rm dS} (\xi) = \Lambda \sin \left( \frac{\xi}{\Lambda} \right) ,
\label{eq:CDL_dS}
\end{equation}
with $\Lambda \equiv \sqrt{3 M_p^2 / U_\text{fv}}$. The traditional CDL bounce that describes false vacuum decay is characterized by an internal region --- the inside of the bubble --- where $\phi$ lies in the vicinity of the true vacuum, and a transition to an exterior region where $\phi$ is near its value in the metastable phase. If false and true vacua are nearly degenerate, the bounce is well described within the ``thin-wall" approximation, where the bubble wall thickness (the region over which the field transitions from false to true vacuum) is much smaller than its overall radius.
For a Minkowski true vacuum, the curvature radius of  a thin-wall bounce is given by
\begin{equation}
	\bar \rho \simeq \frac{\rho_0}{ 1 + (\rho_0 / 2 \Lambda)^2 } , \qquad {\rm with} \qquad \rho_0 = \frac{3 \sigma}{U_\text{fv}} ,
\label{eq:rho_CDL}
\end{equation}
where $\sigma$ is the bubble wall tension given by $\sigma \simeq \int_{\phi_\text{tv}}^{\phi_\text{fv}} \! d\phi \sqrt{ 2 (U(\phi) - U(\phi_\text{fv}) ) }$ \cite{Coleman:1980aw}.\footnote{The thin-wall approximation is applicable provided $U_\text{fv} - U_\text{tv} \ll U_{\rm top} - U_\text{fv}$, with $U_{\rm top}$ the value of the scalar potential at the top of the potential barrier. To leading order, $\sigma \propto \sqrt{U_{\rm top}}$, and corrections involving the difference $U_\text{fv} - U_\text{tv}$ are dropped within the thin-wall approximation.}

The tunneling rate is controlled by the Euclidean action of the bounce. The action of a scalar field minimally coupled to Einstein gravity in four dimensions is given by
\begin{align}
	\label{eq:SE_general}	S_E	& =  \int d^4 x \sqrt{g} \left\{ - \frac{m^2_{\rm Pl}}{2} \mathcal{R} + \frac{1}{2} g^{\mu \nu} \partial_\mu \phi \partial_\nu \phi + U(\phi) \right\} \\
	\label{eq:SE_onshell}		& = - 2\pi^2 \int_0^{\xi_{\rm max}} d \xi \, \rho^3 U \qquad \qquad \text{(on-shell)} .
\end{align}
The second line only applies ``on-shell," that is, when evaluating the action on a solution to the Euclidean field equations consistent with the ansatz of Eq.~(\ref{eq:cdlmetric}).
In the thin-wall limit, the tunneling exponent of the CDL bounce is given by \cite{Coleman:1980aw}
\begin{equation}
	\Delta S_{\rm CDL} 	\equiv S_E \Big|_{\rm CDL} -  S_E \Big|_{\rm dS}
					\simeq \frac{B_0}{[1 + (\rho_0/ 2 \Lambda)^2 ]^2} ,
\label{eq:SE_CDL_general}
\end{equation}
with $B_0 \equiv 27 \pi^2 \sigma^4 / (2 U^3_{\rm fv})$. For a small false vacuum energy density, $U_\text{fv} \ll \sigma^2 / M_p^2$,
\begin{equation}
	\Delta S_{\rm CDL} \simeq \frac{24 \pi^2 M_p^4}{U_\text{fv}} .
\label{eq:SE_CDL}
\end{equation}

The CDL bounce is guaranteed to exist provided the scalar potential satisfies
\begin{align}
\left| \frac{\partial^2 U(\phi)}{\partial \phi^2}  \right|_{\phi = \phi_\text{top}} > \frac{4}{\Lambda_\text{top}^2} ,
\label{eq:necessary}
\end{align}
where $\phi_{\rm top}$ is the location of the potential barrier, and $\Lambda_{\rm top} \equiv \sqrt{3 M_p^2 / U_{\rm top}}$. If this condition is not satisfied, the CDL bounce may fail to be present \cite{Jensen:1983ac,Jensen:1988zx,Hackworth:2004xb,Batra:2006rz}. However, even in the absence of a CDL solution, there is a second type of instability, corresponding to the homogeneous solution where $\phi$ sits at the top of the potential barrier. This instanton was first discussed by Hawking and Moss (HM) \cite{Hawking:1981fz}, and it is given by
\begin{equation}
	\phi_{\rm HM} \equiv \phi_{\rm top} , \qquad {\rm and} \qquad
	\rho_{\rm HM} (\xi) = \Lambda_{\rm top} \sin \left( \frac{\xi}{\Lambda_{\rm top}} \right) .
\label{eq:phirhoHM}
\end{equation}
In this case, the corresponding tunneling exponent reads
\begin{equation}
	\Delta S_{\rm HM} 	\equiv S_E \Big|_{\rm HM} - S_E \Big|_{\rm dS}  = 24 \pi^2 M_p^4 \left( \frac{1}{U_\text{fv}} - \frac{1}{U_\text{top}} \right) .
\label{eq:SHM}
\end{equation}

A common feature of the CDL and HM instabilities is that their  tunneling exponents diverge in the limit $U_\text{fv} \rightarrow 0$, as can be seen from Eqs.~(\ref{eq:SE_CDL}) and (\ref{eq:SHM}). As the false and true vacua become degenerate, the CDL and HM decay channels become ineffective, and the metastable vacuum becomes exponentially long-lived.

We finish this section noting that Eq.~(\ref{eq:EOM_phi}) can be rewritten in integral form as follows:
\begin{equation}
	\left( \rho^3 \phi' \right) \Big|_{\xi = 0}^{\xi = \xi_{\rm max}} = \int_0^{\xi_{\rm max}} d \xi \, \rho^3 \frac{\partial U}{\partial \phi} .
\label{eq:bouncecondition}
\end{equation}
In most cases, the left-hand side of Eq.~(\ref{eq:bouncecondition}) is known. For example, it vanishes for the CDL bounce, on account of the boundary conditions $\phi'(0) = \phi'(\xi_{\rm max}) = 0$, as well as on the de Sitter and HM solutions where $\phi$ is constant. This feature permits the construction of exact solutions to the equations of motion in certain simple cases, as we will illustrate in Section~\ref{sec:modelpot}, where \eqref{eq:bouncecondition} supplies the initial conditions for the bounce solution.

\subsection{Bubble of nothing as a Coleman-De Luccia problem}
\label{sec:BONasCDL}

\begin{figure}[t]
\centering
\includegraphics[align=c,height=0.455\textwidth]{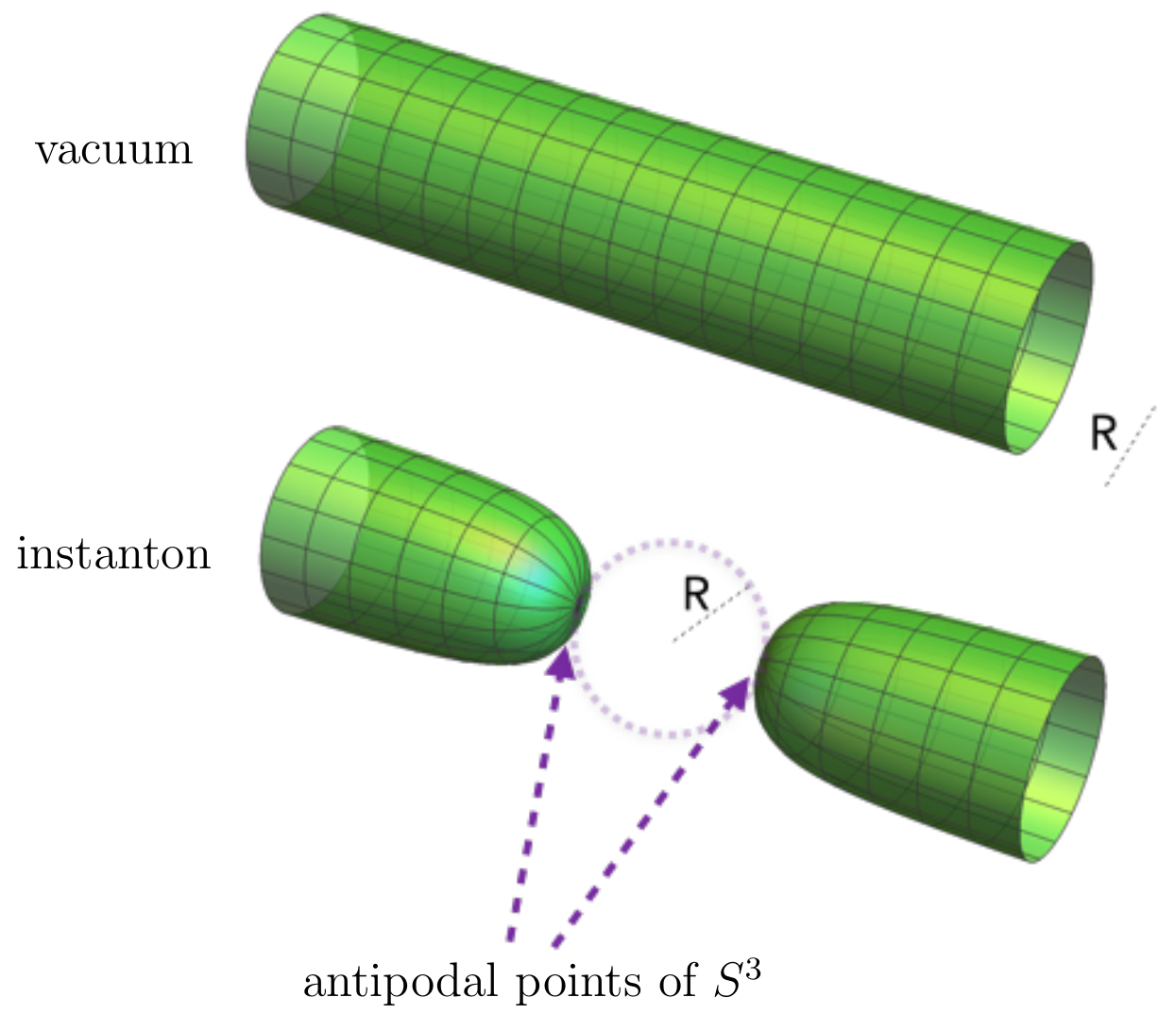}
\hspace{0mm}
\includegraphics[align=c,height=0.455\textwidth]{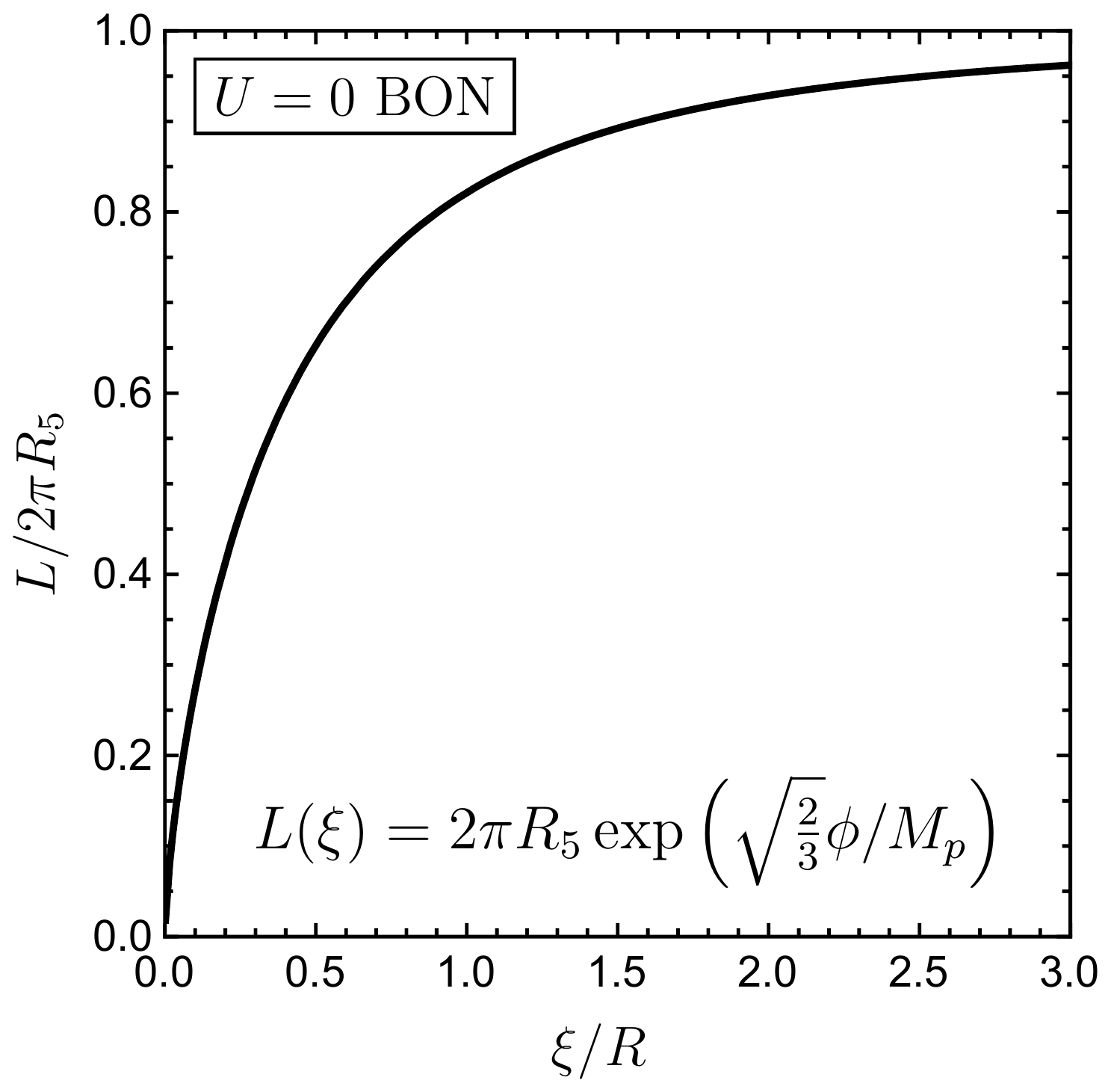}
\caption{\textbf{Left:} Diagram showing the geometry of the KK false vacuum (``vacuum") and BON (``instanton") solutions. The $r$ direction lies along the symmetry axis, with the periodic $y$ coordinate along the azimuth. In the false vacuum, $\phi \equiv 0$, and the KK radius is independent of $r$. For Witten's bubble, the KK radius approaches zero near $r \gtrsim R_5$, whereas far away from the bubble wall the geometry asymptotically approaches that of the false vacuum.
\textbf{Right:} After dimensional reduction, the scalar modulus $\phi$ describes the proper length of the KK circle, $L$. As a function of the CDL coordinate, defined in Eq.~(\ref{eq:def_xi}), $L$ vanishes at $\xi = 0$, corresponding to the location of bubble wall, whereas $L \simeq 2 \pi R_5$ in the asymptotic region $\xi \gg R_5$. }
\label{fig:cucumber}
\end{figure}

The bubble of nothing discovered by Witten is an unusual instability of the KK vacuum $\mathbb{M}^4 \times S^1$ \cite{Witten:1981gj}. The decay can be described semiclassically, in terms of a gravitational instanton equivalent to the five-dimensional Euclidean Schwarzschild solution:
\begin{equation}
	d s_5^2 = dr^2 \left( 1 - \frac{R_5^2}{r^2} \right)^{-1} + r^2 d \Omega_3^2 + \left( 1 - \frac{R_5^2}{r^2} \right) dy^2 .
\label{eq:es5geometry}
\end{equation}
In these coordinates, $r \in [R_5, \infty)$ and $y \sim y + 2 \pi R_5$. As usual, the relationship between the $y$-coordinate periodicity and the Schwarzschild radius ensures the absence of a conical singularity at $r = R_5$. The near-horizon metric is of the form 
\begin{equation}
	d s_5^2 \simeq d \lambda^2 + \lambda^2 d \tilde y^2 + R_5^2 d \Omega_3^2 ,
	\label{eq:bubbleds5}
\end{equation}
where $\lambda \equiv \sqrt{(2R_5)(r-R_5)}$, and $\tilde y \equiv y / R_5$. Indeed, the geometry remains smooth near $r=R_5$, and  the topology of the solution is $\mathbb{R}^2 \times S^3$. As depicted in the left panel of Fig.~\ref{fig:cucumber}, the bubble geometry ends on a smooth cap at $r=R_5$, which we will typically refer to variously as the bubble wall or core. For this BON solution, the scale $R_5$ characterizes three distinct physical scales: the size of the bubble (the area radius of the transverse $S^3$); the magnitude of local five-dimensional curvatures near its wall; and the asymptotic radius of the KK circle.

As described in \cite{Witten:1981gj}, the initial conditions for the nucleated bubble, as well as its Lorentzian-time evolution, can be obtained by analytic continuation of Eq.~(\ref{eq:es5geometry}). Static time slices asymptote to the false vacuum, but near the origin the geometry appears as though a hole has been carved out of the original manifold: from the four-dimensional perspective, spacetime ends on the growing three-sphere ${\bf x}^2 - t^2 = R_5^2$. Nevertheless, the presence of a fifth dimension ensures that the solution remains smooth: although asymptotically the size of the KK radius is $2 \pi R_5$, it smoothly shrinks to zero at the bubble wall located at ${\bf x}^2 - t^2 = R_5^2$.
Unlike in the ordinary picture of tunneling in quantum field theory, the bubble that gets nucleated is not a bubble of true vacuum, but rather it has `nothing' inside.

\begin{figure}
\centering
\includegraphics[width=\textwidth]{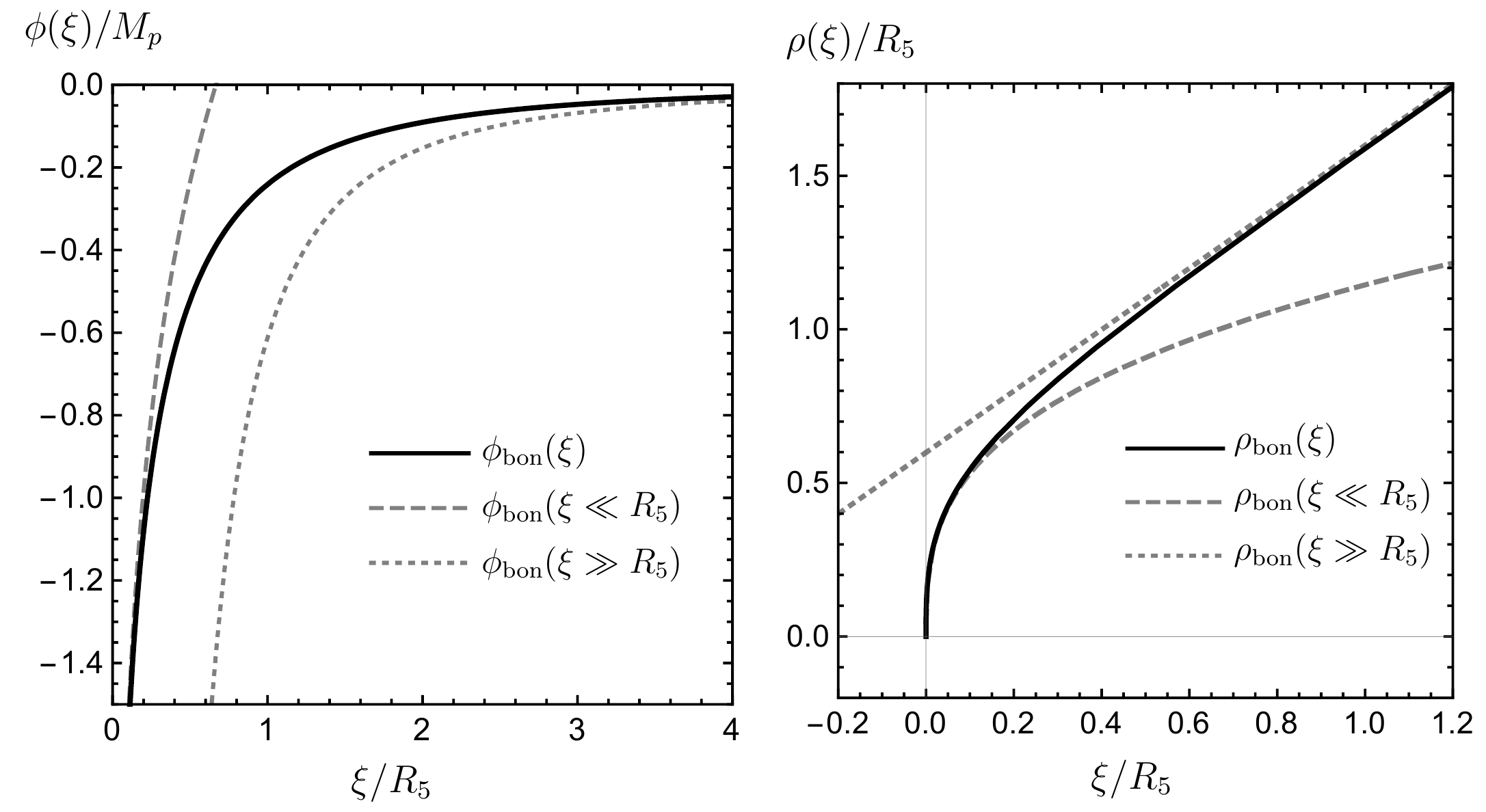}
\caption{The five-dimensional Euclidean Schwarzschild gravitational instanton, or Witten's BON, expressed in terms of the degrees of freedom $\phi_{\rm bon} (\xi)$ ({\bf left}) and $\rho_{\rm bon} (\xi)$ ({\bf right}) of the corresponding CDL problem. The behavior of the solutions in the near-horizon ($\xi \ll R_5$) and asymptotic ($\xi \gg R_5$) regimes is indicated in the figure, following \eqref{eq:appxwitten} and \eqref{eq:WittenLarge}, respectively.}
\label{fig:BON}
\end{figure}

At first sight, Witten's bubble appears to be a rather different instability from CDL tunneling, taking place in a purely gravitational higher dimensional theory. However, it was observed in~\cite{Dine:2004uw} that the BON can be rewritten as a solution to a four-dimensional CDL problem via dimensional reduction. In the dimensionally reduced theory, the radial modulus responsible for setting the size of the KK circle is mapped onto the CDL scalar $\phi$. The Einstein frame, canonically-normalized reduction is obtained via the following parametrization of the five-dimensional metric:
\begin{equation}
	d s_5^2 = e^{ - \sqrt{\frac{2}{3}} \frac{\phi}{M_p}} d s_4^2 + e^{ 2 \sqrt{\frac{2}{3}} \frac{\phi}{M_p}} dy^2 .
\end{equation}
The  Euclidean Schwarzschild solution~(\ref{eq:es5geometry}) can be written in this form, with
\begin{gather}
	\label{eq:phiBON_r} \phi_{\rm bon} (r) = \frac{M_p}{2}\sqrt{\frac{3}{2}} \log \left( 1 - \frac{R_5^2}{r^2} \right), \\
	\label{eq:rhoBON_r} ds_{\rm bon}^2 = dr^2 \left( 1 - \frac{R_5^2}{r^2} \right)^{-1/2} + \left( 1 - \frac{R_5^2}{r^2} \right)^{1/2} r^2 d\Omega_3^2.
\end{gather}
The 4D metric in Eq.~(\ref{eq:rhoBON_r}) can be further rewritten as in Eq.~(\ref{eq:cdlmetric}), with the CDL radial coordinate related to the Schwarzschild radial coordinate as\footnote{ An exact expression for $\xi$ can be obtained in terms of a hypergeometric function, given in  Appendix~\ref{sec:appxExact}. }
\begin{equation}
	\xi (r) \equiv \int_{R_5}^r \frac{d \hat r}{(1 - R_5^2 / \hat r^2)^{1/4}} .
\label{eq:def_xi}
\end{equation}
The corresponding CDL degrees of freedom satisfy Eq.~(\ref{eq:EOM_phi}) and (\ref{eq:EOM_rho}) with $U(\phi) \equiv 0$, and the profile of $\phi_{\rm bon} (\xi)$ and $\rho_{\rm bon}(\xi)$ is shown in Fig.~\ref{fig:BON}.
Near the bubble wall, at $\xi = 0$,
\begin{equation}
	\phi_{\rm bon} (\xi \ll R_5) \simeq M_p \sqrt{\frac{2}{3}} \log\left(\frac{3  \xi}{2R_5}\right), \qquad {\rm and} \qquad
	\rho_{\rm bon} (\xi\ll R_5) \simeq R_5 \left(\frac{3  \xi}{2R_5}\right)^{1/3} ,
	\label{eq:appxwitten}
\end{equation}
whereas asymptotically far from the bubble,
\begin{equation}
	\phi_{\rm bon} (\xi \gg R_5) \simeq - \frac{M_p}{2} \sqrt{\frac{3}{2}} \left( \frac{R_5}{\xi} \right)^2, \qquad {\rm and} \qquad
	\rho_{\rm bon} (\xi \gg R_5) \simeq \xi + \gamma R_5 ,
	\label{eq:WittenLarge}
\end{equation}
with numerical constant
\begin{align}
\gamma \equiv \sqrt{\pi} \Gamma\left(\tfrac{3}{4}\right) \Gamma\left(\tfrac{1}{4}\right)^{-1} \simeq 0.6.
\label{eq:gammadefinition}
\end{align}
Exact solutions for $\phi_{\rm bon}$ and $\rho_{\rm bon}$ can be found in Appendix~\ref{sec:appxExact}.

Eq.~(\ref{eq:appxwitten}) reflects an important difference between Witten's BON and an ordinary CDL bounce: the former appears singular at $\xi=0$, while the latter satisfies $\phi'(0)=0$. Instead, the BON must be thought of as a CDL problem with singular boundary conditions, describing tunneling from $\phi=0$ to $\phi= - \infty$ in the absence of a potential barrier.

Dimensionally reducing the five-dimensional Einstein-Hilbert action one finds
\begin{align}
	\Delta S =	& \int d^4 x \sqrt{g} \left\{ - \frac{M_p^2}{2} {\cal R} + \frac{1}{2} g^{\mu \nu} \partial_\mu \phi \partial_\nu \phi - \frac{M_p}{\sqrt{6}} \Box \phi \right\} \\
			&+\frac{M_p}{\sqrt{6}} \sqrt{g^{rr}} \int_{\partial \mathcal{M}} d^3x \sqrt{h} \partial_r \phi + \Delta S_{\rm GHY} ,
	\label{eq:SE_5Dred}
\end{align}
where $\partial \mathcal{M}$ refers to the boundary manifold defined as the hypersurface $r \equiv r_0$, and $\Delta S_{\rm GHY}$ is the (appropriately regularized) Gibbons-Hawking-York boundary term evaluated on the boundary. On the BON solution, the combination $- M_p^2 {\cal R} +(\partial \phi)^2$ vanishes, and $\Delta S_{\rm GHY} \rightarrow 0$ in the limit $r_0 \rightarrow \infty$.\footnote{The boundary term appropriate for the BON metric can be written as $S_{\rm GHY} = -M_p^2 n^\mu \partial_\mu \text{Vol} (\partial \mathcal{M}) = - 6 \pi^2 M_p^2 (r_0^2 - R_5^2/2)$, whereas the boundary term of the flat metric with the same boundary geometry reads $S_{\rm GHY}^{(\text{flat})} = - 6 \pi^2 M_p^2 r_0^2 \sqrt{1 - R_5^2/r_0^2}$. In total, $\Delta S_{\rm GHY} \equiv S_{\rm GHY} - S_{\rm GHY}^{(\text{flat})}$, which vanishes in the limit $r_0 \rightarrow \infty$.} Integrating by parts the $\Box \phi$ term and combining it with the first term of Eq.~(\ref{eq:SE_5Dred}), the BON action can be written as
\begin{align}
	\label{eq:SE_BON}	\Delta S_{\rm bon} 	&= \pi^2 M_p \sqrt{\frac{2}{3}} \rho_{\rm bon}^3 (\xi) \phi'_{\rm bon} (\xi) \Big|_{\xi = 0} \\
									&= \pi^2 M_p^2 R_5^2 .
\end{align}
Although we have chosen to evaluate the right-hand side of Eq.~(\ref{eq:SE_BON}) at $\xi = 0$, it follows from inspecting Eq.~(\ref{eq:EOM_phi}) that the combination $\rho^3 \phi'$ is a constant on the BON. In particular,
\begin{equation}
	\rho_{\rm bon}^3 (\xi) \phi'_{\rm bon} (\xi) = \sqrt{\frac{3}{2}} M_p R_5^2 \qquad \forall \ \xi .
\end{equation}

Compared to the five-dimensional perspective, the dimensionally-reduced formulation of the BON appears rather pathological. The four-dimensional geometry of Eq.~(\ref{eq:rhoBON_r}) exhibits a naked singularity at $r=R_5$ with diverging curvatures and scalar gradients. However, the combination $- M_p^2 {\cal R} +(\partial\phi)^2$ that enters into the four-dimensional action identically vanishes, since five-dimensional Euclidean Schwarzschild is a solution of Einstein's equations in vacuum. All 5D curvatures remain sub-Planckian, provided $R_5$ is larger than the Planck length. The role of $\phi$ as a geometric modulus restricts the terms in the reduced action, organizing them into finite five-dimensional quantities.

On the other hand, as was noted in \cite{Dine:2004uw}, the advantage of the four-dimensional formulation is that it makes it straightforward to look for instanton solutions with BON boundary conditions in the presence of a scalar potential that stabilizes the geometric modulus. In the absence of supersymmetry, a runaway scalar potential will be generated by Casimir energies, so modulus stabilization is a necessity in any realistic construction. It is then interesting to ask whether BON solutions still exist and whether the corresponding decay rate is significantly modified.
In the presence of a non-zero $U(\phi)$, the region near the bubble wall will be sensitive to the asymptotic form of the potential in the compactification direction, while the region near the metastable vacuum will depend on the details of the mechanism that stabilizes the modulus.
In Section \ref{sec:bounceofnothing}, this useful factorization will allow us to classify BON instabilities of de Sitter vacua in terms of the behavior of the scalar potential in the compactification limit, while treating the region near the metastable vacuum with toy models.

\subsection{Sources of moduli potentials} \label{sec:sourcespotentials}
\label{sec:potentials}

As described in the introduction, we wish to employ a bottom-up strategy, remaining as agnostic as possible about the details of the stabilization mechanism leading to a four-dimensional de Sitter vacuum, and the detailed features of the effective potential relevant for the moduli. Nevertheless, it will be useful to keep in mind some of the sources of moduli potentials that appear naturally in theories with extra dimensions, especially as it pertains to their behavior in the compactification limit. This will allow us to gain some intuition about the regime of validity of some of the conclusions we draw in later sections. Sources of potentials for geometric moduli in theories with extra dimensions have been widely discussed in the literature (see e.g.~\cite{Giddings:2004vr}).

For simplicity, let us first consider $4+n$-dimensional gravity, with the $n$ extra dimensions compactified. Allowing for a non-zero cosmological constant (CC), the higher-dimensional Einstein-Hilbert action reads
\begin{equation}
	S_E = \int d^{4+n} x \sqrt{g_{4+n}} \left\{ - \frac{M^{2+n}}{2} \mathcal{R}_{4+n} + M^{2+n} \Lambda^\text{CC}_{4+n} \right\} ,
\end{equation}
where $M$ is the fundamental Planck mass.
The Einstein frame dimensional reduction is obtained through the following parametrization:
\begin{equation}
	ds_{4+n}^2 = e^{ - \sqrt{\frac{2n}{n+2}} \frac{\phi}{M_p}} d s_4^2 + e^{ 2\sqrt{\frac{2}{n(n+2)}} \frac{\phi}{M_p} } d s_n^2 ,
\label{eq:Eframe}
\end{equation}
and the reduced action reads
\begin{equation}
S_E = \int d^4 x \sqrt{ g} \left\{
		-\frac{M_p^2}{2} \mathcal{R}
		+ \frac{1}{2} g^{\mu\nu} \partial_\mu \phi \partial_\nu \phi
		- \frac{M_p}{2} \sqrt{\frac{2n}{n+2}} \Box \phi
		+ U(\phi) \right\} ,
\label{eq:Sreduced}
\end{equation}
with
\begin{equation}
	U(\phi) = -\frac{M_p^2}{2} \mathcal{R}_n e^{- \sqrt{\frac{2n + 4}{n}} \frac{\phi}{M_p}} + M_p^2 \Lambda^\text{CC}_{4+n} e^{- \sqrt{\frac{2n}{n+2}} \frac{\phi}{M_p}} .
	\label{eq:sourcesofmodulipotentials}
\end{equation}
For example, if the internal manifold is an $n$-sphere with curvature radius $R_n$, then $\mathcal{R}_n = n(n-1)/R_n^2$. In the compactification limit, the contribution to the scalar potential from this internal curvature always dominates over the contribution from the higher-dimensional CC, and therefore $U\rightarrow - \infty$ as $\phi \rightarrow - \infty$.

More generally, the scalar potential of geometric moduli might receive contributions from other non-trivial field configurations present in the higher-dimensional theory, such as branes and fluxes. For example, an $n-$form flux field wrapped around the internal manifold leads to a contribution that scales as \cite{Giddings:2004vr}
\begin{equation}
	U_{\rm flux} (\phi) \propto e^{ - 3 \sqrt{\frac{2n}{n+2}} \frac{\phi}{M_p} } .
\end{equation}
In the compactification limit, a contribution of this kind would dominate over that coming from internal curvature.

Some simple (and non-exhaustive) possibilities for the behavior of the scalar potential in the compactification limit are therefore of the form
\begin{itemize}
\item $U(\phi)\sim 0$ (no flux, higher-dimensional CC, or internal curvature); 
\item $U(\phi)\sim M_p^2 \Lambda^\text{CC}_{4+n} e^{ - \sqrt{\frac{2n}{n+2}} \frac{\phi}{M_p}}$ (no flux or internal curvature; CC dominated); 
\item $U(\phi)\sim -\frac{M_p^2}{2} \mathcal{R}_n e^{ - \sqrt{\frac{2n+4}{n}} \frac{\phi}{M_p}}$  (no flux; internal curvature dominated). 
\end{itemize}

In Section~\ref{sec:bubbleexist} we will discuss what sources of moduli potentials are compatible with the existence of a BON instability of four-dimensional de Sitter vacua.

\subsection{Classical vs effective potentials}
\label{sec:potentials_comments}

We have primarily discussed classical sources of moduli potentials, such as
higher dimensional flux, the cosmological constant, and geometric curvature. These sources can produce strange, asymptotically divergent potentials upon dimensional reduction. If the  potential is singular in the compactification limit, we should view the singularity similarly to the way we view the singularities of the lower-dimensional Ricci scalar and modulus kinetic term in those limits: they are artificial, induced by the choice of field-space coordinates, and  they are absent in the higher-dimensional theory.

What of quantum corrections? Casimir potentials, for example, may play some important role in the stabilization  of the modulus, as in~\cite{Kachru:2003aw}. In general any time there is an accidental zero mode, like the modulus, ``higher order effects" like quantum corrections can lift it.  One might wonder whether Casimir potentials, which are singular as $R\rightarrow 0$, might induce significant backreaction on the BON geometry.

In fact, approximating quantum corrections to the bubble by adding the na\"ive Casimir potential to the dimensionally reduced classical equations of motion is qualitatively incorrect. Casimir potentials are computed on field configurations that are slowly varying in spacetime and only make sense in that context. Near the bubble the modulus is varying rapidly in space --- the 5D classical solution looks more like a finite size (hemi)sphere than a cylinder of slowly varying radius. Because of this,  quantum effects cannot be computed systematically  without computing the full effective action for the 5D metric to some order in $\hbar$ --- there is no systematic truncation in a derivative expansion near the wall. 
However, generically there are also no zero mode fluctuations localized near the bubble wall apart from those associated with translation symmetries. All other modes of the fluctuation operator have eigenvalues set by the finite local curvature scale $R_3$. Then one could compute the first quantum corrections to the effective action via the usual functional determinant in this background. After suitable renormalization the corrections to the effective action are generically suppressed by loop factors compared to classical action. Thus they can be systematically neglected in the leading approximation, along with any backreaction on the classical solution. (This is exactly the same as the story for an ordinary CDL bounce or any other saddle point solution without accidental zero modes.)

In lieu of performing such a complete computation, we model the stabilization part of the potential, and assume that the asymptotic potential,  relevant near the bubble wall, is dominated by classical effects. As argued above, this is typically a reasonable approximation even if quantum effects are important for the stabilization.  Near the bubble wall at $\xi = 0$, field gradients are of order the KK scale, and they dominate the equations of motion (along with the classical potential, if it grows rapidly enough).

Our classical BON solutions could become invalid if the EFT breaks down. Two ways this can happen are if (higher-dimensional) curvatures or stress energy become large compared to the Planck or string scales, or if a large tower of heavy states becomes light.  

The first possibility generally does not occur in BON geometries if the asymptotic volume modulus is larger than the Planck or string scales. The second possibility, the appearance of a tower of light states, is at first sight more concerning. Bubbles of nothing probe an infinite distance in moduli space, and the swampland distance conjecture states that under such an excursion an infinite tower of states should descend below the cutoff~\cite{Draper:2019utz}. In ordinary KK theory, these states correspond to wound string modes, which become light as the radius of the KK circle goes to zero. However, this pathology does not actually arise in BON geometries, because the excursion is highly localized. Ref.~\cite{Draper:2019utz} studied wound strings in the vicinity of a static BON and found that, while they become less massive than asymptotic wound states, they do not become lighter than the string scale. This has a simple explanation: near the smooth cap marking the bubble wall, wound strings simply slip off, becoming ordinary unwound states. There is no localized tower of light states, and string scale effects are expected to be small.

\section{Bounce of Nothing}
\label{sec:bounceofnothing}

In this section, we focus on the description of BON instabilities of four-dimensional de Sitter vacua in the presence of a potential for the moduli. Our starting point is the Euclidean action of Eq.~(\ref{eq:SE_5Dred}), but now supplemented by a scalar potential $U(\phi)$:
\begin{equation}
	S_E = \int d^4 x \sqrt{g} \left\{ - \frac{M_p^2}{2} {\cal R} + \frac{1}{2} g^{\mu \nu} \partial_\mu \phi \partial_\nu \phi - \frac{M_p}{\sqrt{6}} \Box \phi + U(\phi) \right\} .
	\label{eq:SE_5Dred_V}
\end{equation}
Since our choice of false vacuum is the four-dimensional de Sitter solution, no boundary terms need to be included in Eq.~(\ref{eq:SE_5Dred_V}), as the manifolds relevant to our discussion are always compact. Provided that the four-dimensional metric is compatible with the ansatz of Eq.~(\ref{eq:cdlmetric}), the corresponding equations of motion are Eq.~(\ref{eq:EOM_phi}) and (\ref{eq:EOM_rho}).

Therefore our goal is  to find solutions to the CDL field equations that asymptote to the de Sitter false vacuum, but that feature BON boundary conditions at the center of the bounce. In this section, we will focus on obtaining analytic approximations for both the bounce solution and its Euclidean action that determines the decay rate of the false vacuum.
Rather than being comprehensive, we aim to establish the circumstances under which BON solutions can exist in the presence of a moduli potential, and to study the properties of the  decay rate in the limit of small vacuum energy.

In Section~\ref{sec:bubbleexist}, we establish the conditions that moduli potentials need to satisfy in the compactification limit to be compatible with a bubble of nothing solution. We discuss the structure of  solutions describing the decay of a $\text{dS}_4 \times S^1$ false vacuum into a BON in Section \ref{sec:dS4S1}. In Section~\ref{sec:modelpot}, we introduce a toy model for the  potential of the radial modulus that allows us to obtain an analytic description of the bounce in Section~\ref{sec:analyticbonfv}. Finally, in Section \ref{sec:analyticbonaction} we compute the leading contributions to the tunneling exponent, paying special attention to its behavior in the limit $U_\text{fv} \rightarrow 0$. In  Section \ref{sec:applicationgeneric} we briefly describe the application of our analytic methods to more general potentials.

\subsection{Bubble existence conditions} 
\label{sec:bubbleexist}

In order for BON-like instantons to exist, certain conditions must be satisfied by the scalar potential $U(\phi)$ in the compactification limit. We can establish these conditions by examining the behavior of the CDL equations, Eqs.~(\ref{eq:EOM_phi}) and (\ref{eq:EOM_rho}), near $\xi =0$. We concentrate on moduli potentials with the asymptotic behavior $U(\phi) \sim U_0 e^{a \phi / M_p}$ in the compactification direction, corresponding to $\phi \rightarrow - \infty$. We emphasize, however, that we make no prior assumption regarding the sign of $a$: it will turn out that some potentials that grow exponentially in the compactification limit also admit BON solutions.

\begin{figure}
\centering
\includegraphics[width=0.6\textwidth]{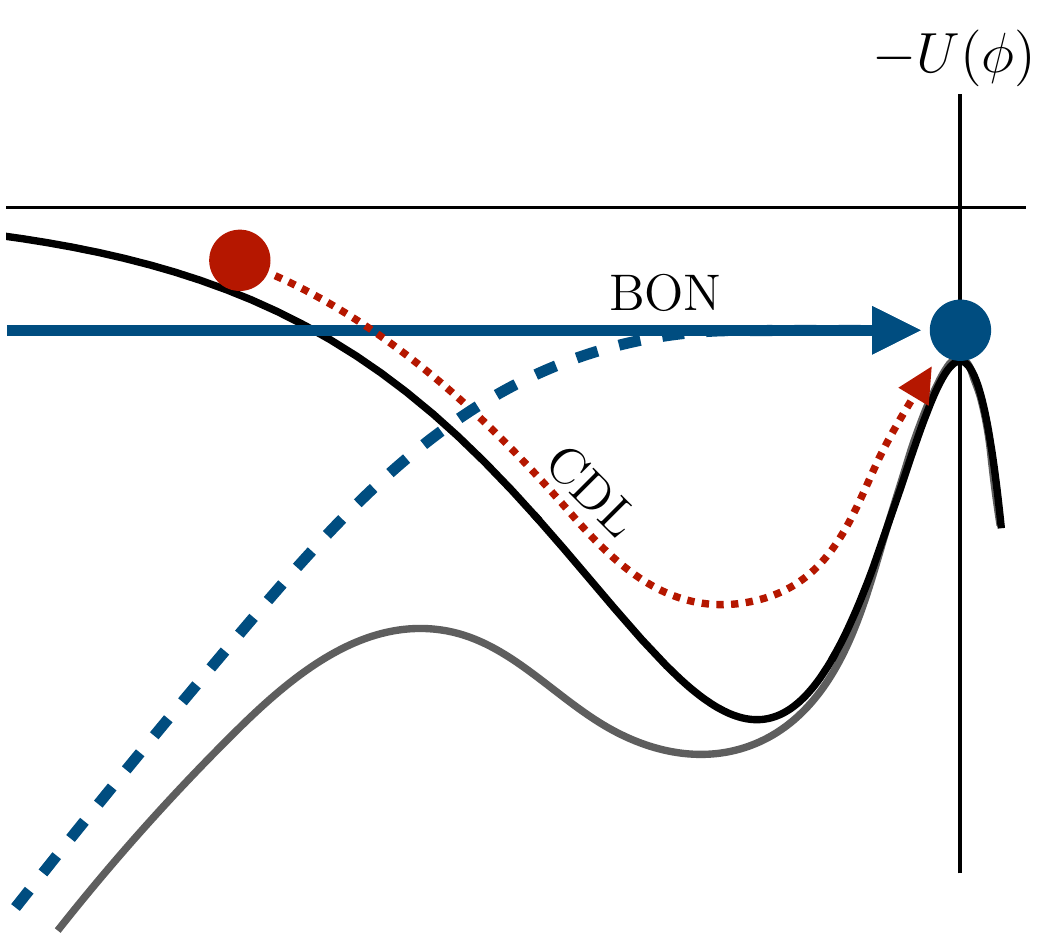}
\caption{ A diagram comparing the BON and CDL boundary conditions. The BON initial conditions $\phi(0) \rightarrow - \infty$, $\phi' \rightarrow + \infty$ ensure that $\phi(\xi)$ quickly approaches the false vacuum. Its approach to $\phi(\xi) \rightarrow \phi_\text{fv}$ is determined by the details of the potential, which is approximately quadratic in the neighborhood of $\phi \approx \phi_\text{fv}$. This direct approach to $\phi_\text{fv}$ is indicated by the straight blue line in the figure.
The CDL solution, on the other hand, begins at finite $\phi(0) = \phi_\text{CDL}$ with  $\phi'(0) = 0$. It subsequently ``rolls'' in the inverted potential, as indicated by the dashed red line, until it approaches $\phi_\text{fv}$. 
In the dashed blue line we show a second BON example, one in which the potential $U(\phi \rightarrow - \infty)$ diverges (with $-U(\phi)$ shown in dark gray). The $\phi'(0) \rightarrow \infty$ BON initial conditions can compensate for the divergence in $U(\phi)$, so that there is a BON solution even if there is no corresponding CDL solution for that potential.
}
\label{fig:cartoon}
\end{figure}

Near the bubble wall, we look for generalizations of the Witten solutions for $\phi(\xi)$ and $\rho(\xi)$  compatible with $dS_4 \times S^n \rightarrow S^3 \times \mathbbm{R}^{1+n}$ decay. For $n \geq 1$, the small $\xi$ solutions are consistent with the following ansatz:
\begin{equation}
	\phi (\xi) \simeq M_p \alpha \log \left( \frac{\xi}{\ell} \right) , \qquad {\rm and} \qquad \rho(\xi) \simeq \eta R_5 \left( \frac{\xi}{\ell} \right)^\beta ,
\label{eq:BON_ansatz}
\end{equation}
with parameters $\alpha$, $\beta$, $\ell$, and $\eta$ to be determined self-consistently. The space of solutions to the CDL equations depends on the dimensionality of the internal manifold. We focus on cases where the internal manifold contains an $n$-sphere submanifold that shrinks to zero size at the bubble wall. It is convenient to discuss the cases $n=1$ and $n \geq 2$ separately.

\subsubsection*{\boldmath{$n = 1$}}

If it is to be free of singularities, in the $\xi \rightarrow 0$ limit the BON metric must have the form
\begin{align}
ds_5^2 \simeq  d\lambda^2 + \lambda^2 \frac{dy^2}{R_5^2} + R_3^2 d\Omega_3^2,
\label{eq:BONmetric}
\end{align}
for some function $\lambda(\xi)$. Unlike \eqref{eq:bubbleds5}, the area radius of the bubble $R_3$ need not be identical to the asymptotic KK radius $R_5$. 
Assuming that $\lambda(\xi)$ and $\rho(\xi)$ have power series expansions in small $\xi \approx 0$, the solutions for $\phi(\xi)$ and $\rho(\xi)$ take the form of \eqref{eq:BON_ansatz}:
\begin{equation}
	\phi (\xi) \simeq M_p \sqrt{\frac{2}{3}} \log \left( \frac{3\xi}{2R_5} \right) , \qquad {\rm and} \qquad \rho(\xi) \simeq \eta R_5 \left( \frac{3\xi}{2R_5} \right)^{1/3}, \qquad {\rm with} \qquad \eta R_5 = R_3 .
	\label{eq:BON_eta_xismall}
\end{equation}
Eq.~(\ref{eq:BON_eta_xismall}) is identical to the near-horizon geometry of the original Witten bubble, given in Eq.~(\ref{eq:appxwitten}), except we now have an additional parameter $\eta > 0$, $\eta = R_3 / R_5$.  
It is easy to show that these $\phi$ and $\rho$ satisfy the equations of motion, Eqs.~(\ref{eq:EOM_phi}) and (\ref{eq:EOM_rho}), as long as $U(\phi)$ does not diverge as severely as $\phi''(\xi) \propto \xi^{-2}$ as $\xi \rightarrow 0$. In fact, the potential $U$ can grow exponentially large as $\phi \rightarrow -\infty$,
\begin{align}
U(\phi) \sim U_0 \exp \left( \frac{a \phi}{M_p} \right) ,
\label{eq:expgrowthU}
\end{align}
subject to the constraint $a > - \sqrt{6}$, and its effect on $\phi(\xi)$ and $\rho(\xi)$ remains negligibly small in the $\xi \rightarrow 0$ limit.

Despite the fact that $U \rightarrow + \infty$ in the compactification limit when $a < 0$, these potentials can still be subject to BON instabilities as long as $a > - \sqrt{6}$. 
For example, as discussed in Section~\ref{sec:potentials}, potentials dominated by a five-dimensional cosmological constant fall within this range, with $a_{\rm CC}  = - \sqrt{2/3}$.
If $a \leq - \sqrt{6}$, on the other hand, the BON decay cannot proceed: although the equations of motion can still be solved for more severely diverging potentials, the solutions for $\phi(\xi)$ and $\rho(\xi)$ are not compatible with  the metric  \eqref{eq:BONmetric}.
As a physically motivated example, axion flux wrapping the $S^1$ yields $a_\text{flux}=-\sqrt{6}$, so  the flux must be screened in order for the $S^1$ to shrink to zero size at the bubble wall.
An example of this type was studied in~\cite{BlancoPillado:2010df}.

\subsubsection*{\boldmath{$n \geq 2$}}
The discussion above can be easily generalized to cases where a higher dimensional sphere shrinks to zero size at the center of the bounce. (The appropriate parametrization of the higher-dimensional metric was given in Eq.~(\ref{eq:Eframe}).) There are no solutions to the field equations 
compatible with \eqref{eq:BON_ansatz}
when $a < - \sqrt{2(n+2)/n}$, and if $a > - \sqrt{2(n+2)/n}$ the absence of a conical singularity is only possible in the case $n=1$, corresponding to our previous discussion. However, when $n \geq 2$, the marginal value $a = - \sqrt{2(n+2)/n}$ admits a different class of BON solutions. In this case, the ansatz of Eq.~(\ref{eq:BON_ansatz}) corresponds to a smooth solution to the Euclidean field equations provided
\begin{equation}
	\alpha = \sqrt{\frac{2 n}{n+2}} , \qquad \beta = \frac{n}{n+2}, \qquad {\rm and} \qquad \ell = \frac{2R_n}{n+2},
\end{equation}
where $R_n$ is the radius of curvature of the $n$-sphere, and only if the overall scale of the potential is given by
\begin{equation}
	U_0 = - n (n-1) \frac{M_p^2}{2 R_n^2} .
\end{equation}
Inspecting the near-horizon metric reveals that the topology of these solutions is of the form $\mathbb{R}^{1+n} \times S^3$. Overall, for this class of solutions to exist, the behavior of the scalar potential in the compactification limit must therefore be of the form
\begin{equation}
	U(\phi) \simeq  - n (n-1) \frac{M_p^2}{2 R_n^2} e^{- \sqrt{\frac{2 (n+2)}{n} } \frac{\phi}{M_p}} .
\end{equation}
Interestingly, this is precisely the behavior induced by the internal curvature of the $n$-sphere, as reviewed in Section~\ref{sec:potentials}. Examples in 6D Einstein-Maxwell theory with a shrinking $S^2$ were studied in \cite{BlancoPillado:2010et,Brown:2011gt}.

Notice that the contribution to the scalar potential from $n$-form flux wrapping the internal manifold scales with an exponent $a_{\rm flux} = -3 \sqrt{2n/(n+2)}$, as discussed in Section~\ref{sec:potentials}. When $n \geq 2$, $a_{\rm flux} < - \sqrt{2(n+2)/n}$, and therefore flux must either be absent, screened, or otherwise disappear near the bubble wall in order for  BON solutions to exist, in keeping with Gauss' law. It would be interesting to generalize our analysis further to admit, for example, pointlike screening charges along the lines of~\cite{Brown:2011gt}.

\bigskip

In the remainder of this manuscript, we restrict our attention to potentials that satisfy $a > - \sqrt{6}$ in the compactification direction, in which case the internal manifold that resolves the apparent singularity at $\xi = 0$ must have dimension $n=1$.

 \subsection{Structure of  bubble solutions with $\text{dS}_4 \times S^1$ vacua}
 \label{sec:dS4S1}
 
 As discussed in the Introduction, our focus on this paper is on the existence of BON instabilities of four-dimensional de Sitter vacua. Let us assume the existence of a false vacuum of the form ${\rm dS}_4 \times S^1$, with Euclidean metric
 \begin{equation}
 	 ds_5^2 = d \xi^2 + \rho_{\rm dS} (\xi)^2 d \Omega_3^2 + d y^2,
 \end{equation}
 with $y \sim y +2 \pi R_5$, and $\rho_{\rm dS} (\xi)$ as given in Eq.~(\ref{eq:CDL_dS}). The parameter $R_5$ is the radius of the KK circle in the false vacuum, and it is an input parameter defining the UV theory. Our goal is  to construct instanton solutions that asymptotically approach this false vacuum, but that satisfy boundary conditions at $\xi = 0$ as given in Eq.~(\ref{eq:BON_eta_xismall}). Unlike the original BON corresponding to an instability of $\mathbb{M}^4 \times S^1$, the Euclidean geometry describing this class of instantons will be compact, with $\xi \in [0, \xi_{\rm max}]$ and $\rho (\xi_{\rm max}) = 0$.
 
Eq.~(\ref{eq:BON_eta_xismall}) corresponds to the small-$\xi$ limit of the one-parameter family of BON solutions introduced in \cite{Draper:2021gmq},  given by
\begin{equation}
	\phi_\eta (\xi) \equiv \sqrt{\frac{3}{2}} M_p \log \eta + \phi_{\rm bon} (\xi \cdot \eta^{-3/2}) ,
	\qquad {\rm and} \qquad
	\rho_\eta (\xi) \equiv \eta^{3/2} \rho_{\rm bon} (\xi \cdot \eta^{-3/2}) .
\label{eq:BON_eta}
\end{equation}
When $\xi \ll R_5$, $\phi_\eta$ and $\rho_\eta$ are indeed as in Eq.~(\ref{eq:BON_eta_xismall}), whereas for $\xi \gg R_5$, we find
\begin{equation}
	\phi_\eta (\xi) \simeq \sqrt{\frac{3}{2}} M_p \left\{ \log \eta - \frac{\eta^3}{2} \left( \frac{R_5}{\xi} \right)^2 \right\},
	\qquad {\rm and} \qquad
	\rho_\eta (\xi) \simeq \xi + \gamma \eta^{3/2} R_5 ,
\label{eq:BON_eta_xilarge}
\end{equation}
for $\gamma$ defined in \eqref{eq:gammadefinition}.
Eq.~(\ref{eq:BON_eta}) provides exact solutions to the Euclidean equations of motion when $U(\phi) \equiv 0$. In the presence of a scalar potential, $\phi_\eta$ and $\rho_\eta$ remain approximate solutions for a limited range of $\xi$, as we will soon discuss. As described in \cite{Draper:2021gmq}, it is only in this latter case that the parameter $\eta$ acquires physical significance. Near $\xi = 0$, the five-dimensional metric corresponding to this solution can be written as
\begin{equation}
	ds_5^2 \simeq d \lambda^2 + \lambda^2 \tilde y^2 + \eta^2 R_5^2 d \Omega_3^2 ,
\end{equation}
with $\lambda \equiv R_5(3 \xi / 2R_5)^{2/3}$ and $\tilde y \equiv y / R_5$. This expression clarifies the meaning of $\eta$: it parametrizes the curvature radius of the $S^3$ of the near-horizon geometry, $R_3 \equiv \eta R_5$, and corresponds to the radius of the `hole' that nucleates in spacetime.

Having established that Eq.~(\ref{eq:BON_eta_xismall}) provides acceptable boundary conditions at $\xi = 0$ in the presence of a scalar potential, we now turn to estimate the range of $\xi$ over which the solution behaves approximately like the $U=0$ BON. Comparing the right-hand side of Eq.~(\ref{eq:EOM_phi}) to, say, the $\phi''$ term on the left side, we find (ignoring irrelevant $\mathcal{O}(1)$ factors)
\begin{equation}
	\frac{\partial U / \partial \phi}{\phi ''} \sim a \left( \frac{U_0 R_5^2}{M_p^2} \right) \left( \frac{\xi}{R_5} \right)^{2 + \sqrt{\frac{2}{3}} a} \qquad {\rm for} \qquad {\xi \lesssim R_5} ,
	\label{eq:eomcomparison}
\end{equation}
where we have used $U(\phi) \sim U_0 e^{ a \phi/M_p}$ for the asymptotic form of the scalar potential in the compactification limit.
As discussed in Section~\ref{sec:bubbleexist}, the BON solutions to the EOM are nonsingular only if the potential satisfies $a > - \sqrt{6}$. With this constraint on $a$, and provided that the overall scale of the potential satisfies $U_0 \ll M_p^2 / R_5^2$, the above ratio remains small in the regime $\xi \lesssim R_5$,
implying that $\phi$ and $\rho$ are well approximated by their $U=0$ BON solutions.

For $\xi \gtrsim R_5$, \eqref{eq:eomcomparison} is no longer necessarily small, and the $U$-dependent terms cause $\phi(\xi)$ and $\rho(\xi)$ to transition away from their $U=0$ solutions.  
 If $(U_0 R_5^2 / M_p^2) \sim 1$, then this transition occurs relatively promptly at $\xi \gtrsim R_5$; if instead $(U_0 R_5^2 / M_p^2) \ll 1$, then the potential only becomes important for $\phi$ and $\rho$ at larger values of $\xi \gg R_5$ (subject to the value of $a$). 
 In both cases, $|\phi |/ M_p \lesssim 1$ and $\rho \sim \xi$ in the transition region, and the modulus starts to approach its value in the false vacuum.

In conclusion, if $U_0 R_5^2 / M_p^2 \lesssim 1$ and $a > -\sqrt{6}$, then the BON solutions are largely indifferent to the $\phi \lesssim - M_p$ asymptotic behavior of $U(\phi)$. As far as $\phi(\xi)$ and $\rho(\xi)$ are concerned, the only important part of the potential is a finite interval $\phi_\star \lesssim \phi \leq 0$, for some $|\phi_\star| \lesssim M_p$ determined by the size of $U_0 R_5^2/M_p^2$ and the value of $a$.\footnote{A shift symmetry in the action makes the location of the $\phi=0$ origin arbitrary, so we set $\phi_\text{fv} \equiv 0$ without loss of generality.}
Here, $\phi_\star$ marks the point at which the solution for $\phi(\xi)$ is no longer well approximated by \eqref{eq:BON_eta_xilarge}.

Generic potentials can be sorted into two categories: 
those that can be approximated by their quadratic expansion for $U(\phi_\star \lesssim \phi \leq 0)$, and those that can't. 
For those that can, the scalar potential is given by 
 $U(\phi) \simeq U_\text{fv} + \frac{1}{2} m^2 \phi^2$, so that
\begin{equation}
	\frac{\partial U/ \partial \phi}{\phi ''} \sim m^2 \xi^2  \qquad {\rm for} \qquad {\xi \gtrsim R_5} .
\end{equation}
Self-consistency of our analysis requires that the mass of the modulus in the false vacuum satisfies $m \ll 1/ R_5 \ll M_p$, and we will assume this hierarchy in the remainder of our analytic treatment.

In this limit, if $\phi_\star$ is sufficiently close to the false vacuum for $U(\phi_\star \lesssim \phi \lesssim 0 )$ to be approximately quadratic, then $\rho_\star (\xi) \sim \xi_\star \sim m^{-1}$ marks the value of $\rho$ for which the effects of the potential become important.
For concreteness, we use $\rho_\star \approx m^{-1}$ to define a specific value for $\phi_\star$ in the $m R_5 \ll 1$ limit,
\begin{align}
\phi_\star \equiv - \frac{m^2 R_5^2 }{2} \sqrt{ \frac{3}{2}} M_p.
\end{align}
We show in Section~\ref{sec:analyticbonfv} that $\phi \rightarrow \phi_\star$ does correspond to the potential becoming important.

Whether or not $U(\phi)$ can be approximated by its quadratic expansion depends largely on the value of $U_0 R_5^2$, so for future reference we introduce the dimensionless constant
\begin{align}
u_0 \equiv \frac{U_0 R_5^2}{M_p^2}.
\end{align}
Here $U_0$ is the characteristic scale of the dominant features of the potential around $ -M_p \lesssim \phi$. It may correspond to the height of a potential barrier, $U_0 \sim U_\text{top}$, or to some exponentially growing term, as in \eqref{eq:expgrowthU}.

For $u_0 \gtrsim 1$, the solutions for $\rho(\xi)$ and $\phi(\xi)$ are generally sensitive to the shape of $U(\phi)$ at $\phi \sim -M_p$, and the quadratic approximation is unreliable (assuming a hierarchy $m \ll M_p$).\footnote{If the potential energy density is higher than the KK scale, then the completely dimensionally reduced theory is no longer a systematic effective field theory. However, it is still useful for our purposes, which is to write the equations of motion for highly symmetric configurations in CDL form. We can omit KK modes not because the energy density is low, but because we are looking for classical solutions with a high degree of symmetry. We could include these modes in the action and they would not change the classical solutions of this type.} 
In the $u_0 \ll 1$ limit, $|\phi_\star| \ll M_p$, $U(\phi)$ is more reliably approximated by its quadratic expansion, and our analysis becomes essentially model-independent. The remainder of this section is driven by this insight. Given $m$ and $U_\text{fv}$, we calculate $\phi(\xi)$ and $\rho(\xi)$, the bubble radius $R_3$, and the tunneling exponent $\Delta S$.


\subsection{A toy model}
\label{sec:modelpot}

As discussed in Section~\ref{sec:dS4S1}, as long as the scalar potential is not too large, the BON solutions for $\rho$ and $\phi$ depend primarily on the behavior of the potential near the false vacuum.
By constructing approximate analytic BON bounce solutions for a  quadratic  potential, we can derive predictions for $\rho(\xi)$, $\phi(\xi)$, $R_3$, and $\Delta S_\text{BON}$ that are generic to a wide class of realistic scalar potentials.

\begin{figure}
\centering
\includegraphics[width=0.9\textwidth]{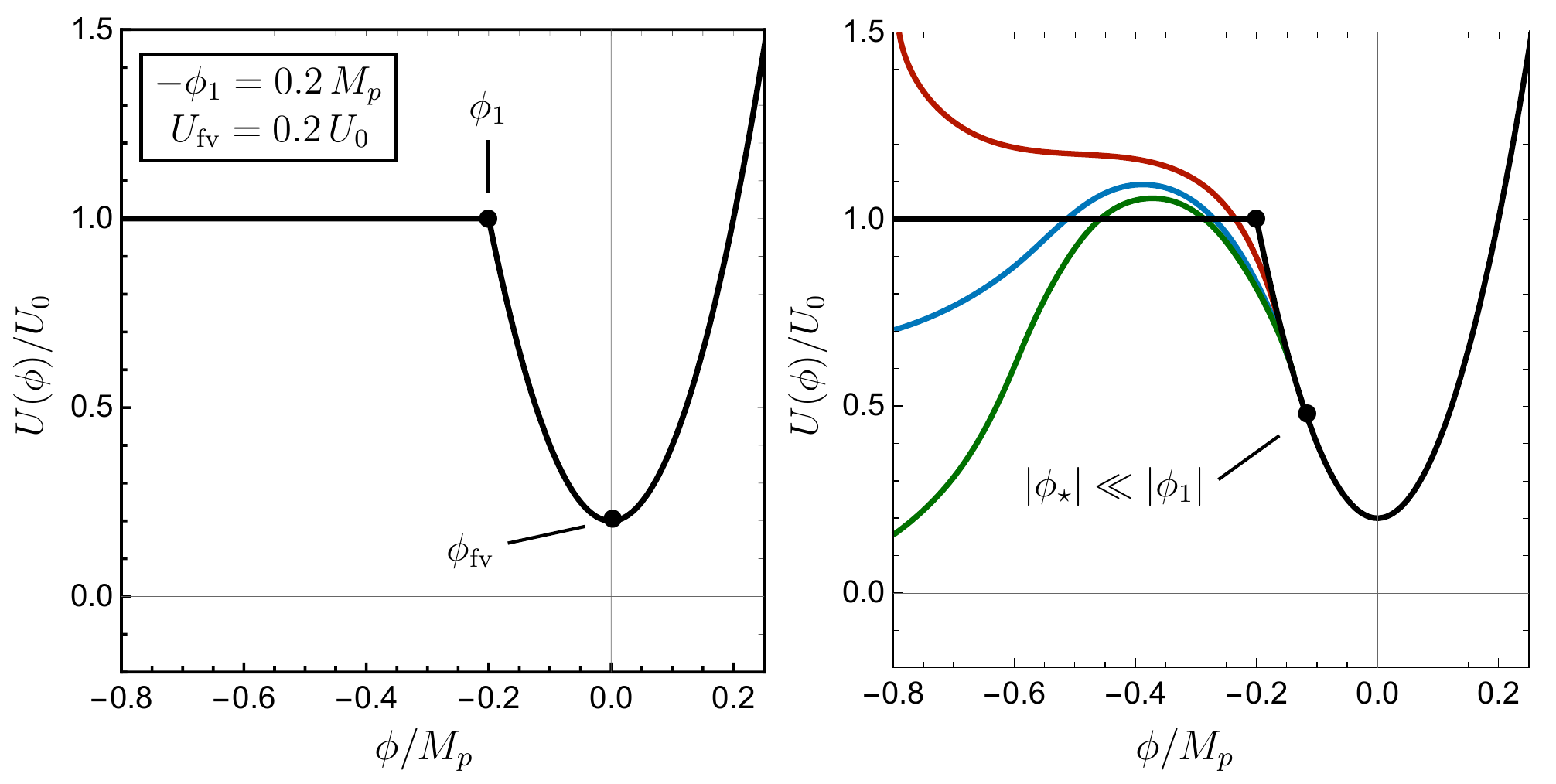}
\caption{
\textbf{Left:} The piecewise potential $U_\text{toy} (\phi)$ from \eqref{eq:Vtoy}, with $\phi_1 = -0.2 \, M_p$ and $U_\text{fv} = 0.2 \, U_0$.
\textbf{Right:}
The quantity $\phi_\star = -\frac{m^2 R_5^2}{2} \sqrt{ \frac{3}{2}} M_p$ indicates the point where the potential $U(\phi)$ begins to have a noticeable effect on the equation of motion for $\phi$ (see the discussion around Eq.~(\ref{eq:phistar})). For $\phi \ll \phi_\star$, $\phi(\xi)$ is well approximated by the Witten BON solution, even if the potential $U(\phi)$ is large. This feature greatly enhances the utility of the piecewise quadratic toy model. All three of the ``realistic'' smooth potentials will have similar solutions for $\phi(\xi)$ and $\rho(\xi)$, despite the significant differences in their asymptotic $\phi \rightarrow - \infty$ limits, as long as $m R_5 \ll1$ is small enough that $U(\abs{\phi} \leq \abs{\phi_\star})$ remains approximately quadratic.
}
\label{fig:potentialExamples}
\end{figure}

As a particularly simple toy potential,  
consider the piecewise quadratic $U_\text{toy}$,
\begin{equation}
	U_\text{toy} (\phi) \equiv	\begin{cases}	U_0 & {\rm for} \ \phi \leq \phi_1  \\
									U_\text{fv} + \frac{1}{2} m^2 \phi^2 & {\rm for} \ \phi_1 \leq \phi ,
						\end{cases}
\label{eq:Vtoy}
\end{equation}
with $U_\text{fv} + \frac{1}{2} m^2 \phi_1^2 = U_0$. $U_\text{toy} (\phi)$ is depicted on the left of Fig.~\ref{fig:potentialExamples}. Near $\phi = 0$, the potential is exactly quadratic and admits a de Sitter vacuum.  As we will see in Section~\ref{sec:analyticbonfv}, although Eq.~(\ref{eq:Vtoy}) does not admit an ordinary CDL decay --- the vacuum at $ \phi = 0$ appears na\"ively stable --- it is unstable to nucleation of a BON.

The constant branch that extends toward the compactification limit ($\phi \rightarrow - \infty$) is motivated by a property of Witten's bubble discussed in Section~\ref{sec:BONasCDL}, namely that  the quantity $\rho^3 \phi'$ remains constant (following \eqref{eq:bouncecondition}) if $\partial U/\partial \phi = 0$.
 Since the  solutions we are looking for behave like Witten's bubble near the center of the bounce, and transition towards the false vacuum at larger $\xi$, this toy potential will allow us to construct analytic expressions for both the bounce and its Euclidean action that will be representative of more realistic models. Following the discussion in Section~\ref{sec:dS4S1}, this toy model will provide approximate solutions for potentials that behave as $U(\phi) \sim U_0 e^{a \phi/M_p}$ in the compactification regime, so long as $a > - \sqrt{6}$ and $U_0 \lesssim M_p^2 / R_5^2$.

More intricate models can be devised that are also analytically tractable, e.g.~by concatenating other quadratic functions to construct a smooth local maximum in $U(\phi)$, and we discuss some of these possibilities and their implications in Section~\ref{sec:exotic}. In this section, however, we focus on Eq.~(\ref{eq:Vtoy}) because it successfully captures the  features of more generic potentials most relevant for the description of BON solutions.

\subsection{A bounce of nothing}
\label{sec:analyticbonfv}

We now proceed to find an analytic solution to the equations of motion in the presence of the potential given in Eq.~(\ref{eq:Vtoy}), with the boundary conditions at $\xi = 0$ specified in Eq.~(\ref{eq:BON_eta_xismall}).
It will be convenient to distinguish three spatial regions, according to the different behaviors of the bounce:
\begin{itemize}
\item A core region near the bubble wall, $0 \leq \xi \leq \xi_1$, where the effects of the scalar potential are subdominant and the solution behaves approximately like Witten's BON;
\item a transition region $\xi_1 \leq \xi \leq \xi_2$ that interpolates between the BON and false vacuum solutions;
\item and an asymptotic region $\xi_2 \leq \xi$, far away from the bubble wall, where $\phi(\xi)$ is exponentially close to the metastable de Sitter vacuum.
\end{itemize}
For the toy potential, the $\xi_1$ that separates the core and transition regions is implicitly defined by $\phi(\xi_1) = \phi_1$. Within the core region, the potential $U(\phi \leq \phi_1) = U_0$ is constant, and the quantity $(\rho^3 \phi')$ is exactly conserved as in Witten's bubble.\footnote{Both $\rho(\xi)$ and $\phi(\xi)$ include small perturbations, of order $R_5^2 U_0 / M_p^2 \ll 1$ compared to a bona-fide BON solution, as shown in Appendix~\ref{appx:Constant}. In the regime $\xi \lesssim R_5$ the corrections are negligibly small.}

In the transition region, $(\rho^3 \phi')$ decreases from its initial value towards $\rho^3 \phi' \rightarrow 0$. Unlike $\xi_1$, the $\xi_2$ outer boundary of the transition region is not sharply defined. Eventually $\phi'$ becomes small enough that $\phi(\xi) \approx \phi_\text{fv}$ is approximately constant, and the spacetime is approximately de~Sitter. Later in this section we demonstrate that $\xi_2 \sim \mathcal O(\text{few}) \times m^{-1}$, and that neither the parameter $\eta$ nor the bounce action depend on the precise value of $\xi_2$.

\begin{figure}
\centering
\includegraphics[width=0.9\textwidth]{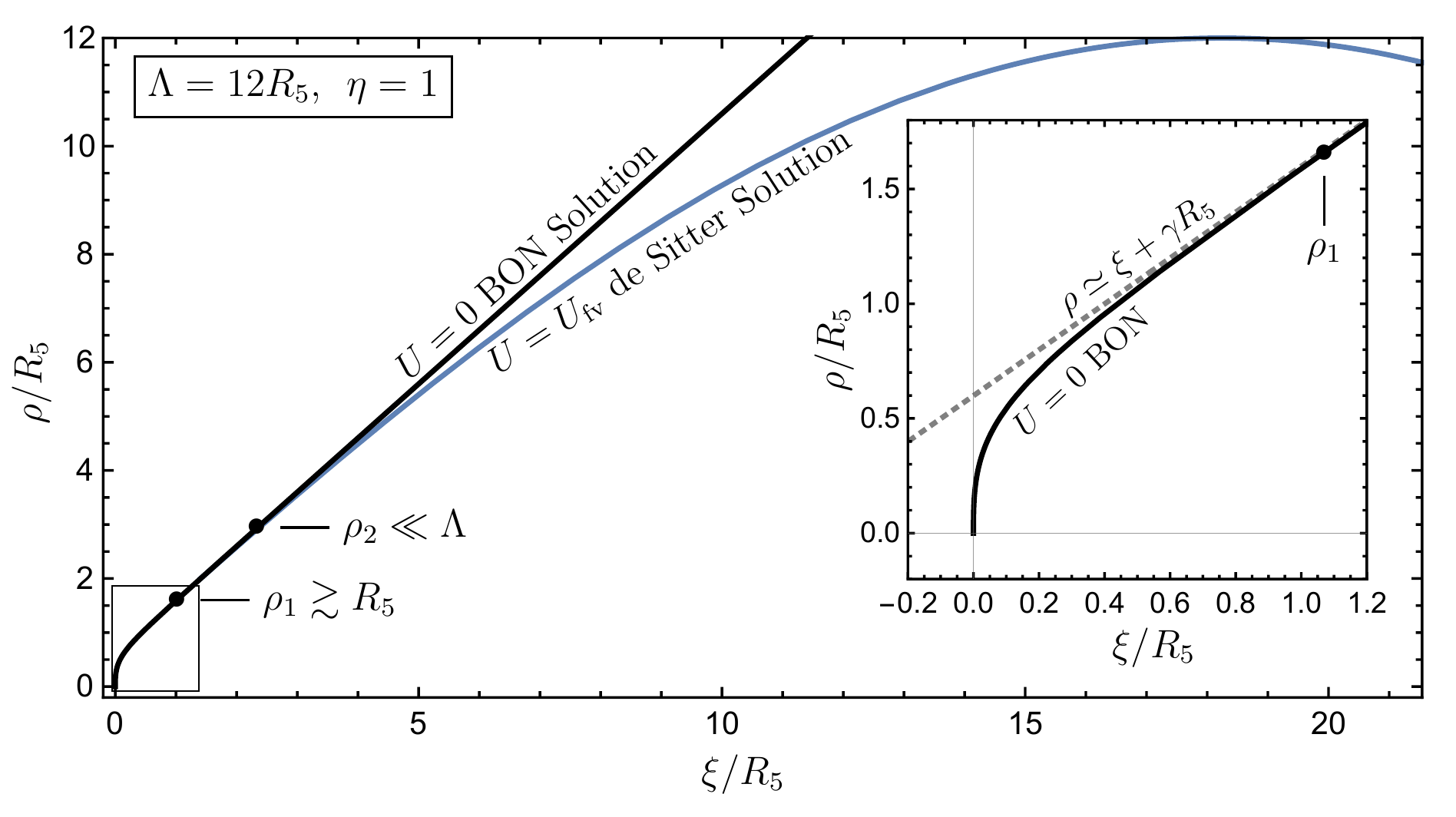}
\caption{We compare the $\rho(\xi)$ profile for the $U=0$ BON (black) to a de~Sitter vacuum with positive $U(\phi_\text{fv})$ (blue). The de Sitter length is $\Lambda = M_p \sqrt{ 3M_p^2 / U_\text{fv}}$, and for this example we have taken $\eta \simeq 1$. The analytic model described in Section~\ref{sec:analyticbonfv} stitches these solutions together by constructing a transition region in which $\rho'\approx 1$ for both of these geometries, corresponding to the region between $\rho_1$ and $\rho_2$ on the plot. Such a region exists if the de Sitter radius is large, $\Lambda\gg R_5$. 
The precise value of $\rho_1$ is determined by solving the equations of motion for a particular potential. As we show in the inset diagram, as long as $\rho_1 \gtrsim R_5$ ($\xi_1 \gtrsim 0.4\, R_5$), the $U=0$ BON solution is well approximated by its $\xi \gg R_5$ expansion with $\rho' \simeq 1$. 
The point $\rho_2$ indicates where $\phi(\rho \gtrsim \rho_2) \approx \phi_\text{fv}$ approaches the false vacuum. Its definition is somewhat arbitrary, but it should satisfy $\phi(\rho_2) - \phi_\text{fv} \ll \phi_1$.
}
\label{fig:rhophidiagram}
\end{figure}

The picture outlined in the bullet points above is sketched in Fig.~\ref{fig:rhophidiagram}, for some potential with de~Sitter radius $\Lambda \gg R_5$.
In the core region, $0 \leq \xi \leq \xi_1$, $\rho(\xi)$ is well approximated by the  $U=0$ BON solution, which becomes approximately linear for $\rho \gtrsim R_5$. This $\rho' \simeq 1$ behavior continues throughout the transition region, $\xi_1 \leq \xi \lesssim \xi_2$, as long as $\rho \ll M_p / \sqrt{U(\phi)}$.
For $\xi \gtrsim \xi_2$, the modulus $\phi(\xi) \approx \phi_\text{fv}$ is nearly constant, and $\rho(\xi)$ is well approximated by a de~Sitter solution.

\subsubsection*{Core region}

In the region $\phi \leq \phi_1$, our choice of potential ensures that the quantity $\rho^3 \phi'$ remains constant, and set by the boundary conditions of Eq.~(\ref{eq:BON_eta_xismall}). In this region, the CDL equations Eqs.~(\ref{eq:EOM_phi}) and (\ref{eq:EOM_rho}) can be rewritten as
\begin{equation}
	\rho^3 \phi' = M_p \sqrt{\frac{3}{2}} R_5^2 \eta^3 , \qquad {\rm and} \qquad
	\rho' = \sqrt{1 + \frac{R_5^4}{4 \rho^4} - \frac{\rho^2}{\Lambda_0^2}} , 
\end{equation}
with $\Lambda_0 \equiv \sqrt{3 M_p^2/{U_0}}$. The last term inside the square root satisfies $\rho \ll \Lambda_0$. This is not an additional assumption on the properties of our potential, but rather a consequence of those already discussed in Section~\ref{sec:modelpot}. Namely, the requirement that the overall scale of the scalar potential away from the asymptotic compactification regime remains $\lesssim M_p^2 / R_5^2$ allows us to neglect this last term, including into the regime where $\xi \gtrsim R_5$ and $\rho' \simeq 1$.

In this core region, not far from the bubble wall, $\rho(\xi)$ and $\phi(\xi)$ are given by Eq.~(\ref{eq:BON_eta_xismall}) and Eq.~(\ref{eq:BON_eta_xilarge}) in the regimes where $\xi \ll R_5$ and $\xi \gg R_5$, respectively.
Appendix~\ref{appx:Constant} provides the complete solutions for $\rho$ and $\phi$, including the leading perturbations from the constant $U=U_0$ term in the equations of motion, and Appendix~\ref{appx:SmallXi} lists the subleading terms in the $\xi \ll R_5$ series expansions.

\subsubsection*{Transition region}

Once $\phi$ enters the quadratic region of the potential, $\phi \geq \phi_1$, Eq.~(\ref{eq:EOM_phi}) can be conveniently written as
\begin{equation}
	\frac{d^2 \phi}{d \rho^2} + \frac{3}{\rho} \frac{d \phi}{d \rho} \simeq m^2 \phi . 
\label{eq:EOM_phi_trans}
\end{equation}
where we have taken $\rho' \simeq 1$, which is a self-consistent assumption.
The relevant solution is given by\footnote{ In full generality, the solution to Eq.~(\ref{eq:EOM_phi_trans}) is of the form $\phi(\rho) = c_1 \frac{K_1(m \rho) }{m \rho} +  c_2 \frac{I_1(m \rho) }{m \rho}$. The coefficient $c_2$ is exponentially suppressed compared to $c_1$, and the approximation $c_2 \simeq 0$ is appropriate for our purposes.}
\begin{equation}
	\phi (\rho) \simeq C \frac{K_1(m \rho) }{m \rho} .
\label{eq:phirhoBessel}
\end{equation}
Demanding continuity of $\phi'$ and $\phi$ across the transition allows us to obtain an expression for both $C$ and $\eta$:
\begin{equation}
	C \simeq - M_p \sqrt{\frac{3}{2}} \frac{\eta^3 R_5^2}{\rho_1^2 K_2 (m \rho_1)} ,
\end{equation}
\begin{equation}
	\log \eta \simeq \eta^3 \frac{m^2 R_5^2}{4} \log \left( \frac{2 e^{-\gamma_E}}{m \rho_1} \right) ,
	\label{eq:logeta}
\end{equation}
where $\rho_1 \equiv \rho(\xi_1) \simeq \xi_1$. Provided $m R_5 \ll 1$, $\log \eta \ll 1$, and we may expand the previous expression around $\eta = 1$:
\begin{equation}
	\eta = \frac{R_3}{R_5} \simeq 1 + \frac{m^2 R_5^2}{4} \log \left( \frac{2 e^{-\gamma_E}}{m \rho_1} \right) = 1 + \frac{m^2 R_5^2}{4} \left\{ \log\frac{1}{m R_5} + \mathcal{O} (1) \right\}.
\label{eq:eta}
\end{equation}
In the regime $m \rho \ll 1$, Eq.~(\ref{eq:phirhoBessel}) is of the form
\begin{equation}
	\phi (\rho) \simeq - M_p \sqrt{\frac{3}{2}} \frac{R_5^2}{\rho^2} \qquad \qquad (m \rho \ll 1),
\end{equation}
which coincides with the leading behavior of the $U=0$ BON solution, up to corrections suppressed by further powers of $R_5 / \rho$. On the other hand, in the asymptotic regime $m \rho \gg 1$, we find
\begin{equation}
	\phi (\rho) \propto - \frac{e^{-m \rho}}{(m \rho)^{3/2}} \qquad \qquad (m \rho \gg 1),
	\label{eq:phirhoexp}
\end{equation}
and the solution approaches the false vacuum exponentially fast.
As  anticipated in Section~\ref{sec:dS4S1}, $\rho \sim m^{-1}$ signals the beginning of the asymptotic regime where the bounce solution quickly approaches the false vacuum.
Motivated by the $\rho \gg m^{-1}$ functional form of $\phi(\rho)$, we define
\begin{align}
\rho_2 \equiv \mathcal O(\text{few}) \times m^{-1}
\label{eq:rho2def}
\end{align}
as the end of the transition region, where $(\rho^3 \phi') \ll R_5^2 M_p$ is much smaller than its initial value at $\xi=0$.

\subsubsection*{False vacuum region}

Once $\rho \gtrsim \rho_2$, $\phi$ is exponentially close to its value in the false vacuum, and $\rho$ is given by
\begin{equation}
	\rho(\xi) \simeq \Lambda \sin\left( \frac{ \xi +  \gamma \eta^{3/2} R_5  }{\Lambda} \right),
\label{eq:rho_dS_BON}
\end{equation}
with $\gamma\simeq 0.6$, as defined in Eq.~(\ref{eq:gammadefinition}). Indeed, this is just the de Sitter solution of Eq.~(\ref{eq:CDL_dS}), shifted by $\mathcal{O} (R_5)$ towards smaller $\xi$.

\subsubsection*{Comment on exponentially growing potentials}

The reader may be concerned that the approximate solution constructed above will break down in the presence of a potential that grows exponentially in the compactification limit. We now show that this is not the case, provided the rate of growth is not too close to the limiting value of the exponent, $a = - \sqrt{6}$, from Section~\ref{sec:bubbleexist}.

Recall that Eq.~(\ref{eq:bouncecondition}), using the relevant $\xi=0$ boundary conditions given in Eq.~(\ref{eq:BON_eta_xismall}), can be written as
\begin{equation}
	\sqrt{\frac{3}{2}} M_p R_5^2 \eta^3 = - \int_0^{\xi_{\rm max}} d\xi \, \rho^3 \frac{\partial U}{\partial \phi}.
\label{eq:bouncecondition_eta}
\end{equation}
This expression allows us to estimate the change in $\eta$ due to the region near the center of the bounce, where the potential is growing exponentially fast. 
Expanding around $\eta = 1$, and ignoring irrelevant $\mathcal{O}(1)$ factors, the shift in $\eta$ can be approximated using the small~$\xi$ expansions of Appendix~\ref{appx:SmallXi}, with the result
\begin{equation}
	\delta \eta \sim - \frac{1}{M_p R_5^2} \int_0^{R_5} d\xi \, \rho^3 \frac{\partial U}{\partial \phi} .
\end{equation}
Combining Eq.~(\ref{eq:BON_eta_xismall}) with the asymptotic form of the scalar potential, $U \sim U_0 e^{a \phi / M_p}$, we find
\begin{equation}
	\delta \eta \sim \frac{U_0 R^2}{M_p^2} \frac{-a}{1 + a / \sqrt{6}} .
\end{equation} 
In the approach to the critical value of the exponent, $a \rightarrow -\sqrt{6}$, the small $\xi$ expansions of Appendix~\ref{appx:SmallXi} break down, and $\eta(R_5)$ is no longer approximated by $\eta \approx 1$.
However, for negative values of $a$ that are not too close to this limiting case, $\delta \eta$ remains small provided $U_0 \lesssim M_p^2 / R_5^2$. Moreover, if $U_0 \lesssim M_p^2 m^2 \ll M_p^2 / R_5^2 $, $\eta$ remains well approximated by Eq.~(\ref{eq:eta}).

\subsection{de Sitter decay rate}
\label{sec:analyticbonaction}

We now turn to the evaluation of the Euclidean action for the bounce solution obtained in Section~\ref{sec:analyticbonfv}. Evaluating Eq.~(\ref{eq:SE_5Dred_V}) on a solution to the Euclidean field equations, and subtracting the action of the de Sitter false vacuum, the tunneling exponent governing the decay rate is given by
\begin{equation}
	\Delta S = \pi^2 M_p \sqrt{\frac{2}{3}} \rho(\xi)^3 \phi'(\xi) \Big|_{\xi = 0} - 2 \pi^2 \int_0^{\xi_{\rm max}} d \xi \, \rho^3 U + 2 \pi^2 \int_0^{\pi \Lambda} d \xi \, \rho_{\rm dS}^3 U_\text{fv} .
\label{eq:DeltaS_bonfv}
\end{equation}
The first term arises after integrating by parts the $\Box \phi$ term of Eq.~(\ref{eq:SE_5Dred_V}), together with the restriction that $\rho(\xi_{\rm max})=0$, since the geometry is  compact. The second term is obtained by evaluating the bulk terms in Eq.~(\ref{eq:SE_5Dred_V}) on a solution to the Euclidean field equations, and the last term corresponds to minus the Euclidean action of the metastable de Sitter vacuum.

The first term in Eq.~(\ref{eq:DeltaS_bonfv}) is given by
\begin{align}
	\Delta S \Big|_{\xi = 0} 	& \equiv \pi^2 M_p \sqrt{\frac{2}{3}} \rho(\xi)^3 \phi'(\xi) \Big|_{\xi = 0} \\
						& = \pi^2 M_p^2 R_5^2 \eta^3 \\
						& = \pi^2 M_p^2 R_5^2 \left\{ 1 + \frac{3}{4} m^2 R_5^2 \log\frac{1}{mR_5} + \mathcal{O} (m^2 R_5^2) \right\},
\label{eq:DeltaS0}
\end{align}
where in the last step we have used Eq.~(\ref{eq:eta}) to expand around $\eta = 1$.

By breaking the last two terms in \eqref{eq:DeltaS_bonfv} into the different contributions coming from the core, transition, and false vacuum regions identified at the beginning of Section~\ref{sec:analyticbonfv}, the bounce action can be written as
\begin{equation}
	\Delta S = \pi^2 M_p^2 \eta^3 R_5^2 + S_\text{core} + S_\text{trans} + S_\text{out} .
\end{equation}
Let us discuss these three different terms separately.

The contribution to the bounce action from the core region is given by
\begin{align}
S_\text{core} 	& \equiv - 2 \pi^2 \int_0^{\xi_1} d \xi \, \rho(\xi)^3 U + 2 \pi^2 \int_0^{\xi_1 + \gamma R_5} d \xi \, \rho_{\rm dS}(\xi)^3 U_{\rm fv} , \\
			& \approx - \frac{3\pi^2}{16} \frac{M_p^2}{\phi_1^2} R_5^4 \left(U_0 - U_\text{fv} \right) + \mathcal O(\pi^2 U_0 R_5^4) \\
			& = \mathcal{O} (\pi^2 M_p^2 R_5^4 m^2) + \mathcal O(\pi^2 U_0 R_5^4) ,
\label{eq:Score}
\end{align}
where in the last step we have used the fact that $(U_0 - U_\text{fv}) / \phi_1^2 \sim m^2$ (see Eq.~(\ref{eq:Vtoy})). From the transition region, we find
\begin{align}
S_\text{trans}	& \equiv -2\pi^2 \int_{\xi_1}^{\xi_2}\! d\xi \, \rho^3 U + 2\pi^2 \int_{\xi_1 + \gamma R_5}^{\xi_2 + \gamma R_5} \! d\xi\, \rho_\text{dS}^3 U_\text{fv}, \\
			& \approx - \frac{3 \pi^2}{8} M_p^2 m^2 R_5^4 \left\{ \log\frac{1}{mR_5} + \mathcal O(1) \right\} ,
\label{eq:Strans}
\end{align}
where the integral is dominated by the region near $\rho \sim m^{-1}$ where the potential is approximately quadratic. Finally, similar to the ordinary CDL bounce solution, the contribution to the vacuum-subtracted action from the asymptotic region where the bounce is  close to the false vacuum is negligibly small, 
\begin{equation}
S_\text{out} \equiv  - 2 \pi^2 \int_{\xi_2}^{\xi_{\rm max}} d \xi \, \rho(\xi)^3 U_\text{fv} + 2 \pi^2 \int_{\xi_2 + \gamma R_5}^{\pi \Lambda} d \xi \, \rho_{\rm dS}(\xi)^3 U_\text{fv} \simeq 0 .
\label{eq:Sout}
\end{equation}

Provided $U_0 \lesssim M_p^2 m^2$, and that $m\ll 1/R_5$, the leading contributions to the bounce action come from Eq.~(\ref{eq:DeltaS0}) and the log-enhanced terms in \eqref{eq:Strans}. In total, 
\begin{equation}
	\Delta S \simeq \pi^2 M_p^2 R_5^2 \left\{ 1 + \frac{3}{8} m^2 R_5^2 \log\frac{1}{mR_5}  + \mathcal O(m^2 R_5^2)  \right\} .
\label{eq:DeltaSbon}
\end{equation}
\eqref{eq:DeltaSbon} provides the leading correction to the BON tunneling exponent for the nucleation of a BON within a four-dimensional de Sitter vacuum in the limit of small energy density. Unlike the CDL and HM instantons of Section~\ref{sec:CDL}, the BON action remains finite as $U_\text{fv} \rightarrow 0$. Indeed, although the solutions for $\phi(\xi)$ and $\rho(\xi)$ are sensitive to the value of $U_\text{fv}$, \eqref{eq:DeltaSbon} is not. This implies that the BON instability persists even in the case of degenerate vacua, $U_\text{fv} \approx U_\text{tv} \rightarrow 0$, where the CDL instanton action \eqref{eq:SE_CDL} diverges.

Finally, let us comment on the contribution to the bounce action from potentials that grow exponentially in the compactification limit. Using Eq.~(\ref{eq:bouncecondition_eta}), we can write the change in the bounce action coming from the first and second terms in Eq.~(\ref{eq:DeltaS_bonfv}) as follows:
\begin{equation}
	\delta(\Delta S) \approx -2 \pi^2 \int_0^{R_5} d\xi \, \rho^3 \left( U + \frac{M_p}{\sqrt{6}} \frac{\partial U}{\partial\phi} \right) .
\end{equation}
With $U \sim U_0 e^{a \phi / M_p}$ in the compactification regime, we find
\begin{equation}
	\delta(\Delta S) \sim - \pi^2 U_0 R_5^4 .
\end{equation}
This $\delta(\Delta S)$ 
is subdominant compared to the leading correction to the BON action in Eq.~(\ref{eq:DeltaSbon}), provided $U_0 \lesssim M_p^2 m^2$.
This estimate of the change in $\Delta S$ is verified by the more careful treatment of the asymptotically exponential potential of Appendix~\ref{appx:SmallXi}.

\subsection{Application to more generic potentials} \label{sec:applicationgeneric}

For $\rho \lesssim m^{-1}$, both $\phi(\xi)$ and $\rho(\xi)$ match the Witten BON solutions to leading order, whether $U(\phi) = U_0$ or $U(\phi) = U_\text{fv} + \frac{1}{2} m^2 \phi^2$.
This indifference to the shape of the potential in the $\xi \ll R_5$ limit extends even to exponentially growing potentials of the form $U(\phi) = U_0 \exp( a \phi/M_p)$ for $ - \sqrt{6} < a < 0$, as we discuss in Section~\ref{sec:bubbleexist} and Appendix~\ref{appx:SmallXi}.
Once $\rho \sim m^{-1}$, on the other hand, the shape of the potential has a direct impact on the equations of motion, affecting how quickly $\phi(\xi)$ approaches the false vacuum.

Given two potentials with matching quadratic expansions about the false vacuum, but different $\phi \rightarrow - \infty$ asymptotic behaviors, there is a small-$R_5\sqrt{U_0}/M_p$  limit in which the solutions for $\phi(\xi)$ are expected to match. The value of $\phi$ at $\rho \sim m^{-1}$,
\begin{align}
\phi_\star \equiv - \frac{m^2 R_5^2}{2} \sqrt{ \frac{3}{2}} M_p,
\label{eq:phistar}
\end{align}
provides a helpful diagnostic. If $U(\phi)$ is approximately quadratic throughout the range $\phi_\star < \phi < \phi_\text{fv}$, then the solutions for $\rho(\xi)$ and $\phi(\xi)$ should be approximated by the piecewise quadratic model, and the leading  $m^2 R_5^2 \log (1/(m R_5))$ fractional corrections to $\eta$ and the action are expected to match, even if the $\phi \ll \phi_\star$ part of the potential is substantially different.
Possibilities of this nature are shown in the right panel of Figure~\ref{fig:potentialExamples}.

\section{Other Exotic Bounces}
\label{sec:exotic}

The bounce discussed in Section \ref{sec:bounceofnothing} effectively concatenates Witten's BON solution near the center of the bounce (where the effects of the scalar potential are subdominant) to the de Sitter false vacuum in the asymptotic regime. This class of bounces may be present even when the potential is monotonically increasing to the left of the false vacuum ($\phi < \phi_\text{fv}$), and therefore no other instabilities are present in the compactification limit.

Depending on the detailed features of the potential in the compactification regime, there may be additional saddle point solutions. For example, if $U(\phi)$ exhibits a local maximum to the left of $\phi = 0$, an ordinary HM solution will also be present. If a second, deeper, minimum is present to the left of the false vacuum, then an ordinary CDL solution might also exist. For instance, this would be the case if the scalar potential features similar behavior in the compactification and decompactification regime (e.g.~as indicated by the black curve in Fig.~\ref{fig:compdecomp}), as might be the case in constructions where the higher-dimensional theory exhibits some equivalence between the limits of small and large radius \cite{Angelantonj:2006ut}.

It is not hard to see how these additional CDL or HM instantons (if present) will also have a BON counterpart. Just like the instability discussed in Section  \ref{sec:bounceofnothing} ``stitches" together the BON and false vacuum solutions, other instantons can be built by stitching the BON to the CDL/HM solutions. The interpretation of this class of ``hybrid" solutions is, however, not immediately clear: they have larger action than the other solutions, and they probably admit two negative fluctuation modes. We discuss their physical meaning further below.

\begin{figure}
\centering
\includegraphics[width=\textwidth]{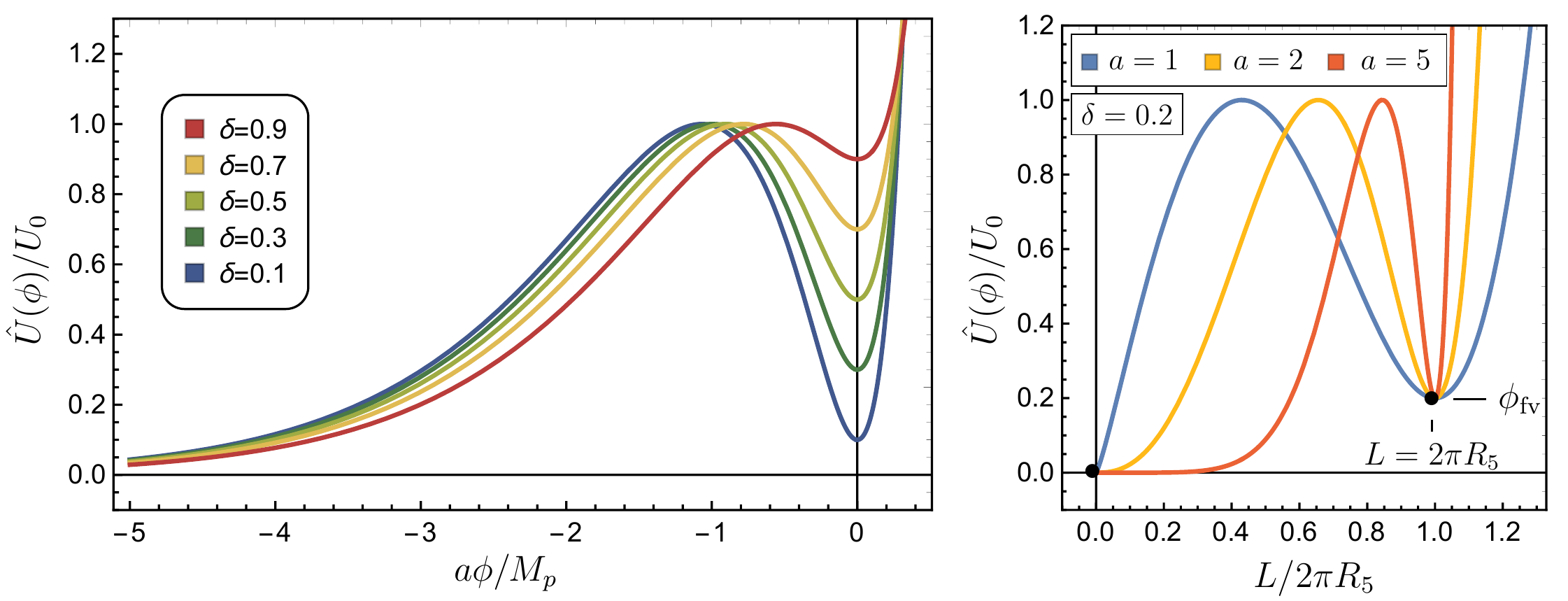}
\caption{
\textbf{Left:}
The scalar potential $\hat U(\phi)$, as given in Eq.~(\ref{eq:hatU}), for several values of $\delta \equiv U_{\rm fv} / U_0$, as indicated in the figure. A potential of this kind, exhibiting a de Sitter vacuum that is separated by a finite barrier from the compactification limit, admits  hybrid BON-CDL and BON-HM instanton solutions.
\textbf{Right:} 
To illustrate the physical meaning of $a$, we show $\hat U$ with fixed $\delta=0.2$ as a function of the proper length of the KK direction, $L = 2\pi R_5 \exp\left( \sqrt{ \frac{2}{3}} \phi / M_p \right)$. Larger values of $a$ create a steeper potential near the false vacuum at $L = 2\pi R_5$, while also flattening $U(L)$ at $L \ll \pi R_5$.
}
\label{fig:hatU}
\label{fig:potentialU}
\end{figure}

In this section, we briefly describe the properties of this hybrid class of instantons. For simplicity, we will restrict our discussion to a scalar potential with the qualitative features of the black curve in Fig.~\ref{fig:compdecomp}, and given by the following toy model
\begin{equation}
	\hat U(\phi) = U_0 \beta_0 \left( \frac{1}{3} e^{3 a \phi/M_p} - \frac{1 + \beta_1}{2} e^{2a \phi/M_p} + \beta_1 e^{a \phi/M_p} \right)
	\qquad {\rm with} \qquad a > 0 .
\label{eq:hatU}
\end{equation}
The parameters $\beta_0$ and $\beta_1$ are determined by demanding that $\hat U(\phi)$ exhibits a local minimum at $\phi_{\rm min} = 0$, as well as a local maximum such that $\hat U(\phi_{\rm top}) = U_0$. 
Exact expressions for $\beta_0$ and $\beta_1$ involve the solution of a cubic equation, with the results given in Appendix~\ref{appx:potential}.
As an expansion in $\delta \equiv U_{\rm fv} / U_0$ for $\delta\ll1$, the coefficients are approximated by
\begin{equation}
	\beta_0 = \frac{81}{4} + \mathcal{O} (\delta), \qquad {\rm and} \qquad \beta_1 = \frac{1}{3} + \frac{8 \delta}{81} + \mathcal{O} (\delta^2) ,
\end{equation}
and the position of the barrier lies at $\phi_{\rm top}/M_p = - a^{-1} (\log 3 + \mathcal{O} (\delta))$.
The mass of the scalar is approximately
\begin{align}
m^2 \equiv \left. \frac{\partial^2 \hat U}{\partial \phi^2} \right|_{\phi_\text{fv}} 
\approx \frac{U_0}{M_p^2} \frac{27 a^2}{2} \left( 1 + \frac{19}{27} \delta + \mathcal O(\delta^2) \right) .
\label{eq:massU}
\end{align}
Figure~\ref{fig:hatU} shows $\hat U (\phi)$ for several values of $\delta$. It also shows $\hat U (\phi)$ for several values of  $a$ on $\hat{U}(L)$, as a function of $L(\phi) = 2\pi R_5 \exp\left(\sqrt{ \frac{2}{3}} \phi/M_p \right)$, demonstrating the impact of $a$ on the potential barrier.

As in Section~\ref{sec:bounceofnothing}, we will make the assumption that $U_0 \lesssim M_p^2 / R_5^2$, and we will often restrict our analysis to the thin-wall limit in which $\delta \ll 1$, with the goal of deriving approximate analytical expressions. Deviations from the thin-wall limit are hard to treat analytically, but they can  be explored numerically, as we do in Section~\ref{sec:numerics}.

\subsubsection*{BON-CDL Instanton}

When $\delta = U_{\rm fv} / U_0 \ll 1$, the potential in Eq.~(\ref{eq:hatU}) admits a thin-wall CDL bounce, with the properties summarized in Section~\ref{sec:CDL}. Up to order one numbers, and for $\delta$ not too small, the value of $\phi$ inside the CDL bubble approximately satisfies 
\begin{equation}
	\phi_{\rm CDL}/M_p \sim - a^{-1} \log \left( \frac{27}{4} \frac{U_0}{U_{\rm fv}} \right) \sim - a^{-1} \log \left( \frac{U_0}{U_{\rm fv}} \right),
	\label{eq:phicdlestimate}
\end{equation}
so that $\hat U (\phi_{\rm CDL}) \lesssim \hat U (0)$. Following \cite{Coleman:1980aw}, the surface tension is given by
\begin{equation}
	\sigma = \int_{\phi_{\rm CDL}}^0 d\phi \, \sqrt{ 2 \left( U(\phi) - U(0) \right) \Big|_{\delta=0} } .
\end{equation}
Because of the exponential falloff of the potential at large and negative $\phi$, the lower limit of integration may be replaced by $-\infty$, and the integral done exactly. One finds
\begin{align}
	\sigma \simeq 2 \sqrt{6} a^{-1} M_p U_0^{1/2} + \mathcal{O} (\delta) \sim a^{-1}M_p U_0^{1/2} .
\end{align}
The bounce radius and corresponding tunneling exponent are as given in Eqs.~(\ref{eq:rho_CDL}) and (\ref{eq:SE_CDL_general}). For instance, in the limit of small energy density, $U_{\rm fv} \ll \sigma^2 / M_p^2$, $\bar \rho \simeq 4 M_p^2 / \sigma$, and $\Delta S$ is given by Eq.~(\ref{eq:SE_CDL_general}).

In analogy with the CDL instanton, let us define
\begin{equation}
	\frac{\phi_0}{M_p} \equiv \sqrt{\frac{3}{2}} \log \eta .
\label{eq:phi0eta}
\end{equation}
For the bounce of Section~\ref{sec:bounceofnothing}, where $\eta \approx 1$, $\phi_0 \approx 0$ up to small corrections, approximately matching the value of $\phi_\text{fv}$ of the false vacuum. A hybrid BON-CDL solution may be obtained by instead choosing $\phi_0 \simeq \phi_{\rm CDL}$. With this choice, $\phi$ approaches $\phi_{\rm CDL}$ in the regime $\xi \gtrsim R_5$, at which point the solution transitions into the familiar CDL instanton. The ratio of the BON radius to that of the KK circle in the false vacuum is now given by
\begin{equation}
	\eta 	= \frac{R_3}{R_5} = e^{\sqrt{\frac{2}{3}} \frac{\phi_0}{M_p} }
		\simeq e^{\sqrt{\frac{2}{3}} \frac{\phi_{\rm CDL}}{M_p} }
		\sim \left(   \frac{U_{\rm fv}}{U_0} \right)^{\frac{\sqrt{2/3}}{a}} 
\label{eq:eta_CDL}
\end{equation}
where the final inequality again uses the estimate Eq.~(\ref{eq:phicdlestimate}) which is only valid if $\delta$ is not too small; otherwise, it must be determined by solving the shooting problem numerically.

The Euclidean action of this solution is approximated by Eq.~(\ref{eq:SE_CDL_general}), combined with the contribution from the BON boundary term at $\xi = 0$; that is,
\begin{align}
	\Delta S_\text{BON-CDL} 	& \simeq \pi^2 M_p^2 R_5^2 \eta^3 + \Delta S_{\rm CDL}
	\label{eq:Sboncdl} \\
						& \simeq \pi^2 M_p^2 R_5^2 e^{\frac{\sqrt{6} \phi_{\rm CDL}}{M_p} } + \frac{24 \pi^2 M_p^4}{U_{\rm fv}}.
\label{eq:DeltaS_BONCDL}
\end{align}
where in the last step we have used the expression for $\Delta S_{\rm CDL}$ appropriate in the limit $U_{\rm fv} \rightarrow 0$.

\subsubsection*{BON-HM Instanton}

The ordinary HM instanton corresponds to the homogeneous solution $\phi_{\rm HM} = \phi_{\rm top}$. As before, choosing BON boundary conditions at $\xi = 0$ with $\phi_0 \simeq \phi_{\rm HM}$ allows us to stitch together the BON and HM solutions. In this case, we have
\begin{equation}
	\eta 	= e^{\sqrt{\frac{2}{3}} \frac{\phi_0}{M_p} }
		\simeq e^{\sqrt{\frac{2}{3}} \frac{\phi_{\rm top}}{M_p} }
		\approx 3^{-\frac{\sqrt{2/3}}{a}} ,
\label{eq:eta_HM}
\end{equation}
where we have used  $\phi_{\rm top}/M_p \simeq - a^{-1} \log 3$, as appropriate for the $\delta \ll 1$ limit of our toy model.
The Euclidean action of this solution is approximately given by
\begin{align}
	\Delta S_\text{BON-HM} 	& \simeq \pi^2 M_p^2 R_5^2 \eta^3 + \Delta S_{\rm HM} \\
						& \approx \pi^2 M_p^2 R_5^2 \, 3^{-\frac{\sqrt{6}}{a}} + 24 \pi^2 M_p^4 \left( \frac{1}{U_{\rm fv}} - \frac{1}{U_{\rm top}} \right).
\label{eq:DeltaS_BONHM}
\end{align}

\bigskip

\subsubsection*{Physical Interpretation of the Hybrid Solutions}
As can be seen from Eqs.~(\ref{eq:DeltaS_BONCDL}) and (\ref{eq:DeltaS_BONHM}), the action of the hybrid class of instantons is always larger than that of the pure BON solution discussed in Section~\ref{sec:bounceofnothing}, as well as that of the ordinary CDL or HM bounce.  Thus, these saddle points never provide the leading contribution to the semiclassical expansion. Furthermore, on physical grounds, it would seem that a better interpretation of these solutions is that they are already  counted by (coincidence regions of) the product of the dilute gas sums for the BON and ordinary CDL/HM bubbles. They almost certainly possess two negative fluctuation modes, corresponding to the fact that they are best thought of as two superposed bounces rather than one. The fact that they provide an exact solution to the Euclidean equations of motion is, in this sense, something of a curiosity, a consequence of the CDL ansatz of a spherically symmetric instanton. On the other hand, when we analyze the space of solutions numerically, we will find that there is a limit in which the distinction between the BON and CDL/HM parts of the solutions dissolves, and the hybrid and pure BON branches of solutions match smoothly onto each other. Thus, somewhere in this regime we expect a transition in the number of negative modes.

\section{Numerical Construction of Bounce Solutions} \label{sec:numerics}

In this section, we perform a numerical exploration of BON bounces of four-dimensional de Sitter vacua. Our numerical analysis complements and extends the discussion of Sections \ref{sec:bounceofnothing} and \ref{sec:exotic}.
In Section \ref{sec:prel} we introduce and discuss the main features of the overshoot-undershoot algorithm that we implement to numerically explore these bounces.
In Section~\ref{sec:expfall} we present numerical results, for the bounce and its action, for potentials that vanish in the compactification regime. For concreteness, we focus on the toy model of Eq.~(\ref{eq:hatU}). As discussed in Section~\ref{sec:exotic}, this example features hybrid BON-CDL and BON-HM solutions in addition to the ``pure'' BON.
Finally, in Section~\ref{sec:cc} we focus instead on potentials that grow in the compactification regime, taking as an example the type of exponential growth that would arise if the five-dimensional theory featured a positive CC.

\subsection{Preliminaries}
\label{sec:prel}

The analytic methods of Section~\ref{sec:bounceofnothing} rely on the following assumptions:
\begin{itemize}
\item at small $\xi \lesssim \xi_\star$, the scalar potential $U(\phi)$ can be treated as a small perturbation to the $U=0$ Witten solutions for $\phi$ and $\rho$;
\item for $\xi \gtrsim \xi_\star$, where $U$ becomes important, $U(\phi)$ is approximately quadratic, $U(\phi) \approx U_\text{fv} + \frac{1}{2} m^2 \phi^2$. In this case, $\xi_\star \sim m^{-1}$ sets the size of $\rho_\star \sim \xi_\star$, and $\phi_\star \sim -m^2 R_5^2 M_p$.
\end{itemize}
These assumptions are self-consistent in the $m \ll 1/R_5 \ll M_p$ limit as long as the potential satisfies $U_0 \ll M_p^2 /R_5^2$, for a scale $U_0$ that characterizes the dominant features of the potential in the $- M_p \lesssim \phi \lesssim \phi_\star$ region.
In this regime, where the solution is well-approximated by Witten's bubble until $\phi$ reaches the quadratic basin near the false vacuum, the bounce and its action can be estimated using the results of the piecewise quadratic model of Eq.~(\ref{eq:Vtoy}). This estimate is  independent of the asymptotic behavior of $U(\phi)$. 

For larger values of $m R_5$, where $U(\phi)$ ceases to be approximately quadratic near $\phi \sim \phi_\star$, our piecewise toy model is no longer a valid tool for predicting the bounce action $\Delta S$, or the value of $\eta = R_3/R_5$. 
We must instead follow a numerical approach.\footnote{If it is still the case that $U_0 \ll M_p^2 / R_5^2$, then it may be possible to apply the $\rho' \simeq 1$ approximation to solve the equations of motion for a different piecewise model, constructed to more closely resemble $U(\phi)$. We investigate another building block for such an approach in Appendix~\ref{appx:Linear}. However, even these more complicated analytic models cease to be reliable in regions of parameter space where $U_0 \gtrsim M_p^2 / R_5^2$.}
We employ a shooting method to construct numeric bounce profiles for $\phi$ and $\rho$. The initial data for $\rho$, $\phi$, and their derivatives are defined at some $\xi = \xi_\text{init}$ close to zero ($\xi_\text{init} \ll R_5$). 
Solutions for $\rho(\xi)$ and $\phi(\xi)$ are then obtained  from an iterative Runge--Kutta method on some finite grid of points ($\xi_\text{init}, \xi_\text{init} + \delta\xi, \xi_\text{init} + 2 \delta\xi, \ldots$) with some grid spacing $\delta\xi \ll R_5$. 
In our case the BON initial conditions are fully determined by $R_5$ and $\eta$, following \eqref{eq:BON_eta_xismall}; or, if greater precision is required, Eqs.(\ref{eq:rhoxiU0}--\ref{eq:phixiU0}).

For generic values of $(\eta, R_5)$, the solution for $\phi(\xi)$ can be classified into one of three possible categories:
\begin{itemize}
\item \textbf{undershoot}: there is some turnaround point where $\phi'(\xi_\text{turn}) = 0$, where $| \phi(\xi_\text{turn}) | > | \phi_\text{fv} |$, and $\phi(\xi \gg \xi_\text{turn})$ diverges towards $- \infty$. There is no point at which $\phi(\xi) = \phi_\text{fv}$;
\item \textbf{overshoot}: the solution for $\phi(\xi)$ crosses the false vacuum at some $\phi(\xi_\text{cross}) = \phi_\text{fv}$, and diverges to $+ \infty$ at $\xi \gg \xi_\text{cross}$;
\item \textbf{bounce}: bounce solutions lie on the boundaries between undershoot and overshoot solutions on the $(\eta, R_5)$ plane. They approach $\phi'(\xi) \rightarrow 0$ as $\xi \rightarrow \xi_\text{max}$, where for de~Sitter vacua $\xi_\text{max}$  is defined as the point where $\rho(\xi_\text{max}) = 0$ ($\xi_\text{max} > 0$).
\end{itemize}
Only the bounces are acceptable solutions, as any physical solution must satisfy $\phi'(\xi) \xrightarrow{\xi \rightarrow \xi_\text{max}} 0$.
Bounce solutions can be found numerically using a recursive bisection approach, starting with an undershoot and an overshoot solution  and testing points in the $(\eta, R_5)$ plane lying between the two solutions, until the interval between $(\eta, R_5)|_\text{under}$ and $(\eta, R_5)|_\text{over}$ becomes sufficiently small.
Given that $R_5$ is a property of the vacuum, we keep it fixed, and vary $\eta$ to find $R_3 = \eta R_5$ for the bounce solution. Thus we may think of the bubble size $R_3$ as the shooting parameter for BON geometries.

If a potential admits a CDL solution in the compactification direction, we employ the same recursive numeric technique to find $\phi_\text{CDL}$. In this case the boundary conditions at the center of the bubble are defined by $\phi' = 0$ and $\phi = \phi_\text{CDL}$, with $\rho \simeq \xi$; furthermore, the radial coordinate $\xi$ now measures from the center of the bubble, rather than the $r = R_3$ radius of the BON.

As discussed in Section \ref{sec:bounceofnothing}, even a potential $U \sim U_0 \exp(a \phi/ M_p)$ with exponential growth in the $\phi \rightarrow - \infty$ limit can have a limited, even negligible effect on the solutions for $\phi$ and $\rho$ in the region $\xi \ll R_5$, as long as $a > - \sqrt{6}$. We investigate the exponentially growing potentials in greater detail in Appendix~\ref{appx:SmallXi}, and find that 
the leading potential-dependent corrections to $\rho$ and $\phi$ are suppressed by factors of $(U_0 R_5^2  / M_p^2) (\xi_\text{init} / R_5)^p$, where $p = 2 \left( 1 + a/\sqrt{6} \right)$.
For $a > - \sqrt{2/3}$ this is small even compared to the subleading $(\xi/R_5)^{4/3}$ term in the series expansion of the $U = 0$ Witten bubble. As $a \rightarrow - \sqrt{6}$ the potential-dependent terms become more important, especially if $U_0 R_5^2 / M_p^2$ is not small. 
As long as we keep comfortable distance away from $a \approx - \sqrt{6}$, however, an exponentially growing potential is no impediment to numerically solving the equations of motion.

In the remainder of this section, we explore the full set of bounce solutions in the $(\eta, R_5)$ plane,  from the  $U_0 \lll M_p^2 / R_5^2$ limit to larger values $U_0 \gtrsim M_p^2 / R_5^2$.
As it is this combination of dimensionful quantities that dictates whether the results are compatible with the analytic model of Section~\ref{sec:bounceofnothing}, we define it as a single parameter
\begin{equation}
	u_0 \equiv \frac{U_0 R_5^2}{M_p^2}.
\end{equation}
In the $U_0 \rightarrow 0$ limit where $u_0 \rightarrow 0$ (keeping $R_5 \gg M_p^{-1}$ appropriately large), the BON bounce solutions should be well described by the analytic approximations of Section~\ref{sec:bounceofnothing}, and the hybrid BON-CDL and BON-HM solutions (if they exist) can be understood as the superposition of a small BON onto the center of a larger CDL or HM bubble.
As $u_0$ is increased towards $1$, the shape of the potential $U(\phi)$ away from the false vacuum becomes more important, and the approximations of Section~\ref{sec:bounceofnothing} become less accurate. For $u_0 \gtrsim 1$, we must rely on the numerics. We may find that at large enough values of $u_0$, there cease to be any BON solutions at all.

Before beginning our investigation of particular potentials, let us highlight a simplifying feature of the equations of motion, Eqs.~(\ref{eq:EOM_phi}--\ref{eq:EOM_rho}). It is possible to simultaneously remove the dependence on $M_p$ and rescale the potential by some overall constant, $U_0$, by defining dimensionless versions of all of the length scales:
\begin{align}
\xi \rightarrow \xi \sqrt{ U_0}/M_p,
&&
\rho \rightarrow \rho \sqrt{ U_0}/M_p,
&&
R_{3(5)} \rightarrow R_{3(5)} \sqrt{ U_0}/M_p.
\end{align}
Under this coordinate redefinition, the action \eqref{eq:SE_5Dred_V} is simply rescaled by an overall factor of $M_p^4 / U_0$.
This saves us from the hassle of conducting independent scans over different values of $U_0$ and $R_5$: as far as the equations of motion are concerned, the solutions depend only the combination $u_0 \propto U_0 R_5^2$.
As a result, many of the plots in this section include $\sqrt{u_0}$ on one of the axes, or otherwise show length scales in units of $M_p/\sqrt{U_0}$.

\subsection{Exponentially falling potentials}
\label{sec:expfall}

In this section, we focus on potentials of the form of \eqref{eq:hatU} (depicted in Fig.~\ref{fig:potentialU}), with a single de Sitter vacuum at $\phi_\text{fv} = 0$; a local maximum at $\phi_\text{HM} = \phi_\text{top}$; and a global Minkowski vacuum at $\phi \rightarrow - \infty$. This class of potentials admit CDL or HM solutions in addition to the  BON solutions we construct. 
When $u_0 \gtrsim 1$ we find that the BON and CDL/HM branches of solutions eventually merge together, with $\Delta S$ approaching the action of the CDL solution, $\Delta S \approx \Delta S_\text{CDL}$. 
In these cases we find that there is a maximum value of $u_0$ such that potentials with $u_0 > u_0^\text{(max)}$ always lead to overshoot solutions. This implies that the corresponding KK theories are stable with respect to BON formation, unless the potential permits additional  branches of bounce solutions at larger values of $u_0$.\footnote{For example, Ref.~\cite{Rajantie:2017ajw} explores potentials where the CDL thin-wall limit does not apply, which lead to multiple CDL solutions. These models would have multiple branches of BON-CDL hybrid solutions, with various values of $u_0^\text{max}$.}

\subsubsection{Parameter space of BON solutions}

\begin{figure}[t]
\centering
\includegraphics[width=\textwidth]{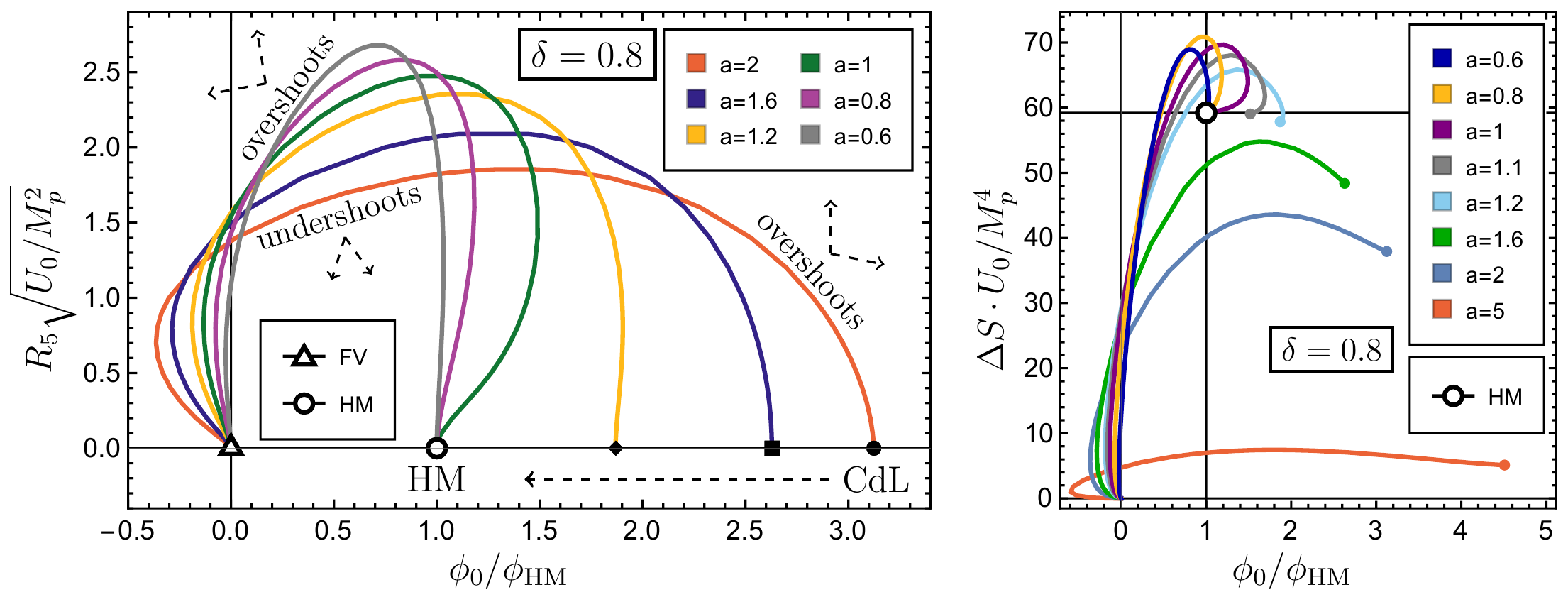}
\caption{\textbf{Left:} A plot of the initial condition parameter space for the bounce solutions. Each curve corresponds to a family of BON bounce solutions determined by the input $R_5$ and the  toy potential $\hat{U}(\phi)$ of \eqref{eq:hatU}. The abscissa shows the solutions for the shooting variable $\phi_0$ in units of $\phi_\text{HM}$, the Hawking--Moss solution. For each given value of $R_5$ there are zero or two solutions; in the latter case, the smaller value of $|\phi_0|$ corresponds to the ``pure" BON, or ``BON-FV" branch, while the larger value of $|\phi_0|$ corresponds  to the BON-HM/BON-CDL branch. 
In this plot we fix $\delta \equiv U_\text{fv}/U_0 = 0.8$ and we vary the parameter $a$ to demonstrate the transition between BON-HM and BON-CDL solutions on the right-hand branch.
In the $u_0 \rightarrow 0$ limit ($R_5^2 U_0 \rightarrow 0$), the $a \geq 1.1$ bounce solutions have BON-CDL branches with $\phi_0 \rightarrow \phi_\text{CDL}$, shown on the plot with solid circle, square, and diamond markers at ``$R_5 = 0$''. 
For smaller $a \leq 1.0$, the condition \eqref{eq:necessary} is not satisfied, and there is no CDL solution. In this case the $\phi_0  \neq  0$ branches of $u_0 \rightarrow 0$ hybrid BON solutions converge instead on the Hawking--Moss value $\phi_0 = \phi_\text{HM}$, shown  with an open circle marker. 
Outside each curve lies the region of overshoot solutions for each potential. The regions labeled ``undershoots''  include the typical undershoot solutions satisfying $\phi(\xi \rightarrow \infty) \rightarrow +\infty$, as well as any solutions that oscillate about $\phi_\text{HM}$. 
\\
\textbf{Right:} We show the vacuum-subtracted action $\Delta S$ as a function of shooting parameter $\phi_0$, with fixed $\delta = 0.8$ and varying $a$. The small $a$ examples $a = 0.6, 0.8, 1.0$ with BON-HM solutions have actions $\Delta S \approx \Delta S_\text{HM}$ in the small $u_0$ limit, which for $\delta=0.8$ is approximately $\Delta S_\text{HM} \simeq 59.2 \, M_p^4/U_0$. 
In this limit the BON-HM branches terminate at $\phi_0 \rightarrow \phi_\text{HM}$, as indicated by the large open plot marker at $(\phi_\text{HM}, \Delta S_\text{HM})$.
For $a \geq 1.1$ where \eqref{eq:necessary} is satisfied, the BON-CDL hybrid bounces approach $\phi_0 \rightarrow \phi_\text{CDL}$ at small $u_0$, and have actions $\Delta S \approx \Delta S_\text{CDL} < S_\text{HM}$. For these solutions we show the action of the pure CDL solutions with the smaller colored markers at $(\phi_\text{CDL}, \Delta S_\text{CDL})$.
Each BON-FV branch terminates at $\phi_0 \rightarrow 0$ with $\phi_0 < 0$.
}
\label{fig:paramspace}
\end{figure}

Particularly for  solutions on the BON-CDL/HM branches, it is more convenient to express the shooting parameter $\eta$ in terms of the combination $\phi_0 = M_p \sqrt{3/2} \log \eta$ introduced in \eqref{eq:phi0eta}.
This is the asymptotic value of $\phi(\xi)$ for the $U= 0$ BON in the $\xi \gg R_5$ limit, \eqref{eq:BON_eta_xilarge}. 
From the discussion in Section~\ref{sec:exotic}, in the $u_0 \lll 1$ limit we expect to find two types of solutions: a bona fide BON, with $\eta \approx 1$ and $\phi_0 \approx 0$; and a BON-CDL hybrid solution with $\phi_0 \approx \phi_\text{CDL}$, if there is a CDL solution, or $\phi_0 \approx \phi_\text{HM}$ if not. 
To distinguish these two branches of BON solutions, we hereafter refer to the former type as ``BON-FV,'' to indicate that $\phi_0(\eta) \approx \phi_\text{fv}$ is in the vicinity of the false vacuum, i.e.~$\eta \approx 1$.


Based on these two data points, we can infer that in the $u_0 \rightarrow 0$ limit (e.g. $U_0 \rightarrow 0$), the range of initial conditions $\phi_\text{CDL(HM)} < \phi_0 < 0$ lead to undershoot solutions, while outside this range the $(\phi_0, u_0 \approx 0)$ initial conditions produce overshoots. 
Whether the undershoot region extends to arbitrary large $u_0$ or whether it is capped at some finite value remains to be determined.
Our numeric computations show that in this case, the BON solutions are indeed capped at some $u_0^\text{max}$.

\begin{figure}
\centering
\includegraphics[width=0.55\textwidth]{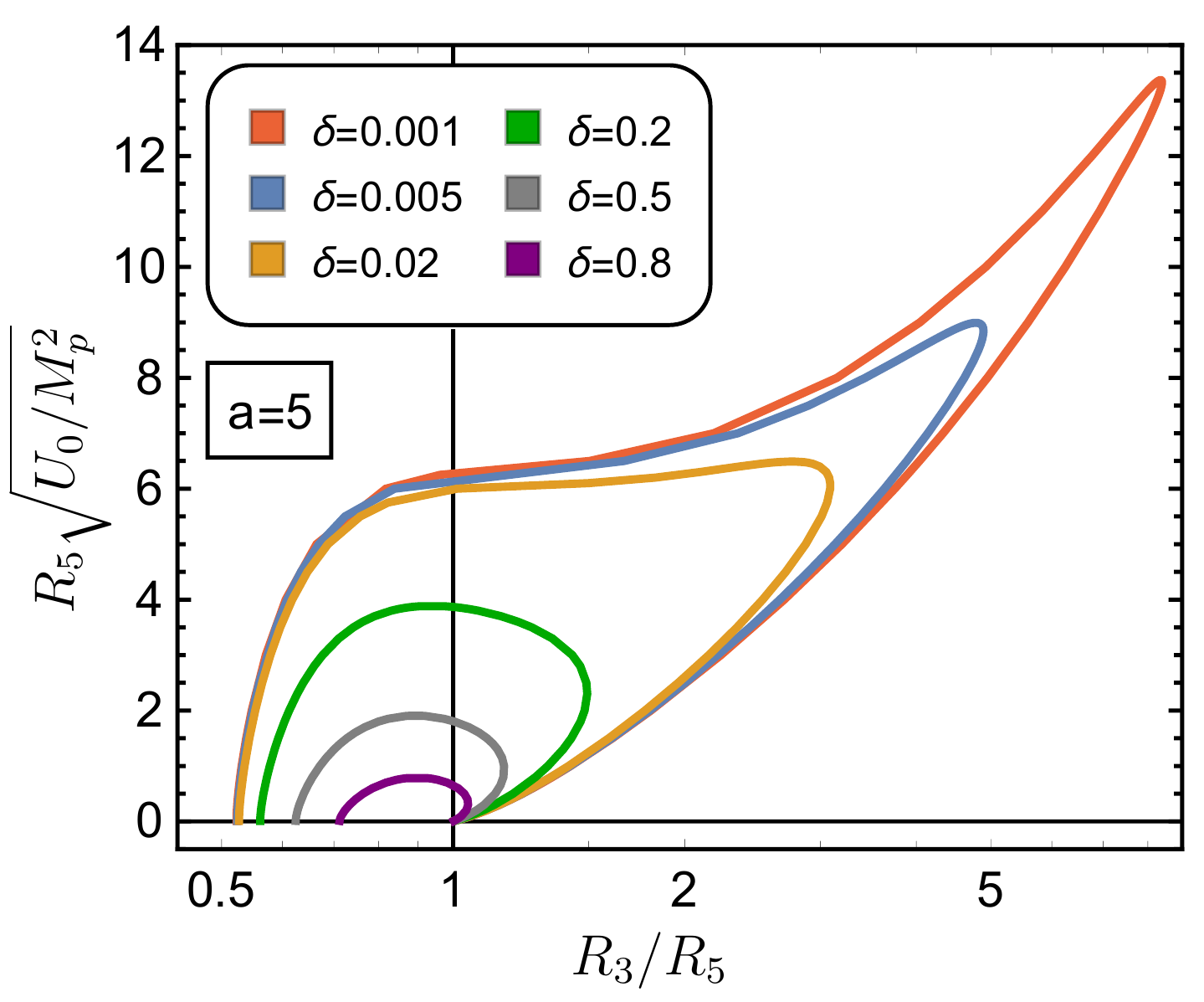}
\caption{ 
The trajectories of the BON bounce solutions in the $(\eta, R_5)$ plane are shown for $a=5$ and various $0 < \delta < 1$. 
Each value of $\delta$ has a BON-FV branch of solutions, with $R_3 \approx R_5$ ($\phi_0 \approx 0$) in the $u_0 \ll 1$ limit where the potential becomes unimportant. 
The BON-CDL branch of solutions has $R_3 < R_5$ ($\phi_0 \approx \phi_\text{cdl}$) in the $u_0 \ll 1$ limit, and an action $\Delta S \approx S_\text{CDL}$ that is large compared to $\pi^2 M_p^2 R_5^2$. 
When $u_0 \gtrsim 1$, the potential $U(\phi)$ is no longer a small perturbation to the Witten bubble, and we find that the BON-FV and BON-CDL branches merge together, with $u_0$ for the bounce solution bounded from above by some $u_0^\text{max}$.
For $u_0 > u_0^\text{max}$, all solutions to the equations of motion are of the overshoot type.
As $\delta \rightarrow 0$ the CDL action diverges as $\delta^{-1}$, and the unification of the BON-FV and BON-CDL branches is pushed to ever-larger values of $u_0\gg 1$.
}
\label{fig:Uofdelta}
\end{figure}

Figure~\ref{fig:paramspace} shows an example of the BON parameter space $(\phi_0, R_5)$, 
for relatively large $\delta \equiv U_\text{fv} / U_0 = 0.8$, and for six values of $a$, chosen to highlight the transition between CDL and HM solutions.
For $(\phi_0, R_5)$ points outside each curve, the initial conditions lead to overshoot solutions; inside each curve, e.g.~for $\phi_\text{HM} < \phi_0 < 0$ with relatively small $U_0 R_5^2 < M_p^2$, the solutions to the EOM undershoot the false vacuum. Bounce solutions are shown along the solid lines.

For $\delta = 0.8$, \eqref{eq:necessary} guarantees the existence of CDL solutions for $a \geq 1.1$. For the three $a > 1.1$ examples, the value of $\phi_\text{CDL}$ is plotted in the left panel of Fig.~\ref{fig:paramspace} as a solid black mark placed at $(\phi_\text{CDL}, R_5 = 0)$ .
As anticipated, we find that the bounce solutions on the right half of the plots approach $\phi_0 \rightarrow \phi_\text{CDL}$ in the $u_0 \rightarrow 0$ limit. 
If $a < 1.1$, on the other hand, there is no CDL solution: instead, the $u_0 \rightarrow 0$ hybrid BON solutions approach the HM value, $\phi_0 \rightarrow \phi_\text{HM}$. To emphasize this convergence, we plot $\phi_0$ in units of $\phi_\text{HM}$ for each curve, where the exact value of $\phi_\text{HM}(a, \delta)$ for the potential \eqref{eq:hatU} is given in \eqref{eq:exactphihm}.

In all six examples, the left edge of each curve ends at $\phi_0 \rightarrow 0$ in the $u_0 \rightarrow 0$ limit. This is the BON-FV solution, or the ``pure'' BON. As we expect from our analytic model, in the $u_0 \ll 1$ limit, $\eta$ is somewhat bigger than $1$; $\phi_0$ is positive; and $\phi_0 / \phi_\text{HM}$ is negative. This behavior persists until $u_0 \gtrsim 1$, at which point $\phi_0$ changes sign and begins to approach $\phi_\text{HM}$.

At larger $u_0 \gtrsim 1$, we find that the existence of undershoot solutions is bounded from above, where the BON-CDL (or BON-HM) branch merges with the matching BON-FV set of bounce solutions. Thus for a given potential there is an upper limit on $u_0$, $u_0\leq  u_0^\text{max}$. This $u_0^\text{max}$ can be understood as an upper bound on $R_5$ for a fixed barrier height, or an upper bound on $U_0$ given a fixed $R_5$, beyond which there is no longer any BON instability.
This does not mean that the vacuum is stable: it is still subject to the regular CDL (or HM) bubble nucleation, with finite $\phi(0) = \phi_0$ and $\phi'(0) = 0$ boundary conditions at the center of the bubble.

The right panel of Fig.~\ref{fig:paramspace} shows the action $\Delta S$ as a function of $\phi_0$, for the same fixed $\delta = 0.8$. As before, the rightmost edge of each curve ends at a $u_0 \rightarrow 0$ hybrid BON-CDL or BON-HM solution, while the leftmost edges end at BON-FV $\phi_0 \approx 0$ solutions, also with $u_0 \rightarrow 0$. The intermediate portions of each curve correspond to larger values of $u_0 \leq u_0^\text{max}$.
Note that the HM action, $\Delta S_\text{HM} = 24\pi^2 (1/\delta - 1) M_p^4/U_0 \simeq 59.2 M_p^4 / U_0$, is larger than any of the CDL actions, which can be read off of the rightmost endpoints of the $a \geq 1.1$ BON-CDL curves.

Following any of the curves from the rightmost edge towards the center, we see that the hybrid BON actions initially increase with nonzero $u_0$: thus, the tunneling rates for the hybrid BON-CDL or BON-HM bounces is slower than the associated CDL or HM rate, at least for small $u_0$.
This is what we expect from our interpretation of the hybrid bounces of Section~\ref{sec:exotic}: for small $u_0$, the BON-CDL hybrid is essentially just the coincidentally concentric combination of a BON bubble inside a larger CDL solution. It is easier to nucleate a CDL bubble without also putting a BON at its center.

Continuing along each curve towards the BON-FV solutions, $\phi_0 \rightarrow 0$, eventually the BON actions $\Delta S$ decrease below $\Delta S_\text{CDL}$ or $\Delta S_\text{HM}$; and in the extreme $u_0 \rightarrow 0$ limit of the BON-FV branch, its action $\Delta S \approx \pi^2 M_p^2 R_5^2$ becomes vanishingly small compared to the CDL or HM actions, both of which scale as $\Delta S_\text{CDL(HM)} \propto 1/U_0$.

In Fig.~\ref{fig:Uofdelta} we show another example, this time with $a=5$ kept constant and $0.001 \leq \delta \leq 0.8$ varied. For this plot we show the shooting parameter in terms of $\eta = R_3 / R_5$ rather than $\phi_0(\eta)$, while keeping $\sqrt{u_0} \propto R_5 \sqrt{U_0}$ on the vertical axis. 
Thanks to the larger value of $a$ in this example, every value of $\delta \leq 1$ corresponds to a $\hat{U}(\phi)$ that has a CDL solution.
As $\delta \rightarrow 0$, the CDL action diverges as $\Delta S_\text{CDL} \propto 1/\delta$. The BON-FV and BON-CDL branches become increasingly more distinct, allowing solutions to exist for larger values of $u_0$. 
As a consequence of the increasingly large difference between $\Delta S_\text{CDL} \sim \pi^2 M_p^4/ U_\text{fv}$ and $\Delta S_\text{BON} \sim \pi^2 M_p^2 R_5^2$ in the $U_\text{fv} \rightarrow 0$ limit, the BON-FV and BON-CDL branches become ever more distinct, allowing BON solutions to exist at larger values of $u_0$.
As a first approximation one could estimate the upper bound on $u_0$ by setting $\Delta S_\text{CDL} \sim \pi^2 (M_p R_5^\text{max})^2$. This approach works well if $u_0^\text{max} \lesssim 1$, but is less accurate for $u_0^\text{max} \gg 1$, where the BON action is no longer proportional to $R_5^2$.

\subsubsection{Field profiles} \label{sec:profiles}


Now that we have established which values of $\eta$ and $R_5$ correspond to bounce solutions, we can inspect the solutions for $\rho$ and $\phi$. 
To verify our expectations from the analytic model, we begin with a specific example with small $u_0 = 10^{-2}$.

\begin{figure}
\centering
\includegraphics[width=0.8\textwidth]{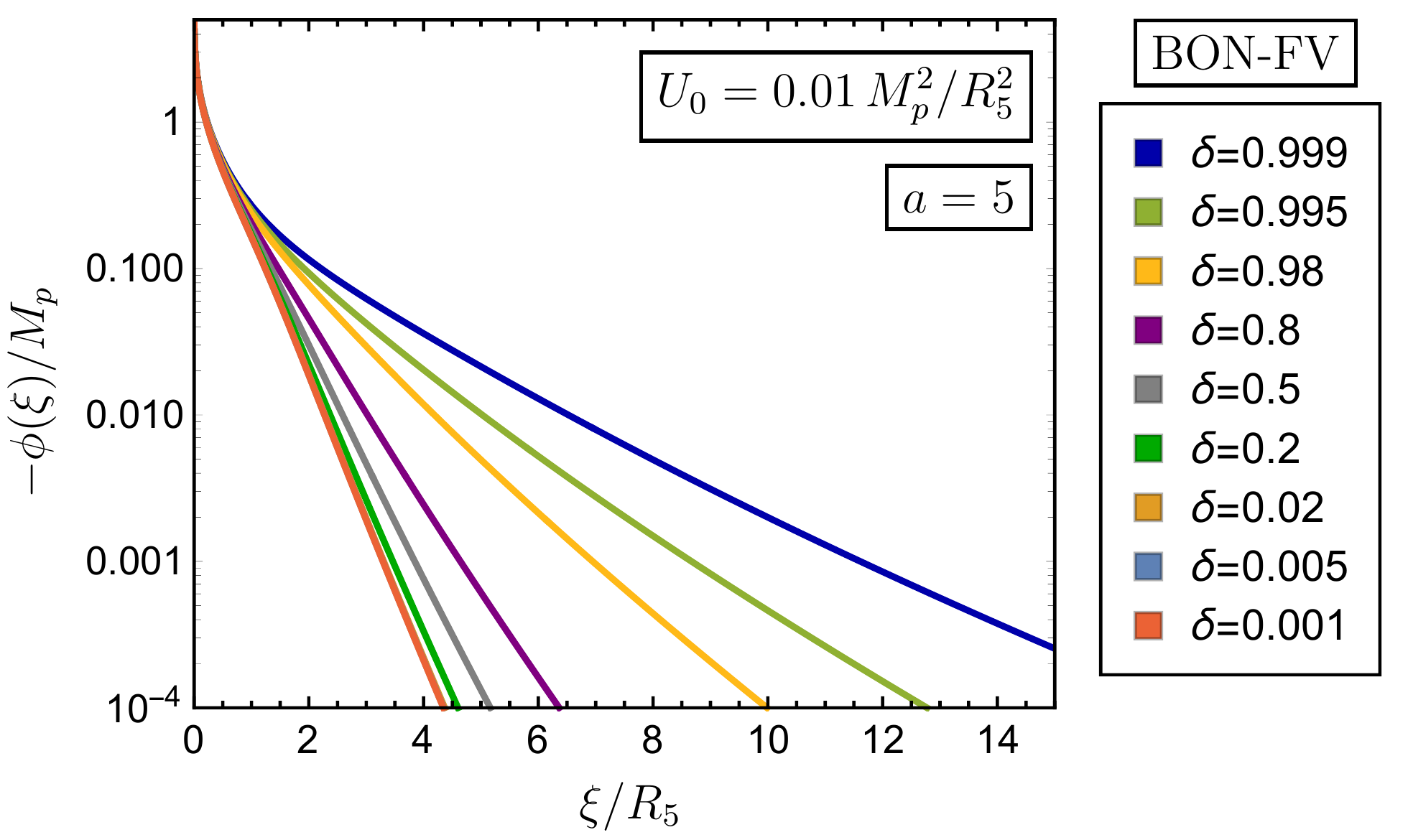}

\hspace{1.4cm} \includegraphics[width=0.87\textwidth]{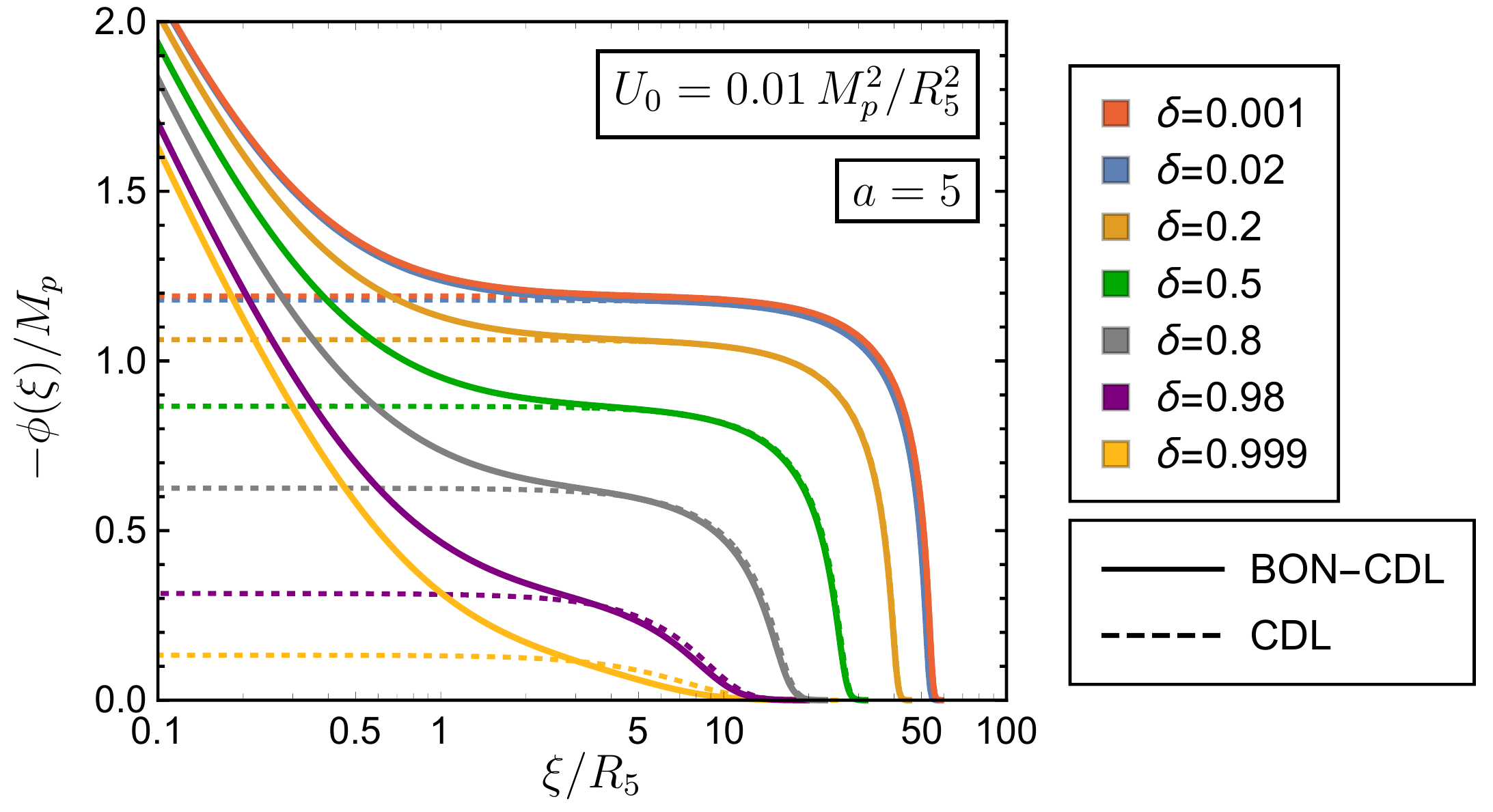}
\caption{ 
Profiles of $\phi(\xi)$ with $U_0 = 0.01\, M_p^2 / R_5^2$ on the BON-FV (top) and BON-CDL (bottom) branches, for $a=5$ and various $0 < \delta < 1$.  
For the BON-FV solutions the $0 < \xi \lesssim R_5$ profiles are nearly identical: only for $\xi \gtrsim R_5$ does the shape of the potential have much impact on the profiles for $\phi(\xi)$.
On the BON-CDL branch with small $\delta \lesssim 0.5$, $\phi(\xi)$ quickly approaches $\phi(\xi) \simeq \phi_\text{CDL}$, where it lingers for a while before passing through the barrier. For $\delta \gtrsim 0.5$ the approach to the false vacuum is more immediate. 
The BON-CDL $\phi(\xi)$ is shown on a log-linear plot, so that the BON-like ($\xi \lesssim R_5$) and CDL-like ($\xi \gg R_5$) parts of the solution are simultaneously visible. For $\delta \leq 0.5$ the transition from $\phi(\xi) \simeq \phi_\text{CDL}$ to $\phi(\xi) \approx \phi_\text{fv}$ is rapid, implying the validity of the thin-wall approximation. As $\delta \rightarrow 1$ there is no intermediate region of constant $\phi(\xi) \approx \phi_\text{CDL}$, so the thin-wall description does not apply. 
}
\label{fig:profilephi}
\end{figure}

In Fig.~\ref{fig:profilephi}, we show the $\phi(\xi)$ profiles for fixed $u_0 = 10^{-2}$ on the BON-FV and BON-CDL branches, for fixed $a=5$ and a variety of $\delta$ values, $0.001 \leq \delta \leq 0.999$. 
On the BON-FV branch,  in the $\xi \lesssim R_5$ region, $\phi(\xi)$ matches the Witten BON profile with $\phi_0 \approx 0$ independently of the value of $\delta$. After $\xi \gtrsim R_5$, the potential  becomes important: for example, at larger $\delta$, the solutions for  $\phi(\xi)$ take longer to approach the false vacuum. 

On the BON-CDL branch, especially for smaller values of $\delta \lesssim 0.5$, the profile for $\phi(\xi)$ has two distinct regions: an inner core, $\xi \lesssim R_5$, where $\phi(\xi)$ closely matches the $U=0$ Witten  solution with $\phi_0 \approx \phi_\text{CDL}$; and a CDL-like region $\xi \gtrsim  R_5$, with $\phi(\xi) \approx \phi_\text{CDL}$ remaining approximately constant until $\xi \gg R_5$.
In the lower plot we also show the CDL solution as a dashed line, to emphasize its similarity with the BON-CDL hybrid. For $\xi \gtrsim 3 R_5$ the BON-CDL and CDL solutions match nearly exactly, particularly for $\delta \ll 1$.

\medskip

\begin{figure}[t]
\centering
\includegraphics[width=\textwidth]{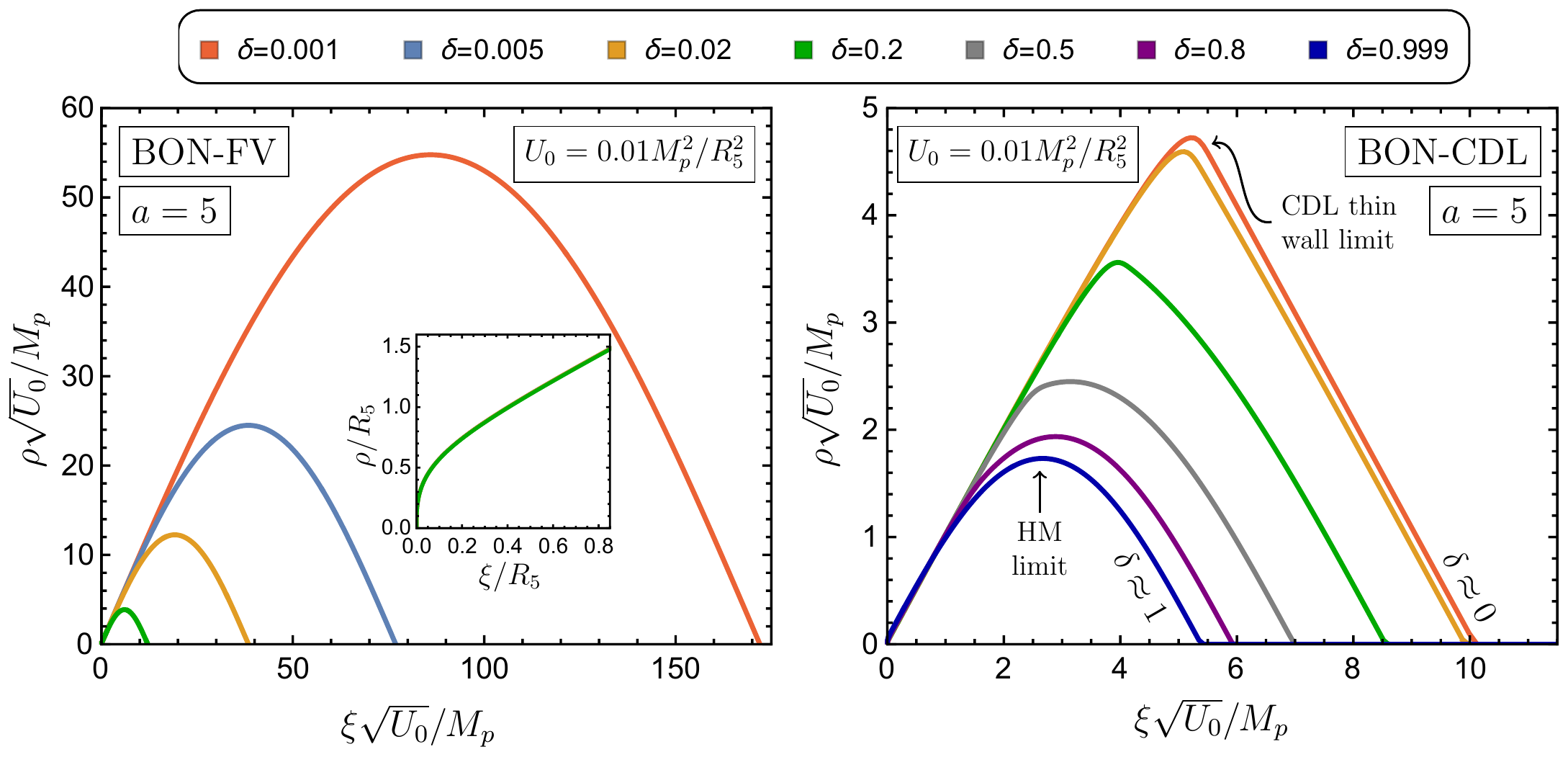}
\caption{ 
Profiles for $\rho(\xi)$ for $U_0 R_5^2 / M_p^2=0.01$ on the BON-FV ({\bf left}) and BON-CDL ({\bf right}) branches.  
For the BON-FV branch the solutions for $\rho(\xi)$ are nearly identical to the false vacuum solution $\rho_\text{fv}(\xi)$ for $\xi \gtrsim R_5$. This encompasses the majority of the domain $0 \leq \xi \leq \xi_\text{max} \approx \pi \Lambda$, which is why we only show $\delta = 0.001 \ldots 0.2$.
For the BON-CDL solutions, the profiles for $\rho(\xi)$ follow the $\xi \gg R_5$ linear growth of \eqref{eq:WittenLarge} out to $\xi \gtrsim 10 R_5$ or larger, with a relatively sharp transition from $\rho' \approx 1$ to $\rho' \approx -1$ in the case of small $\delta$.
As $\delta \rightarrow 1$, the BON-CDL solutions for $\rho(\xi)$ more closely match the Hawking--Moss behavior $\rho(\xi) \approx \rho_\text{HM}(\xi)$ for all $\xi \gtrsim R_5$, even though the $a=5$ CDL and HM solutions for $\phi(\xi)$ are distinct.
}
\label{fig:profilerho}
\end{figure}

In Fig.~\ref{fig:profilerho}, we show the solutions for $\rho(\xi)$ for the same set of potentials $U(a, \delta)$.
On the BON-FV branch the potential barrier has very little effect on $\rho(\xi)$: it is well approximated by 
the $U=0$ Witten BON solution
for $\xi \lesssim \mathcal O(\text{few}) \times R_5$ at small $\xi$, and then by $\rho(\xi) \simeq \rho_\text{dS}(\xi + \mathcal O(R_5) )$ for $R_5 \lesssim \xi \leq \xi_\text{max}$. 
This is the kind of solution shown in Fig.~\ref{fig:rhophidiagram}, where the transition across the potential barrier occurs in a region where $\rho' \simeq 1$ to good approximation.
The behavior on the BON-CDL branch is more varied. At small $\delta \lesssim 0.2$, the solutions are well approximated by the CDL thin-wall limit, where the transition between $\rho' \approx + 1$ and $\rho' \approx -1$ occurs abruptly. Similarly, we can see from Fig.~\ref{fig:profilephi} that $\phi(\xi) \approx \phi_\text{CDL}$ remains constant for $R_5 \lesssim \xi \lesssim \mathcal O(10) R_5$, and that the transition from $\phi(\xi) \approx \phi_\text{CDL}$ to $\phi(\xi) \approx \phi_\text{fv}$ also occurs over a relatively small range of $\xi$.
For the smallest values of $\delta \leq 0.02$, the solutions for $\rho$ are very similar to each other, each having an approximately triangular profile. In the thin-wall approximation the value of $\bar\rho \approx 4.6 \, M_p / \sqrt{U_0}$ at the transition is estimated by \eqref{eq:rho_CDL}.

At larger $\delta \gtrsim 0.5$, the thin-wall results are less accurate. In particular, the thickness of the bubble wall increases, while the bubble itself becomes smaller.
As $\delta \rightarrow 1$ the solution for $\rho$ approaches a different well-understood limit: the sinusoidal Hawking-Moss solution defined in \eqref{eq:phirhoHM},
\begin{align}
\rho_{\delta \rightarrow 1} (\xi \gtrsim \eta^{3/2} R_5) \approx \rho_\text{HM}(\xi + \mathcal O(R_5)).
\end{align}
For $\xi \lesssim R_5$, the solution for $\rho$ at any value of $\delta$ is well approximated by the $U=0$ BON $\rho_\eta(\xi)$, \eqref{eq:BON_eta_xismall}, but with potentially values of $\eta-1$  (see Fig.~\ref{fig:Uofdelta}).

\subsubsection{Instanton action} 

\begin{figure}[t]
\centering
\includegraphics[width=0.9\textwidth]{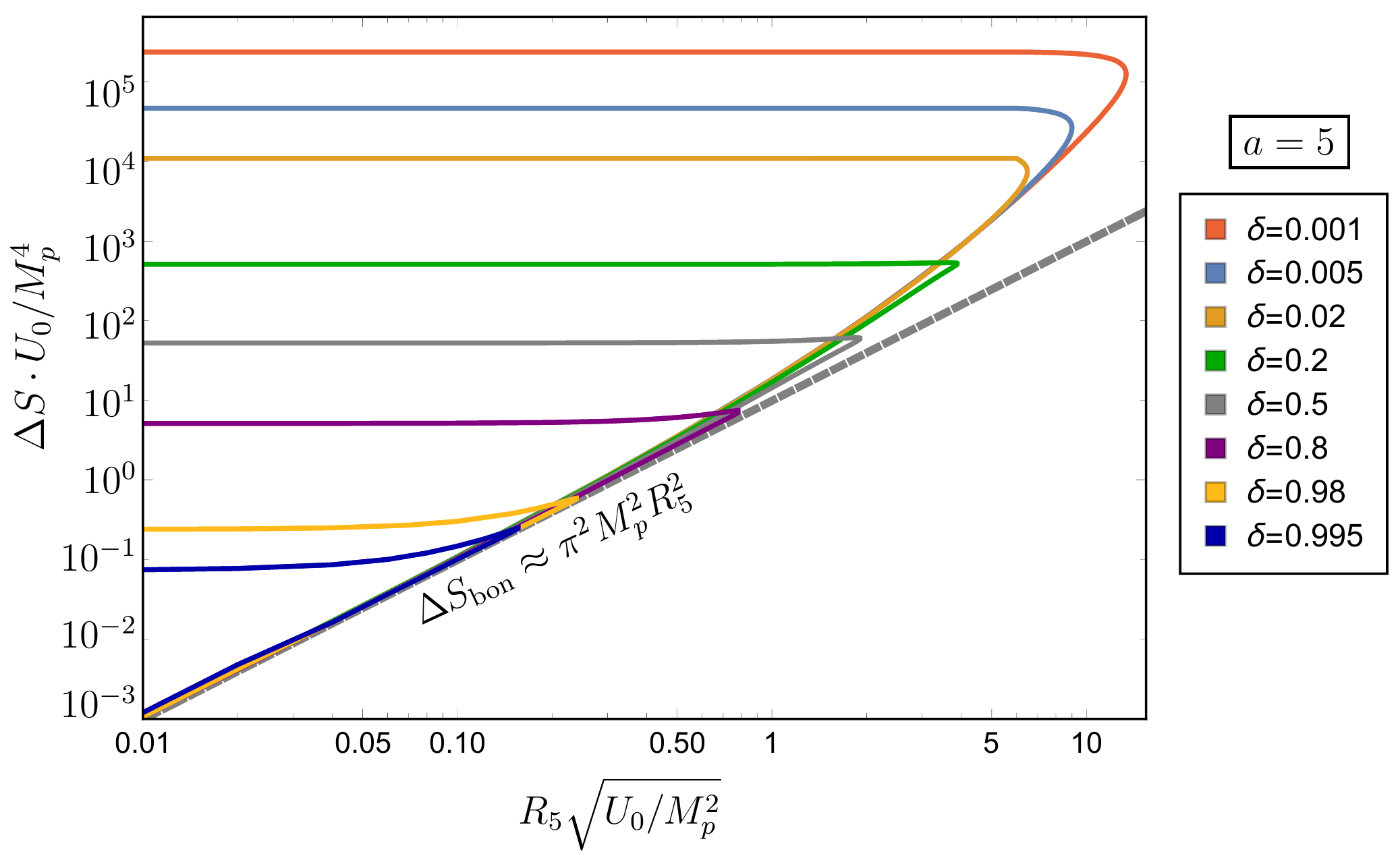}
\caption{
The de Sitter-subtracted Euclidean action $\Delta S$ is shown in units of $M_p^4/U_0$ as a function of $R_5$, for $a=5$ and various $0 < \delta < 1$. The dashed gray line shows the $\Delta S \approx \pi^2 M_p^2 R_5^2$ expectation from the small $u_0 \ll 1$ limit of the BON-FV solutions. It tracks the numerically-computed action closely on this branch for all $u_0\ll 1$. This branch is always of lowest action and determines the tunneling rate.
On the BON-CDL branch the action approaches a constant, the ordinary CDL action $\Delta S \approx S_\text{CDL}$, which is large compared to $M_p^4/U_0$ for $\delta \lesssim 0.5$. For $\delta \ll 1$, $S_\text{CDL} \approx 24 \pi^2 M_p^4/( \delta \cdot U_0)$, grows as $1/\delta$. 
At larger values of $u_0$, where $\pi^2 M_p^2 R_5^2 \sim \Delta S_\text{CDL}$, the BON-FV and BON-CDL branches of bounce solutions meet. (This method of estimating the value of $u_0=U_0 R_5^2 / M_p^2$ at which the two branches meet becomes  imprecise once $u_0 \gtrsim 1$, where $\pi^2 M_p^2 R_5$ is no longer a good approximation to the BON-FV action.)
}
\label{fig:actionSE-R}
\end{figure}

Given numerical solutions for $\phi(\xi)$ and $\rho(\xi)$, the de Sitter-subtracted Euclidean action $\Delta S$ can be obtained by integrating Eq.(\ref{eq:DeltaS_bonfv}).
In Fig.~\ref{fig:actionSE-R}, we plot this $\Delta S$ as a function of $R_5 \sqrt{U_0}$, for the $a=5$, $0.001 \leq \delta \leq 0.995$ examples.
 At small $u_0 \lesssim 1$, the BON-FV action is well approximated by $\Delta S \simeq \pi^2 M_p^2 R_5^2$. Given specific values of $R_5$ and $U_0$ subject to $U_0 R_5^2 \ll M_p^2$, the BON-FV is the lowest-action bubble solution, and is therefore the solution relevant for determining the decay rate. On the BON-CDL branch, the $u_0 \rightarrow 0$ action is approximately constant and equal to the ordinary CDL action, $\Delta S \approx \Delta S_\text{CDL}$, up to corrections of order $M_p^2 R_5^2$ that are negligibly small.
 
In the $\delta \gtrsim 0.5$ examples, thanks to the relatively small values of $\Delta S_\text{CDL}$, the $\Delta S_\text{CDL} \sim \pi^2 M_p^2 (R_5^\text{max})^2 $ method of estimating $R_5^\text{max}$ works reasonably well, as these $R_5^\text{max}$ correspond to $\sqrt{u_0} \lesssim 1$.
Once $u_0^\text{max} \gg 1$, as in the $\delta < 0.5$ examples, this approximation becomes increasingly inaccurate.

The $\Delta S_\text{BON-FV} \ll \Delta S_\text{CDL}$ hierarchy persists up until the largest values of $u_0$: however, for the larger $\delta \approx 1$ examples, the lower-action BON branch can increase above $\Delta S_\text{BON} > \Delta S_\text{CDL}$ as $u_0 \rightarrow u_0^\text{max}$.
For these $\delta \gtrsim 0.8$ potentials, there are values of $U_0$ and $R_5$ where a BON instability exists, but is slower than the competing CDL process.
In the opposite limit, $\delta \rightarrow 0$, the relationship between $\Delta S_\text{BON}(u_0^\text{max})$ and $\Delta S_\text{CDL}$ is flipped: in this case, if the BON exists, it is faster than CDL. The clearest examples are the $\delta \leq 0.02$ curves.


\subsection{Exponentially growing potential: five-dimensional CC} \label{sec:cc}

In Section~\ref{sec:bubbleexist}, we showed that the existence of BON solutions depends on the asymptotic form of the classical scalar potential. For potentials $U(\phi) \sim U_0 e^{a \phi / M_p}$, we found that $a > -\sqrt{6}$ is a necessary constraint, following from the requirement that the metric $ds_5$ should be free of singularities as $\xi \rightarrow 0$.
In the previous section we have focused on potentials with $a >0$, which asymptote to zero in the compactification limit. But the bound $a > -\sqrt{6}$  is weaker, permitting even exponential growth of the potential as long as the rate is not too large.\footnote{Furthermore, a broad class of even faster-growing potentials probably admit a different type of BON solution under some plausible assumptions about the spectrum. Fluxes, for example, violate the inequality above, but can be screened by suitable dynamical sources, leading to a hybrid Schwinger production--BON decay process. In this case the BON core lies inside the region where the flux has been screened, and the modulus moves in a softer potential. A particular example of this type was studied in~\cite{Brown:2010mf}. A general treatment of such cases should be possible along the lines of our study above, but we defer it for future work.}
Now we consider cases where the potential satisfies $a > -\sqrt{6}$ but grows exponentially in the compactification limit. The simplest examples of this type arise in the dimensional reduction of theories with positive higher-dimensional CC, c.f.~\eqref{eq:sourcesofmodulipotentials}. Here the asymptotic growth of the four-dimensional potential is exponential, but slow enough that the appropriate bubble of nothing initial conditions as $\xi\rightarrow 0$ still match \eqref{eq:BON_eta_xismall}.

\begin{figure}
\centering
\includegraphics[width=0.8\textwidth]{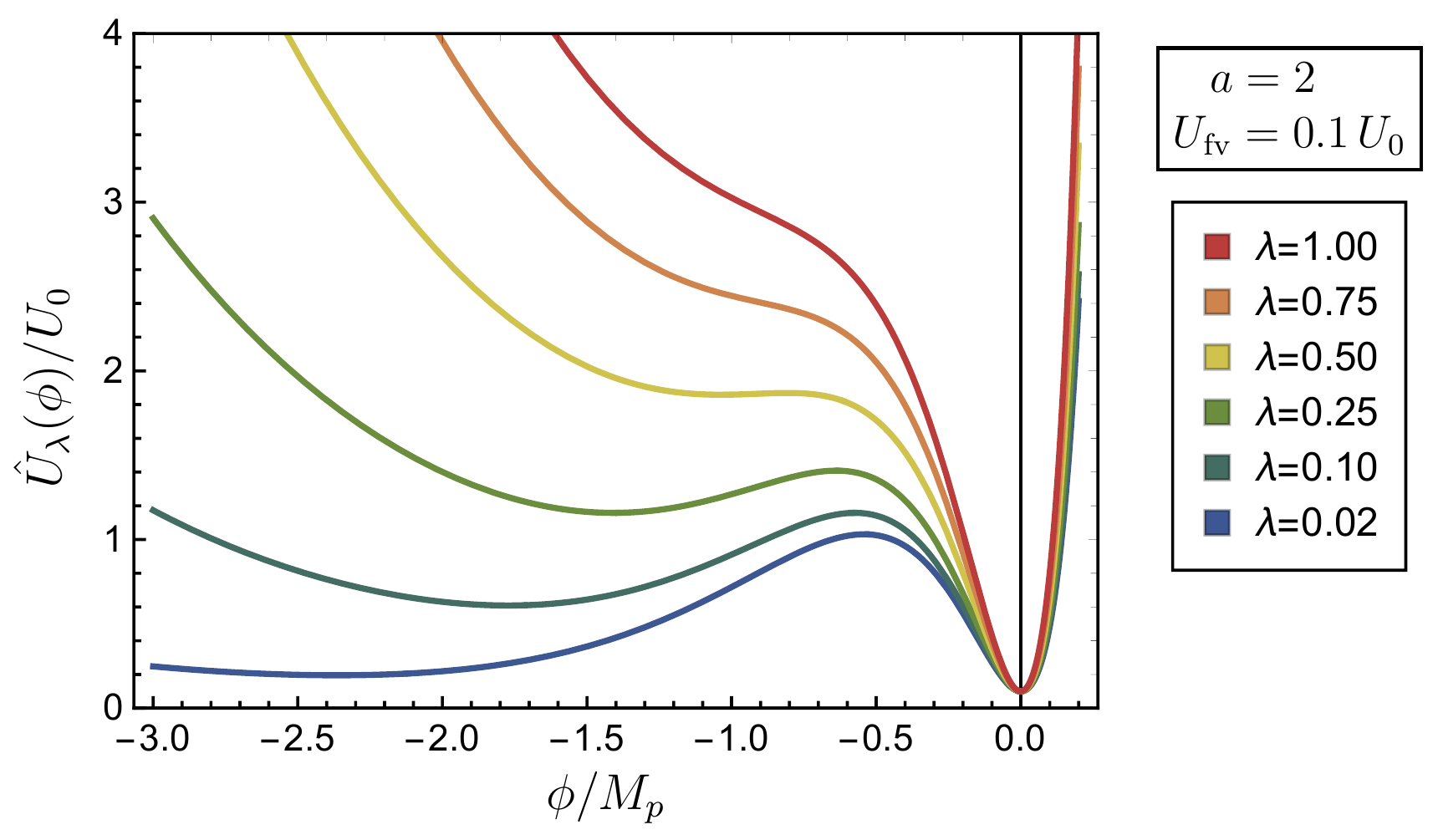}
\caption{The scalar potential $\hat U_\lambda(\phi)$ of \eqref{eq:hatUlambda}, for several values of $\lambda \equiv \Lambda^\text{CC}_5 M_p^2 / U_0$.}
\label{fig:hatULambda}
\end{figure}

Several example potentials of this type are shown in Fig.~\ref{fig:hatULambda}. Specifically, they correspond to the following modification of the $\hat{U}$ of \eqref{eq:hatU}:
\begin{align}
	\hat U_\lambda(\phi) &\equiv  U_0 \left( \frac{\hat U(\phi)}{U_0} + \lambda \exp \left(  -\sqrt{ \frac{2}{3}} \frac{\phi}{M_p} \right) \right),
	&&
	\lambda \equiv \frac{ \Lambda^{\rm\tiny CC}_5 M_p^2}{U_0}.
\label{eq:hatUlambda}
\end{align}
The coefficient for the exponential growth, $a = - \sqrt{2/3}$, is motivated by an $n=1$ scenario in which $U(\phi)$ is dominated by the five-dimensional CC, $\Lambda^\text{CC}_5$, as in \eqref{eq:sourcesofmodulipotentials}. 
This is small enough compared to the limiting value $a = - \sqrt{6}$ that the field solutions for $\xi \ll R_5$  are still closely approximated by the  $U=0$ Witten solutions.
Here $U_0$ is the height of the barrier in the $\lambda \rightarrow 0$ limit: however, when $\lambda \gg 1$ the exponential overwhelms the $\hat U$ contribution, and there ceases to be any local maximum in $\hat U_\lambda(\phi)$. Then it is clear that there is no ordinary CDL solution that tunnels in the compactification direction.\footnote{Although there is probably still one in the decompactification direction; our potentials are not meant to be realistic on this side, since it does not influence the BON solutions in which we are interested.}  Nonetheless, BON solutions can still exist. This is intuitively clear from the shooting problem in the inverted potential --- there may be no point $\phi_0 <0$ where a rolling ball can start from rest and reach the false vacuum at the origin, but it can still reach the false vacuum by shooting in from $-\infty$ at high velocity.

\begin{figure}
\centering
\includegraphics[width=\textwidth]{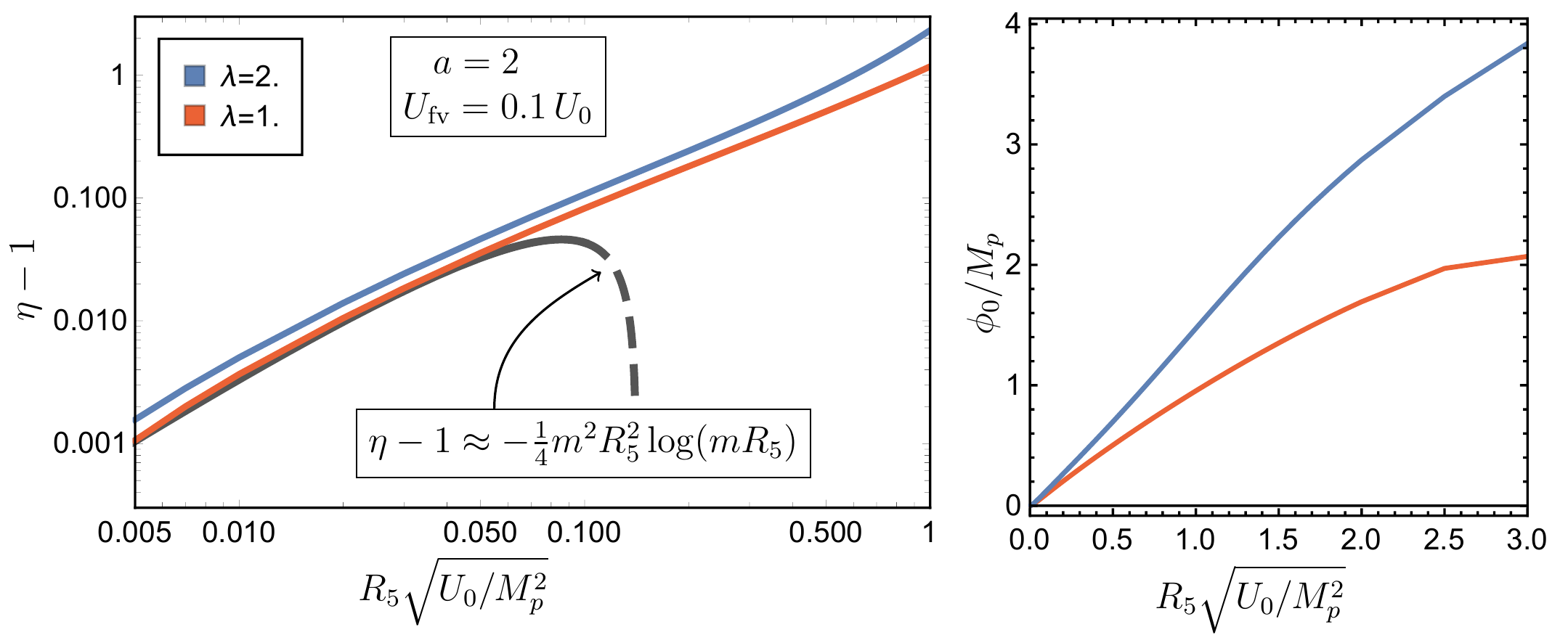}
\caption{Shooting parameter solutions on the   BON-FV branch for potentials dominated in the compactification limit by exponentially growing contributions from a five-dimensional CC. 
\textbf{Left:}  The value of the shooting variable $\eta \equiv R_3/R_5$ is shown as a function of $\sqrt{u_0}$ for two values of dimensionless parameter $\lambda = \Lambda_5^\text{CC} M_p^2 / U_0 \leq 2.0$. The thick dashed line shows the theoretical expectation from the piecewise quadratic model, \eqref{eq:eta}, in the limit $m R_5 \ll 1$, in terms of the $m^2 \sim a^2 U_0 / M_p^2$ of \eqref{eq:massU}. For $\sqrt{u_0} \lesssim 0.1$, the agreement is quite close, indicating that the exponentially growing potential has little effect on the shooting solution in this regime. As $m R_5 \rightarrow 1$ the corrections to  $\eta \approx 1$ are dominated by the $\mathcal O(m^2 R_5^2)$ terms that are not logarithmically enhanced.
\textbf{Right:} Here we show $\phi_0$, the alternative formulation of the initial conditions, as a function of $\sqrt{u_0}$, for the same  values of $\lambda$. 
The $\lambda = 1.0$ and $\lambda = 2.0$ potentials (red and blue in the plot) do not have any CDL or HM solution, or any hybrid bounces.
}
\label{fig:phi0R5Lambda.pdf}
\end{figure}

\begin{figure}
\centering
\includegraphics[width=0.55\textwidth]{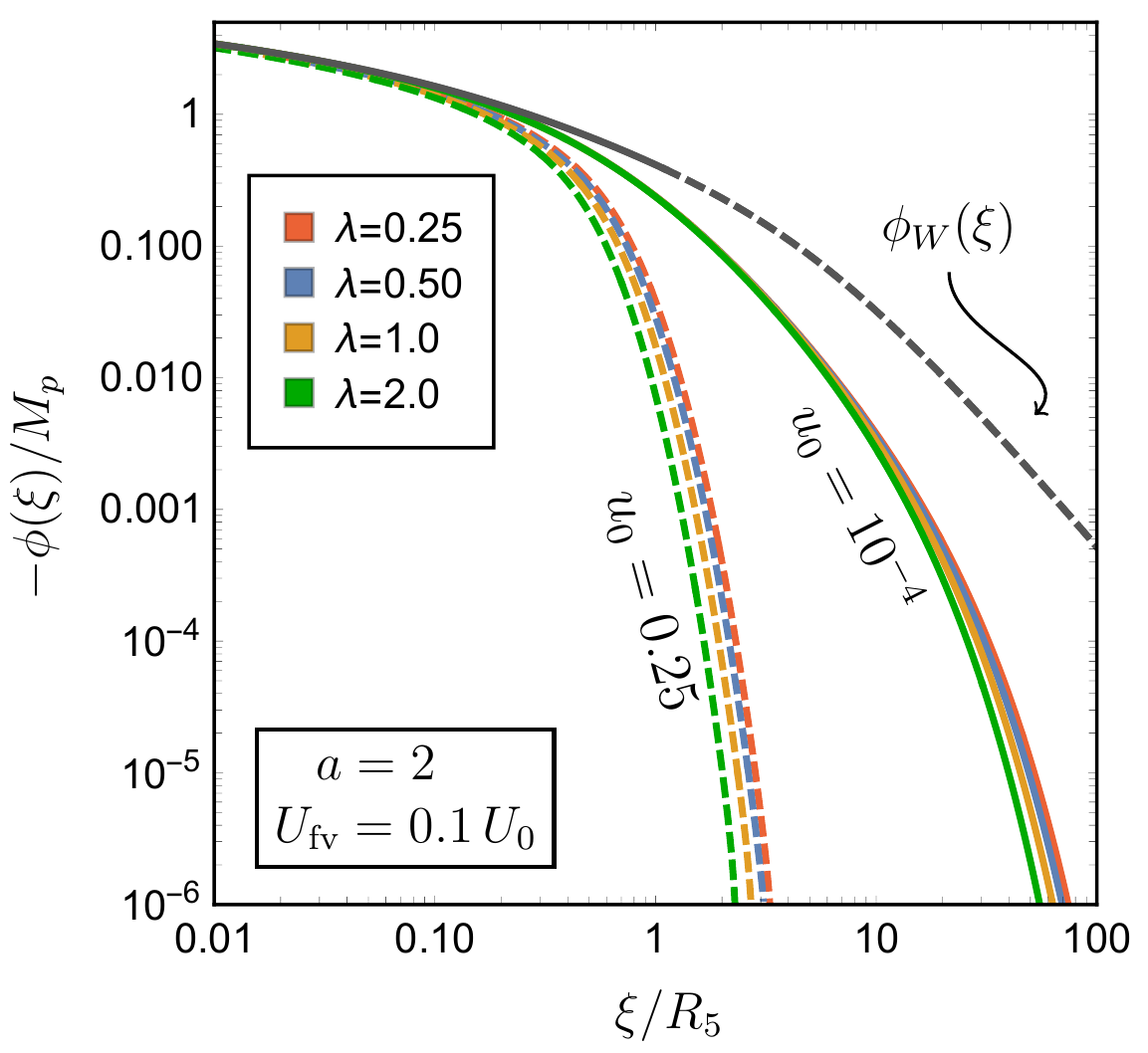}
\caption{The modulus profile on the   BON-FV branch for exponentially growing CC-dominated potentials. 
The solid and dashed lines correspond to $u_0  = 10^{-4}$ and $u_0 = 0.25$, respectively.
At small $\xi$ the profile remains BON-like, despite the rapid growth in $U(\phi)$. As expected from the small $\xi$ series expansions of Appendix~\ref{appx:SmallXi}, the departure from $\phi(\xi) \approx \phi_\text{bon} (\xi)$ occurs as $\xi$ approaches $2 R_5 / 3$. 
In the examples with larger $u_0$ the approach to the false vacuum is fast, occurring here at $\xi \sim \mathcal O(R_5)$, while in the examples with $u_0 = 10^{-4}$, $\phi(\xi)$ only approaches $\phi_\text{fv}$ in the $\xi \gg R_5$ limit.
The $U_0 \rightarrow 0$ limit indicated by $\phi_W$ (dashed, gray) has the slowest approach to the false vacuum, as it is given by the power law $\phi(\xi) \propto 1/\xi^2$ rather than the exponentially damped $\phi(\xi) \propto e^{- m \xi}$ in the $\xi \gg R_5$ limit.
}
\label{fig:phixilambda}
\end{figure}

\begin{figure}
\centering
\includegraphics[width=0.85\textwidth]{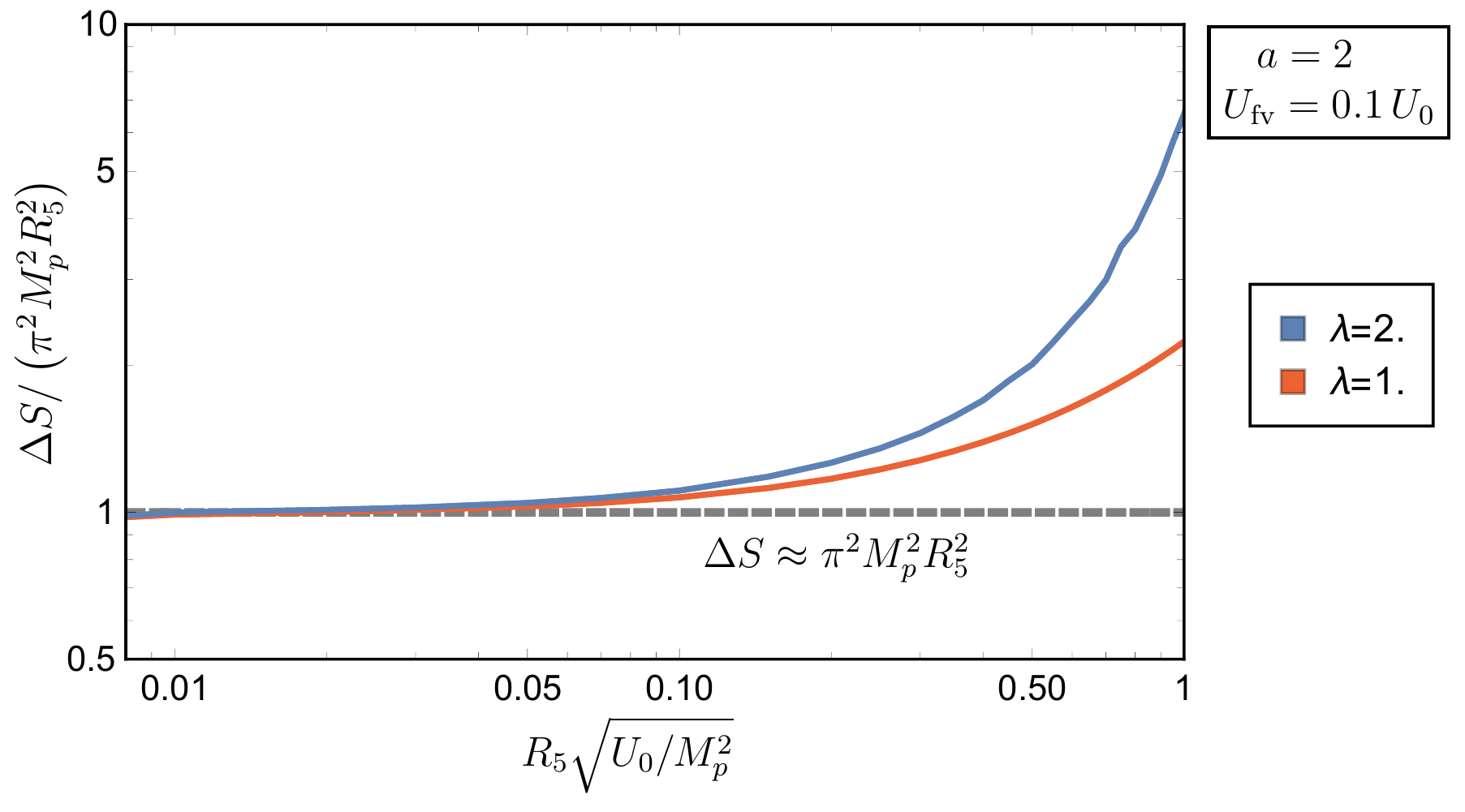}
\caption{The action $\Delta S$ on the  BON-FV branch  for exponentially growing CC-dominated potentials, shown here normalized by a factor of $\pi^2 M_p^2 R_5^2$ to highlight the leading deviations. At small $u_0 \lesssim 1$, the action remains well approximated by $\pi^2 R_5^2 M_p^2$, with corrections to $\Delta S$ that become large as $m R_5 \gtrsim 1$.}
\label{fig:actionLambda.pdf}
\end{figure}

We solve these cases numerically with the same methods described in Section~\ref{sec:prel}. Sample shooting parameter solutions, field profiles, and actions are shown in Figs.~\ref{fig:phi0R5Lambda.pdf}, \ref{fig:phixilambda}, and~\ref{fig:actionLambda.pdf}, respectively.  In Fig.~\ref{fig:phi0R5Lambda.pdf} we provide a comparison with the analytic solution for the shooting parameter derived in Section~\ref{sec:bounceofnothing}. The agreement between the numeric solution for $\eta$ and the prediction from \eqref{eq:eta} is quite good in the small $u_0 \ll 1 $ limit.  The potential, despite its exponential growth, does not grow fast enough to significantly impact the shooting solution for $\eta = R_3/R_5$. Similarly, in  Fig.~\ref{fig:phixilambda} we see that the modulus profile $\phi(\xi)$ continues to match Witten's solution at small $\xi\lesssim R_5$. At larger $\xi \gtrsim R_5$, $\phi(\xi)$ departs from the $U=0$, $\phi_W \propto 1/\xi^2$ solution to instead approach the false vacuum exponentially fast, as in \eqref{eq:phirhoexp}.

Fig.~\ref{fig:actionLambda.pdf} shows that the action continues to be dominated by the $\pi^2 R_5^2 M_p^2$ term in the small $u_0 \ll1$ regime, with corrections that become increasingly large as $u_0 \rightarrow  \mathcal O(1)$.


\section{Conclusions}\label{sec:conclusions}
In this work we have studied bubble of nothing instabilities of four-dimensional de Sitter vacua in theories with compactified extra dimensions. By mapping the problem of finding the relevant gravitational instanton to solving a four-dimensional CDL problem, we have been able to treat the moduli potential in a general, model-independent way, separating the physics of moduli stabilization from the  criteria that determine the existence and qualitative properties of these bubbles.

The asymptotic behavior of the scalar potential in the compactification limit,  as well as the dimensionality of the internal sphere that shrinks to zero volume at the bubble wall, are the primary data determining the existence of BON instabilities. Focusing on the case where the shrinking internal manifold is an $S^1$, as in the original BON instability of Kaluza-Klein theory identified in \cite{Witten:1981gj}, we have shown that BON solutions are compatible with stabilization of the radial modulus even in certain cases where the scalar potential grows exponentially fast in the compactification limit. 
The presence of the potential is in fact typically irrelevant near the center of the bounce (corresponding to the bubble wall), as long as the rate of exponential growth is smaller than the critical value we derive in Section~\ref{sec:bounceofnothing}.
Combining this observation with other properties of the equations of motion, we have constructed approximate analytic solutions for the geometry and action of the BON.
The Euclidean action of the solution is finite, and it remains finite even for vanishing vacuum energy density in the false vacuum, a limit in which other decay channels described by either CDL or HM instantons become irrelevant. When the KK scale $1/R_5$ is the largest mass scale in the problem, the potential provides small, calculable corrections to the tunneling exponent.  

Thus, for phenomenological purposes, the most interesting properties of these bubbles are that they can survive modulus stabilization with  positive cosmological constant, even in cases where the stabilizing potential grows exponentially in the compactification limit; the decay rate remains unsuppressed in the limit of degenerate vacua, so long as this limit does not also restore supersymmetry (as seems likely in our Universe); and, if the potential admits an ordinary CDL decay in either the compactification or decompactification limits, the BON process can be faster, especially if the KK radius $R_5$ is relatively small.
For these potentials with CDL (or HM) solutions, we find from our numeric study that in the limit of large $R_5$, where the BON action approaches or exceeds the CDL or HM action, there is some $R_5^\text{max}$ above which there are no longer any BON solutions to the equations of motion. In these cases the vacuum decay occurs exclusively through the regular CDL (or HM) instanton.

Our methods may be extended to other classes of BON solutions. For example, as we have seen in Section~\ref{sec:bubbleexist}, there is a second class of BON boundary conditions arising when the asymptotic behavior of the potential is equally important to the other terms in the CDL equations. This is not reflective of a fine-tuning, but rather arises naturally in cases where the dominant contribution to the asymptotic potential comes from curvature on the collapsing internal manifold. The techniques described in this paper can be directly applied to such cases. Furthermore, even potentials that na\"ively grow too quickly to admit a BON solution could also be of interest: if other degrees of freedom become active near the bubble wall to tame the potential, bubble solutions may still exist. The most familiar example where this could arise is in the context of flux compactifications. Here, Gauss' law requires that flux wrapped over the collapsing sphere must be either screened by nucleation of dynamical sources or  ``slip off" onto other parts of the internal manifold. In this process the growth of the potential can be softened such that one of the types of BON solutions is allowed. Examples of both types have appeared in the literature in the context of explicit top-down models~\cite{Horowitz:2007pr,BlancoPillado:2010df,Brown:2010mf,BlancoPillado:2010et,Ooguri:2017njy}. The impact of the flux on the decay rate is model-dependent in general, but at least in some cases the contribution of charged sources is subleading~\cite{Horowitz:2007pr}. Bottom-up methods like those used in this work could also be adapted to these cases, most easily in the point charge approximation~\cite{Brown:2010mf}, allowing the study of BON in flux compactifications while remaining agnostic about the precise form and origin of  the potential in the vicinity of the false vacuum. This is an intriguing direction for future work.

What can we conclude about our universe? From the decay rate estimates and the age of the universe we can infer, for example, an upper bound on the size of compactified dimensions~\cite{Draper:2021gmq}: for a collapsing internal $S^1$, a decay rate similar to that of Witten's bubble implies a lower bound on the size of the KK circle, $2 \pi R_5 \gtrsim 50 M_p^{-1}$. 
More generally, the presence of rapid BON instabilities can potentially shrink the region of the string landscape that is capable of giving rise to a vacuum compatible with our universe.

\section*{Acknowledgments}
We thank Nathaniel Craig, Michael Dine, Paddy Fox, Arthur Hebecker, and Matt Reece for comments and discussions.
PD and BL acknowledge support from the US Department of Energy under Grant No.~DE-SC0015655.
The research of IGG is funded by the Gordon and Betty Moore Foundation through Grant GBMF7392, and by the National Science Foundation through Grant No.~NSF PHY-1748958.


\appendix

\addtocontents{toc}{\protect\setcounter{tocdepth}{1}} 

\section{Solutions to Equations of Motion } \label{sec:appxExact}

In this Appendix we provide several exact and semi-exact analytic solutions to the equations of motion. Section~\ref{appx:Witten} reproduces the exact solutions for $\rho(\xi)$ and $\phi(\xi)$ for the $U=0$ Witten bubble~\cite{Witten:1981gj}.
In Section~\ref{appx:Constant} we show how the solutions for $\rho(\xi)$ and $\phi(\xi)$ are modified in the presence of a constant potential, $U = U_0$. This situation arises in the $\phi \leq \phi_1$ region of the toy potential in Section~\ref{sec:modelpot}, shown in Figure~\ref{fig:potentialExamples}.  
Lastly, in Section~\ref{appx:SmallXi} we provide the small $\xi \ll R_5$ solutions for $\rho$ and $\phi$ in the presence of a potential with the asymptotic form $U(\phi) \sim U_0 \exp(a \phi/M_p)$ in the compactification limit. These corrections are particularly relevant to the numeric calculation of the bounce solutions, which depend sensitively on the initial conditions defined at $\xi_\text{init} \ll R_5$.

\subsection{Witten Bubble of Nothing} \label{appx:Witten}

In the absence of a potential, the quantity $(\rho^3 \phi')$ is exactly conserved, and the equations of motion for $\phi$ and $\rho$ can be expressed as
\begin{align}
\rho^3 \phi' = M_p \sqrt{\frac{3}{2} } \eta^3 R_5^2,
&&
\rho'^2 = 1  + \left( \frac{\eta^3 R_5^2}{2 \rho^2} \right)^2,
\label{eq:WittenEOM}
\end{align}
where from the perspective of the four dimensional equations of motion, $(\eta^3 R_5^2)$ is simply an integration constant associated with $(\rho^3 \phi')$.
Both equations can be integrated, using
\begin{align}
d\xi = \frac{d\rho}{+\sqrt{ 1 + \left( \frac{\eta^3 R_5^2}{2 \rho^2} \right)^2 } } ,
\label{eq:dxidrho}
\end{align}
where the $+$ sign is appropriate for initial conditions with $\rho' > 0$. By inspecting the $\rho'$ equation of motion from \eqref{eq:WittenEOM}, it is clear that $\rho'^2$ is bounded from below by $\rho'^2 = 1$, meaning that $\rho'$ cannot change sign for $0 \leq \rho < \infty$.

\eqref{eq:dxidrho} can be integrated to produce an exact solution for $\xi(\rho)$. Imposing $\rho(0) = 0$ so that there is no conical singularity at $\xi = 0$,  
\begin{align}
\xi(\rho)&= \frac{2\rho^3  }{3\eta^3 R_5^2} \, {_2 F_1}\!\!\left.\left( \begin{array}{c} \tfrac{1}{2} , \tfrac{3}{4} \\ \tfrac{7}{4} \end{array} \right| -\frac{4 \rho^4}{\eta^6 R_5^4}  \right) .
\end{align}
In the $\rho\ll R_5$ or $\rho \gg R_5$ limits, the series expansions for $\xi(\rho)$ can be inverted to produce $\rho(\xi)$, which is often more useful.

Rather than finding $\phi(\xi)$, the simpler analytic expression is $\phi(\rho)$, which we derive by integrating
\begin{align}
d\phi = \frac{  M_p d\rho }{\sqrt{ 1 + \frac{\eta^6 R_5^4}{4 \rho^4} } } \sqrt{ \frac{3}{2}} \frac{\eta^3 R_5^2}{\rho^3} ,
&&
\sqrt{ \frac{2}{3} } \frac{\phi - \phi_0}{M_p}  = -\arcsinh \left( \frac{ \eta^3 R_5^2}{2 \rho^2} \right) ,
\end{align}
where we introduce a second integration constant, 
\begin{align}
\phi_0 \equiv M_p \sqrt{ \frac{3}{2}} \log \eta.
\end{align}

\paragraph{Series Expansions:}

Near the center of the bubble core, $\xi \ll R$, the function $\xi(\rho)$ can be inverted through its series expansion, 
\begin{align}
\rho(\xi \ll R_5) &= \eta  R_5 \left( \frac{ 3 \xi}{2  R_5} \right)^{1/3} \left( 1 + \frac{2 \eta^{-2}}{7 } \left( \frac{ 3   \xi}{2  R_5} \right)^{4/3} - \frac{30 \eta^{-4}}{539} \left( \frac{ 3 \xi}{2 R_5} \right)^{8/3} + \mathcal O(\xi^4/R_5^4)  \right) ,
\label{eq:smallxirho}
\end{align}
and $\phi(\rho)$ can be expressed as a function of $\xi$,
\begin{align}
\phi(\xi \ll R_5)  &=   M_p \sqrt{\frac{2}{3}} \left( \log\left( \frac{3 \xi }{2R_5} \right) - \frac{9 \eta^{-2}}{14} \left( \frac{3\xi}{2R_5} \right)^{4/3} + \frac{531 \eta^{-4}}{2156} \left( \frac{3\xi}{2R_5} \right)^{8/3}  + \mathcal O(\xi^4/R_5^4)  \right) .
\label{eq:smallxiphi}
\end{align}
As a consequence of the singular $\xi = 0$ initial conditions, the $(\xi / R_5)^{4/3}$ corrections produce $\mathcal O(1)$ terms in the equations of motion even in the $\xi \rightarrow 0$ limit.

Well away from the bubble center, $\xi \gg R_5$, $\rho(\xi)$ and $\phi(\xi)$ are approximated by
\begin{align}
\rho(\xi \gg R_5) &= \xi + \eta^{3/2} R_5 \frac{\Gamma \left( \frac{3}{4} \right)^2 }{\sqrt{ 2\pi} }  + \frac{ \eta^6 R_5^4}{8 \xi^3} + \mathcal O(R^5/\xi^4)  \\
\phi(\xi \gg R_5) &= \phi_0 - M_p  \sqrt{ \frac{3}{2} } \left( \frac{ \eta^3 R_5^2}{2 \xi^2} - \frac{\Gamma\left( \frac{3}{4} \right)^2 \eta^{9/2} R_5^3 }{\sqrt{ 2\pi} \xi^3 }  + \mathcal O(R_5^4 / \xi^4) \right) .
\end{align}
As $\xi \rightarrow \infty$, $\rho \approx \xi$ grows linearly, while $\phi \rightarrow \phi_0$.

\subsection{Constant Potential} \label{appx:Constant}

As established in \eqref{eq:bouncecondition}, $(\rho^3 \phi')$ is conserved if the potential is constant, even if $U(\phi)$ is nonzero,
\begin{align}
U(\phi) = U_0 .
\end{align}
The equations of motion can be solved with BON initial conditions, this time with
\begin{align}
\rho'^2 = 1 + \left( \frac{\eta^3 R_5^2 }{2 \rho^2} \right)^2 - \frac{ \rho^2}{\Lambda_0^2},
&&
d\xi = \frac{d\rho}{+ \sqrt{  1 + \left( \frac{\eta^3 R_5^2 }{2 \rho^2} \right)^2 - \frac{ \rho^2}{\Lambda_0^2} } } ,
\label{eq:dxidrhoUconstant}
\end{align}
where in analogy with $\Lambda$ for the de~Sitter vacuum we have defined
\begin{align}
\Lambda_0 \equiv \sqrt{ \frac{3 M_p^2}{U_0} } .
\end{align}
The exact solution to the equations of motion can be expressed in terms of elliptic functions. It can also be expanded in powers of $R_5/ \Lambda_0$, which is helpful for capturing the leading perturbations to $\xi(\rho)$ and $\phi(\rho)$ in the limit $\rho \ll \ell$. 
To first order in $(R_5^2 / \Lambda_0^2)$ and $(\rho^2 / \Lambda_0^2)$, 
\begin{align}
&\xi \simeq  \frac{2 \rho^3}{3\eta^3 R_5^2} \, {_2 F_1}\! \left. \left( \begin{array}{c} \frac{1}{2} ~~~ \frac{3}{4} \\ \frac{7}{4} \end{array} \right| - \frac{4 \rho^4}{\eta^6 R_5^4} \right) 
+ \frac{5 \eta^3 R_5^2 }{24 \Lambda_0^2}\rho   \left( \frac{1 + \frac{8 \rho^4}{5 \eta^6 R_5^4} }{\sqrt{ 1 + \frac{4 \rho^4}{\eta^6 R_5^4}} } - \, {_2 F_1}\! \left. \left( \begin{array}{c} \frac{1}{4} ~~~ \frac{1}{2} \\ \frac{5}{4} \end{array} \right| - \frac{4 \rho^4}{\eta^6 R_5^4} \right) \right)
\label{eq:WittenxirhoExact} \\
&\sqrt{ \frac{2}{3} } \frac{\phi - \phi_0}{M_p  } \simeq - \arcsinh\left( \frac{\eta^3 R_5^2}{2 \rho^2} \right)+ \frac{\eta^3 R_5^2}{4\Lambda_0^2} \arcsinh \frac{2 \rho^2}{\eta^3 R_5^2} - \frac{\rho^2}{2 \Lambda_0^2}   \left(1 + \frac{4 \rho^4}{\eta^6 R_5^4} \right)^{-1/2} .
\end{align}
The series expansions in $\rho/\Lambda_0 \ll 1$ can be extended to arbitrarily higher order by integrating the series expansion of \eqref{eq:dxidrhoUconstant}.
For $R_5 \ll \Lambda_0$, the $\rho \gg R_5$ behavior of the solutions is approximately de~Sitter, with $\rho' \approx \sqrt{1 - \rho^2/\ell^2}$.
Only if $R_5 \sim \Lambda_0$ is it necessary to abandon the series expansions in favor of the elliptic function solution for $\xi(\rho)$.

\subsection{Small $\xi$ Limits} \label{appx:SmallXi}

Given some nonzero potential, the small $\xi$ expansions for $\rho(\xi)$ and $\phi(\xi)$ generically include potential-dependent terms not present in \eqref{eq:smallxirho} and \eqref{eq:smallxiphi}. This is clear from \eqref{appx:Constant}. More generically, for a potential with $\phi \rightarrow -\infty$ asymptotic behavior
\begin{align}
U(\phi) \approx U_0 \exp( a \phi/M_p)
\end{align}
with $ a > -\sqrt{6}$ (here we consider only the case of an internal shrinking $S^1$, or $n=1$ in the notation of Section~\ref{sec:bubbleexist}), approximate solutions for $\rho$ and $\phi$ can be derived by linearizing the equations of motion in the $\xi \ll R_5$ limit.
The leading $U_0$ dependent perturbations enter the series expansion at order $(\xi/R_5)^p$ with
\begin{align}
p \equiv 2 \left(1 + \frac{a}{\sqrt{6}} \right),
\end{align}
with the solutions
\begin{align}
\rho(\xi \ll R_5) &\simeq  \eta  R_5 \left( \frac{ 3 \xi}{2  R_5} \right)^{1/3} \Bigg( 1 + \frac{2 \eta^{-2}}{7 } \left( \frac{ 3   \xi}{2  R_5} \right)^{4/3} - \frac{30 \eta^{-4}}{539} \left( \frac{ 3 \xi}{2 R_5} \right)^{8/3}  \nonumber\\& \hspace{6.5cm} +  \frac{4 (1 - p)}{9p (p+1) } \frac{R_5^2 U_0}{M_p^2}  \left( \frac{3 \xi}{2 R_5} \right)^p 
\Bigg) , 
\label{eq:rhoxiU0}\\
\phi(\xi \ll R_5) &\simeq M_p \sqrt{\frac{2}{3}} \Bigg( \log\left( \frac{3 \xi }{2R_5} \right) - \frac{9 \eta^{-2}}{14} \left( \frac{3\xi}{2R_5} \right)^{4/3} + \frac{531 \eta^{-4}}{2156} \left( \frac{3\xi}{2R_5} \right)^{8/3} \nonumber\\& \hspace{6.5cm} + \frac{2(3-p)}{3p(p+1)} \frac{R_5^2 U_0}{M_p^2} \left( \frac{3\xi}{2 R_5} \right)^p 
\Bigg) . \label{eq:phixiU0}
\end{align}
The effect of the potential can be treated as a small perturbation in the $\xi \ll R_5$ limit as long as $a > - \sqrt{6}$ ($p > 0$).
In the small $R_5$ limit, $R_5^2 U_0 \ll M_p^2$, the impact of the potential is further suppressed, which is why we have terminated the series expansion at $R_5^2 U_0 M_p^{-2} (\xi / R_5)^p$ while keeping the $(\xi/R_5)^{8/3}$ contribution from the small $\xi$ expansion of the Witten solution.
As the point-and-shoot numeric solution is highly sensitive to the initial conditions at $\xi = \xi_\text{init}$, the $(\xi/R_5)^{8/3}$ level of precision is needed in some cases even when $\xi_\text{init} \ll R_5$.

Working at linear order in $R_5^2 U_0 / M_p^2$, the small $\xi$ solutions for $\rho$ and $\phi$ are easily extended to potentials of the form
\begin{align}
U(\phi) \approx \sum_j U_j \exp( a_j \phi/M_p),
\end{align}
if the asymptotic form of the potential includes multiple terms with similar $a_j$. In this case, the potential-dependent corrections to the Witten solutions for $\rho(\xi)$ and $\phi(\xi)$ can be collected into
\begin{align}
\delta\rho(\xi \ll R_5) &\simeq \eta  R_5 \left( \frac{ 3 \xi}{2  R_5} \right)^{1/3}  \left( \sum_j   \frac{4 (1 - p_j)}{9p_j (p_j+1) } \frac{R_5^2 U_j}{M_p^2}  \left( \frac{3 \xi}{2 R_5} \right)^{p_j} \right) ,
\\
\delta\phi(\xi \ll R_5) &\simeq   M_p \sqrt{\frac{2}{3}} \left( \sum_j   \frac{2(3-p_j)}{3p_j(p_j+1)} \frac{R_5^2 U_j}{M_p^2} \left( \frac{3\xi}{2 R_5} \right)^{p_j}  \right)
\end{align}
where
\begin{align}
p_j \equiv 2 \left(1 + \frac{a_j}{\sqrt{6}} \right)
\end{align}
for each term in the sum.

\subsubsection*{Perturbation to $\eta$}

At the same level of precision, it is possible to write the leading $U_0$ dependence as a multiplicative correction to the BON solutions for $\rho$ and $\phi'$:
\begin{align}
\rho(\xi) &\approx \rho_\text{bon}^{(\eta)}(\xi) \left( 1 +\frac{4 (1 - p)}{9p (p+1) } \frac{R_5^2 U_0}{M_p^2}  \left( \frac{3 \xi}{2 R_5} \right)^p \right) \\
\phi'(\xi) &\approx \phi'_\text{bon}(\xi) \left( 1 + \frac{2(3-p)}{3(p+1)} \frac{R_5^2 U_0}{M_p^2} \left( \frac{3\xi}{2 R_5} \right)^p \right) \\
\rho^3 \phi' &\approx  \sqrt{ \frac{3}{2}} M_p \eta^3 R_5^2 \left( 1 + \frac{4 - 2p}{3p} \frac{R_5^2 U_0}{M_p^2} \left( \frac{3\xi}{2 R_5} \right)^p \right) .
\label{eq:quantityN}
\end{align}
This is a convenient way to approximate the correction to $\eta$ caused by the presence of the exponentially growing potential in the $\xi \ll R_5$ limit.
In the analytic model of Fig.~\ref{fig:potentialExamples} with $U = U_0$ for $\phi \leq \phi_1$, the constant potential guarantees that $(\rho^3 \phi')$ should be conserved. Incidentally,  $U = U_0$ corresponds to the $p=2$ special case where the leading small $\xi$ correction to $\rho^3 \phi'$ vanishes, as it must.

Given a particular value of $(\rho^3 \phi')$ at $\phi = \phi_1$, and imposing continuity in $\phi(\xi)$ and $\phi'(\xi)$, the solution for $\phi(\xi \geq \xi_1)$ is given in general by the linear combination of Bessel functions
\begin{align}
\phi(\rho \geq \rho_1) \simeq c_1 \frac{K_1(m \rho) }{m \rho} + c_2 \frac{I_1(m \rho) }{m\rho} .
\label{eq:besselphirho}
\end{align}
If $c_2$ is nonzero, then whether this solution is an overshoot or undershoot can be read off from the sign of $c_2$. For the bounce solution, $c_2$ must be exponentially suppressed.
Imposing $c_2 \simeq 0$ as a requirement on the initial conditions, the value of $\eta(R_5)$ for a given $R_5$ must satisfy
\begin{align}
\log \eta & \simeq \left. \left(\sqrt{ \frac{2}{3} } \frac{(\rho^3 \phi') }{M_p} \right) \right|_{\xi = \xi_1} \frac{m^2}{4} \log \left( \frac{ 2 e^{-\gamma_E}}{m \rho_1} \right) \\
\eta - 1 &\approx \eta^3 R_5^2 \left( 1 + \frac{4 - 2p}{3p} \frac{R_5^2 U_0}{M_p^2} \left( \frac{3\xi_1}{2 R_5} \right)^p \right) \frac{m^2}{4} \left( \log\frac{1}{m R_5} + \mathcal O(1) \right)
\end{align}
as in \eqref{eq:logeta}, but where we have left the solution for $\eta$ in terms of the value of $(\rho^3 \phi')$ at $\xi_1$.
The $\eta \simeq 1$ expansion is still appropriate, with the leading terms
\begin{align}
\eta \approx 1 + \frac{m^2 R_5^2}{4} \log \frac{1}{m R_5} + \mathcal O(m^2 R_5^2)
\end{align}
unchanged, assuming the $R_5^2 U_0 (\xi_1 /R_5)^p$ term remains small.
This is achieved when  $\xi_1 \lesssim 2 R_5 / 3$, which in turn requires $\phi_1 \gtrsim 0.36\, M_p$. 
In this limit the transition from the BON-like initial condition to the approximately quadratic $\phi \sim \phi_\text{fv}$ regime occurs promptly. If instead $\xi_1 \gtrsim 2 R_5/3$, then there may be an intermediate region where $\phi(\xi)$ is neither described by the Witten BON nor the Bessel function solution of \eqref{eq:besselphirho}, and the prediction for $(\eta - 1)$ is only reliable if $R_5^2 U_0 / M_p^2$ is very small.

\subsection{Piecewise Linear Model} \label{appx:Linear}

The analytic model of Section~\ref{sec:analyticbonfv} is not the only potential with tractable equations of motion.
If the potential is approximately linear in the region where $\rho' \simeq 1$, analytic solutions for $\rho$, $\phi$ and $\eta$ can be derived in a similar manner as in the piecewise quadratic model of Figure~\ref{fig:potentialExamples}.
Models of this form are relevant especially at larger values of $R_5^2 U_0$, where the potential becomes important before $\phi(\xi)$ is close enough to the false vacuum that $U(\phi)$ is approximately quadratic. 
Labeling the point where the solution for $\phi$ deviates from the Witten solution as $\phi_\star$, in analogy with \eqref{eq:phistar},
we can expect a linear model to be particularly useful when $\phi_\star$ happens to fall in the neighborhood of an inflection point in $U(\phi)$.

A characteristic example of the linear model interpolates between $U(\phi_1) = U_0$ and $U(\phi_\text{fv}) = U_\text{fv}$ with constant slope: for example,
\begin{align}
\left\{ \begin{array}{l l l } 
U(\phi) = U_0 \exp(a (\phi - \phi_1) / M_p)	& \hspace{5mm} &	\phi \leq \phi_1 \\
U(\phi) = U_0 +  (\Delta U / \Delta \phi)  (\phi - \phi_1) 	& & \phi_1 \leq \phi \leq \phi_\text{fv} \\
U(\phi) = U_\text{fv} 	&&  \phi_\text{fv} \leq \phi 
\end{array}
\right.
, 
\label{eq:piecewiselinear}
\end{align}
where
\begin{align}
\Delta U \equiv  U_0 - U_\text{fv},
&&
\Delta \phi \equiv \phi_\text{fv} - \phi_1.
\end{align}
As long as $a \gtrsim - \sqrt{6}$ and $R_5^2 U_0 \lesssim M_p^2$,  \eqref{eq:quantityN} indicates that $(\rho^3 \phi')$ will remain approximately constant for $0 \leq \xi \lesssim R_5$.
For simplicity we specialize to the $a=0$ case, where $U(\phi \leq \phi_1) = U_0$ remains exactly constant, and $(\rho^3 \phi')$ is conserved,
\begin{align}
\rho^3 \phi' = M_p \sqrt{ \frac{3}{2}} \eta^3 R_5^2 ,
&&
\rho' = \sqrt{1 + \frac{\eta^6 R_5^4}{4 \rho^4} - \frac{\rho^2}{\Lambda_0^2}} ,
\end{align}
where for convenience we define 
\begin{align}
\Lambda_0 \equiv \sqrt{ \frac{3 M_p^2}{U_0} } ,
&&
\Lambda \equiv \sqrt{ \frac{3 M_p^2}{U_\text{fv}} } .
\end{align}
Up to corrections of order $R_5^2/\Lambda_0^2$, $\rho$ and $\phi$ are approximately equal to the $U=0$ Witten solutions.
The values of $\xi_1$ and $\rho_1 = \rho(\xi_1)$ defined by $\phi(\xi_1) \equiv \phi_1$ can be obtained from \eqref{eq:WittenxirhoExact}
If $(\phi_\text{fv} - \phi_1) \lesssim M_p$, then $\rho_1 \gtrsim R_5$, and $\xi_1$ can be determined more simply from the large $\xi$ expansions of \eqref{eq:WittenxirhoExact}.

The primary differences between the linear and quadratic models appear, of course, in the transition region $\xi_1 \leq \xi \leq \xi_2$, where $U(\phi)$ and $(\rho^3 \phi')$ change in value.
If $\rho_1 \gtrsim R_5$ and $\rho_2 \ll \Lambda_0$, then the solution for $\rho$ throughout the entire transition is well described by 
\begin{align}
\rho'\simeq 1,
&&
\rho(\xi_1 \leq \xi \leq \xi_2) \approx \xi + \gamma \eta^{3/2} R,
\end{align}
for the $\gamma \approx 0.6$ defined in \eqref{eq:gammadefinition}. 
Here $\xi_2$ refers to the end of the transition, where $\phi(\xi_2) = \phi_\text{fv}$.
Unlike the quadratic model, the interval $\xi_2 - \xi_1$ is precisely defined, because $\phi(\xi)$ approaches the false vacuum in finite time. The value of $\xi_2$ is found by integrating $\phi'$:
\begin{align}
\phi(\xi_1 \leq \xi \leq \xi_2) & \simeq \phi_1 + \frac{ (\xi - \xi_1) (\xi - \xi_1 + 2 \rho_1)}{8 \rho_1^2 (\xi - \xi_1 + \rho_1)^2 } \bigg( 4 \sqrt{ \frac{ 3}{2} }  M_p \eta^3 R_5^2
\nonumber\\ & \hspace{4cm} -  \frac{\Delta U}{\Delta \phi} \rho_1^2 (\xi - \xi_1) \left( \xi - \xi_1 + 2 \rho_1 \right) \bigg) ,
\label{eq:phitransition}
\\
\phi(\xi_2 \leq \xi \leq \xi_\text{max}) & = \phi_\text{fv}.
\end{align}

For the linear model, the bounce condition \eqref{eq:bouncecondition} becomes quite simple:
\begin{align}
\left( \rho^3 \phi' \right) \Big|_{\xi = \xi_1}^{\xi = \xi_2}  &= \frac{\Delta U}{\Delta \phi} \int_{\xi_1}^{\xi_2} \! d\xi\, \rho^3 \\
\sqrt{ \frac{3}{2}} M_p \eta^3 R_5^2 &\simeq  \frac{-\Delta U}{\Delta \phi} \int_{\rho_1}^{\rho_2} \! d\rho\, \rho^3 =  \frac{-\Delta U}{\Delta \phi} \frac{\rho_2^4 - \rho_1^4}{4}.
\end{align}
Assuming a nonzero value of $U_\text{fv}$, the value of $(\rho^3 \phi')$ must vanish as $\xi \rightarrow \xi_2$. Otherwise, $\phi(\xi)$ would continue moving in the positive direction with constant $U(\phi)$, while $\rho \rightarrow \Lambda$. Once $\rho$ decreases from its maximum, the value of $\phi'$ would begin to increase again, ultimately approaching $\phi' \rightarrow \infty$ as $\rho \rightarrow 0$.

In the $U_0 R_5^2 \ll M_p^2$ limit ($R_5 \ll \Lambda_0$), the solution for $\eta$ is approximately
\begin{align}
\log \eta &\approx  \frac{ \eta^{3/2} R_5}{\Lambda_0}   \left( \frac{1}{2} \sqrt{ \frac{3}{2} } \frac{M_p}{\Delta \phi} \right)^{1/2} \left( \sqrt{ 1 - \frac{U_\text{fv}}{U_0} } + \mathcal O(R_5 / \Lambda_0) \right) \\
\eta &\approx 1 + \left( \frac{R_5^2 }{2 \sqrt{6} M_p } \left( \frac{ U_0 - U_\text{fv}}{\Delta \phi} \right)  \right)^{1/2} .
\end{align}
Unlike \eqref{eq:eta} from the quadratic model, where $(\eta - 1) \sim m^2 R_5^2 \log(m R_5)$, this correction to $\eta-1$ is linear in $R_5$.
Calculating the bounce action $\Delta S$, we find that the leading correction to $\Delta S = \pi^2 M_p^2 R_5^2$ is also quadratic in $R_5$:
\begin{align}
\Delta S &\approx \pi^2  M_p^2 \eta^3 R_5^2 \left( 1 - \sqrt{6}\frac{  \Delta \phi}{M_p} + \mathcal O\left(U_0 R_5^2 / M_p^2\right)  \right) \\
\Delta S &\approx \pi^2  M_p^2  R_5^2 \left( 1  -\sqrt{6} \frac{   \Delta \phi}{M_p}  + \left( \frac{3}{2} \right)^{3/4} \frac{R_5}{M_p}  \sqrt{ \frac{U_0 - U_\text{fv} }{\Delta \phi / M_p} } \right) .\label{eq:lineardeltaS}
\end{align}
So, the corrections to $\eta$ and $\Delta S$ in the small $R_5$ limit are both large compared to the analogous terms of the piecewise quadratic model.

The linear model may be useful in cases where $\phi(\xi)$ diverges from $\phi_\text{bon}$ at values of $\phi_\star$ far away from the false vacuum. 
If $U(\phi \approx \phi_\star)$ is approximately linear in this region, then we anticipate corrections to the $\pi^2 M_p^2 R_5^2$ action  that are quadratic in $R_5$, as in \eqref{eq:lineardeltaS}. 
The fact that the $U \sim \text{(linear)}$ equations of motion can be integrated when $\rho' \simeq 1$ also permits the assembly of complicated piecewise models, composed of linear and quadratic sections, that can more accurately reconstruct a given $U(\phi)$. 
As long as $\rho' \simeq 1$ throughout the transition across the potential barrier, solutions for $\phi$ and $\rho$ can be found by stitching together consecutive instances of \eqref{eq:besselphirho} and \eqref{eq:phitransition}.

\section{Analytic Continuation}

Expanding the line element of the unit three-sphere as $d \Omega_3^2 = d \psi^2 + \sin(\psi)^2 d \Omega_2^2$, the $O(4)$-symmetric ansatz of \eqref{eq:cdlmetric} can be written as
\begin{equation}
	ds_E^2	 = d \xi^2 + \rho(\xi)^2 \left( d \psi^2 + \sin(\psi)^2 d \Omega_2^2 \right) .
\end{equation}
Performing the analytic continuation $\psi \rightarrow \pi / 2 + i \chi$, with $\chi \in \mathbb{R}$, we obtain the Lorentzian-signature metric
\begin{equation}
	ds^2 = - \rho(\xi)^2 d \chi^2 + d \xi^2 + \rho(\xi)^2 \cosh(\chi)^2 d \Omega_2^2 .
\label{eq:ds_Lorentz_v1}
\end{equation}
The plane of symmetry of the metric given by $\psi = \pi /2$ plays the role of $t_E = 0$, and the continuation $\psi \rightarrow \pi / 2 + i \chi$ is equivalent to $t_E \rightarrow i t$.

For Witten's BON, in the regime $r \gg R_5$, we have $\rho(\xi) \simeq r \simeq \xi + \gamma R_5$, and the previous equations reads
\begin{align}
	ds^2 	& = - r^2 d \chi^2 + d r^2 + r^2 \cosh(\chi)^2 d \Omega_2^2 \\
		& = - d \tilde t^2 + d \tilde r^2 + \tilde r^2 d \Omega_2^2 ,
\end{align}
where in the second step we have used the new coordinates:
\begin{equation} \begin{cases}
	\tilde r \equiv r \cosh(\chi) , \\
	\tilde t \equiv r \sinh(\chi) .
\end{cases} \end{equation}
As discussed in \cite{Witten:1981gj}, the BON can be regarded as a kind of distorted Minkowski spacetime, subject to the restriction $r^2 = \tilde r^2 - \tilde t^2 \geq R_5^2$.

Our BON solution, on the other hand, asymptotes to the 4D de Sitter vacuum, as given in \eqref{eq:rho_dS_BON}. The appropriate change of coordinates in this asymptotic regime is of the form
\begin{equation} \begin{cases}
	\tilde r \equiv \Lambda \sin \left( \frac{\xi + \gamma R_5}{\Lambda} \right) \cosh(\chi) , \\
	\tilde t \equiv \Lambda \arctanh \left[ \tan \left( \frac{\xi + \gamma R_5}{\Lambda} \right) \sinh(\chi) \right] .
\end{cases} \end{equation}
\eqref{eq:ds_Lorentz_v1} then reads
\begin{equation}
	d s^2 \simeq - \left( 1 - \frac{\tilde r^2}{\Lambda^2} \right) d \tilde t^2 + \left( 1 - \frac{\tilde r^2}{\Lambda^2} \right)^{-1} d \tilde r^2 + \tilde r^2 d \Omega_2^2 ,
\end{equation}
which indeed corresponds to the static description of de Sitter spacetime.

\section{Specific Potential}  \label{appx:potential}

For the numeric evaluation of the equations of motion, we specialize to potentials of the form
\begin{align}
U(\phi) = U_0 \sum_j b_j e^{j \, a \phi/M_p}, 
\label{eq:potentialUform}
\end{align}
where the parameter $a$ controls the overall scaling of the potential with $\phi$, and the coefficients $b_j$ are such that there is a local minimum $U = U_\text{fv} >0$ at $\phi_\text{fv}$, and a local maximum of $U = U_\text{HM}$ at some $\phi_\text{HM} < \phi_\text{fv}$.
The simplest potentials in this family require only three nonzero $b_j$, for example $j=1,2,3$.

The action  is invariant with respect to shifts in $\phi$, and we use this freedom to set $\phi_\text{fv} = 0$.
The equations of motion also do not depend on the overall normalization of $U(\phi)$: a factor of $U_0$ can be absorbed into a redefinition of the coordinates $\xi$ and $\rho$, which rescales the action by an overall constant:
\begin{align}
\xi \rightarrow \sqrt{U_0/M_p^2}\, \xi,
&&
\rho \rightarrow \sqrt{U_0/M_p^2}\, \rho,
&&
S \rightarrow (M_p^4/ U_0)\, S.
\end{align} 
In this way the four independent parameters $(a, b_{1}, b_2, b_3)$ can be reduced to two parametric degrees of freedom: the original $a$, and a parameter $\delta$ defined as
\begin{align}
\delta \equiv \frac{U_\text{fv}}{U_0}.
\end{align}
The coefficients $b_{j}$ in this scheme are given by
\begin{align}
\frac{U(\phi)}{U_0} =  \beta_0 \left( \beta_1 e^{a \phi} - \frac{1 + \beta_1}{2} e^{2 a \phi} + \frac{1}{3}e^{3a\phi}  \right),  
\label{eq:UadeltaC}
\end{align}
where $\beta_0$ and $\beta_1$ satisfy
\begin{align}
\beta_0 &= \frac{3(9 - 5 \delta)}{4} + \frac{9}{4} \left( (3 + \sqrt{\delta} )(1 - \sqrt{\delta} )^{2/3} (1 + \sqrt{\delta})^{1/3} + (3 - \sqrt{\delta} )(1 + \sqrt{\delta} )^{2/3} (1 - \sqrt{\delta})^{1/3}  \right) ,
\nonumber\\
\beta_1 &= \frac{2\delta}{\beta_0} + \frac{1}{3} ,
\end{align}
for $0 \leq \delta \leq 1$. The solutions for $\beta_0$ and $\beta_1$ can in principle be extended simply to $\delta < 0$ via $\sqrt{\delta} \rightarrow i \sqrt{ - \delta}$.

The location of the center of the barrier, $\phi_\text{top}$, is given by 
\begin{align}
\frac{ a\, \phi_\text{top} }{M_p} =  \log(\beta_1) ,
\label{eq:exactphihm}
\end{align}
and the Hawking--Moss solution is simply $\phi(\xi) = \phi_\text{top}$.

\bibliography{dS_decays_refs}
	
\end{document}